\documentclass[oneside]{sn-jnl}
\usepackage[T1]{fontenc}
\usepackage[utf8]{inputenc}
\usepackage{booktabs}
\usepackage{amsmath}
\usepackage{amsthm}
\usepackage{amssymb}
\usepackage{stackrel}
\usepackage{graphicx}
\usepackage[numbers]{natbib}

\makeatletter

\usepackage{xfrac}

\providecommand{\tabularnewline}{\\}

\numberwithin{equation}{section}
\numberwithin{figure}{section}

\usepackage{etoolbox}
\usepackage{hyperref}
\usepackage[nottoc,numbib]{tocbibind}
\newenvironment{widetext}{}{}

\ifdefined\showcaptionsetup
 \PassOptionsToPackage{caption=false}{subfig}
\fi
\usepackage{subfig}
\makeatother

\begin{document}
\title{Numerical investigation of the interior geometry of semiclassical
evaporating spherical charged black holes}
\author{Gil Arad}
\date{January 5, 2026}
\abstract{We developed a numerical code which evolves the semiclassical Einstein's
equation (with the quantum stress-energy contribution added as a source
term) for the spherically symmetric metric inside an evaporating semiclassical
charged black hole. An analytical approximation for the evolving semiclassical
metric was recently developed by Ori and Zilberman (and will be briefly
overviewed here). We seek to numerically check the validity of this
analytical approximation. The Einstein equations in this case are
partial differential equations for the two unknown metric functions
which fully describe the spherically symmetric metric. We begin our
numerical simulation close to the event horizon with regular initial
data specified by a variant of the charged Vaidya metric. We then
evolve the metric functions deep into the neighborhood of the inner
horizon. We explore the results of running this numerical code in
several representative cases. Our numerical simulations confirm the
validity of the above mentioned analytical approximation in all these
cases.}
\maketitle

\section{Introduction}

Kerr and Reissner--Nordström (RN) geometries, describing rotating
and spherically symmetric charged black holes, contain an\emph{ inner
horizon }(IH). This IH is a null hypersurface that also serves as
a \emph{Cauchy horizon} (CH) -- which is the future boundary where
solutions to the field equations, based on initial data provided outside
the black hole, fail to be unique. Therefore predictions based on
the field equations can be made only up to the Cauchy horizon. The
geometry is regular up to and across the IH. There is a unique analytical
extension of the geometry beyond the IH, which describes a smooth
passage to other external universes \citep{carter1966complete,graves1960oscillatory}. 

Imagine an observer falling into the black hole. The geometry of the
black hole interior will necessarily force the observer to cross the
IH (because between the two horizons $r$ is timelike coordinate).
What will such an observer encounter? Can the observer cross the IH?
In the pure Kerr and RN geometries, there is a smooth passage across
the IH, leading to the mentioned other external universes. It turns
out, however, that if the black hole is perturbed by classical or
quantum matter field, the situation might change, as we will now briefly
discuss.

Classical matter fields falling into the black hole may effect the
regularity of the CH. As was previously demonstrated, the regular
geometry at the IH now becomes a null weak curvature singularity that
forms at the\emph{ }Cauchy horizon, both in RN \citep{hiscock1981evolution,poisson1990internal,ori1991inner,brady1995black,hod1998mass,burko1997structure}
and in Kerr \citep{ori1992structure,ori1999oscillatory,brady1998late,dafermos2017interior}
black holes. 

Quantum perturbation are also expected to have analogous effects on
the geometry. These may be studied using the theoretical framework
of semiclassical general relativity, considering quantum fields propagating
in classical curved spacetime. In semiclassical forumlation, the stress-energy
tensor to be inserted in the Einstein equation is the renormalized
expectation value of the stress-energy tensor (RSET) associated with
the quantum field (in addition to any classical stress-energy tensor,
if such exists--such as electromagnetic stress-energy tensor in the
case of RN background). The resulting equation is called the \emph{semiclassical
Einstein equation}. 

This equation needs to be handled self-consistently. Note that the
curvature of spacetime leads to non-zero RSET of the quantum fields,
and this RSET in turn deforms the spacetime, an effect called \emph{backreaction}. 

Solving the semiclassical Einstein equation self-consistently is obviously
a challenging task. A natural first step towards tackling this problem
is to first compute the RSET for a \emph{fixed} classical background
metric, such as Schwarzchild or RN or Kerr. However, since we are
interested in the effect of quantum fields on the IH, we shall not
consider here the Schwarzchild case. The near IH RSET, In particular
its $T_{uu}$ and $T_{vv}$ component was previously computed for
the RN case in Ref. \citep{Zilberman_2020} and for the Kerr case
in Ref. \citep{Zilberman_2022,McMaken_2024}. For closely related
papers see also \citep{hollands2020quantum,alberti2025quantumeffectsnearextremalcharged,hollands2020quantum2,klein2021renormalized,klein2021quantum,klein2024infinite,klein2024long}. 

For simplicity we restrict our attention here to spherical symmetry,
hence we shall focus in this work on the case of spherical charged
black hole. We shall investigate the backreaction effect of the RSET
in the interior of the black hole, with special attention to the near-IH\footnote{Whenever we refer to the evolving backreacted geometry, by ``near-IH''
we mean the region of very large Eddington-like null coordinates $u$
and $v$.} geometry. 

Recently an approximate analytical solution was developed by Amos
Ori and Noa Zilberman (still unpublished). This approximation is based
on the fact that astrophysical black holes have very large masses
(compared to the Planck mass) and therefore their evaporation rate
is very small. We briefly present this analytical approximation for
the semiclassically evolving metric functions below. 

The mentioned analytical approximation was developed under a key assumption
about the behavior of the RSET in the evolving metric:  our assumption
is that the asymptotic behavior found near the IH in RN is naturally
generalized to the evolving backreaction geometry, in a certain manner
that we specify at the end of subsection \ref{subsec:Choice-of-source}
below. 

Our main goal in this work is to numerically explore this issue --
more specifically, to check the validity of the above mentioned analytical
approximation. To this end, we developed a numerical code (published
as \citep{SphericallySymmetricEinsteinEqSolver}) which allows us
to evolve the semiclassical Einstein equation in this model of spherically
symmetric metric that evolves inside the evaporating semiclassical
charged black hole. We work in double-null coordinates, using the
outgoing null coordinate $u$ and the ingoing null coordinate $v$.
The resulting equations are partial differential equations for the
two unknown functions $R(u,v)$ and $S(u,v)$ (to be defined later),
which fully define the evolving spherically symmetric metric. We begin
our numerical simulation close to the EH, and evolve the Einstein's
evolution equations towards the future, until we get deep into the
near-IH region. The numerical simulation assumes a specific form for
the mass $M(v)$ of the evaporating black hole as a function of the
ingoing Eddington coordinate $v$, and that the charge-to-mass ratio
$\frac{Q}{M}$ at the EH remains constant upon evaporation -- as
well as a form for the source terms which incorporate the semiclassical
contributions to $T_{uv}$ and $T_{\theta\theta}$, and for the current
$J_{u}$ \footnote{In practice, we choose a function $Q(u,v)$ which determines the current
components $J_{u}$ and $J_{v}$}. Our numerical simulation confirms the validity of the mentioned
analytical approximation. 

\part{Theoretical overview}

\section{Theoretical background}

\subsection{Spherically symmetric Einstein equations}\label{subsec:Spherically-symmetric-Einsteins}

Begin with the following spherically-symmetric metric:

\begin{equation}
ds^{2}=-2e^{S(u,v)}\frac{L_{0}}{r(u,v)}du\;dv+r^{2}(u,v)d\Omega^{2}\label{eq:spherically symmetric metric}
\end{equation}
where $d\Omega^{2}$is defined as:

\begin{equation}
d\Omega^{2}=d\theta^{2}+\sin^{2}\theta d\phi^{2}
\end{equation}
We'll use geometrized unit system where $c=1$ and $G=1$. In this
unit system, there is an equivalence between distance, time and mass
units. This metric depends on two unknown functions: $S(u,v)$ and
$r(u,v)$. $r(u,v)$ has units of length, and so has $L_{0}$. $S(u,v)$
is dimensionless. The coordinates $u$,$v$ have dimensions of length.
We're free to choose $L_{0}$ as we wish. We define 
\begin{equation}
R(u,v)=r^{2}(u,v)
\end{equation}

Calculating Einstein's tensor for this metric, using \emph{Ricci package}
in Mathematica we get:

\begin{equation}
G_{uu}=\frac{1}{R}\left(R_{,u}S_{,u}-R_{,uu}\right)\label{eq:Guu}
\end{equation}

\begin{equation}
G_{vv}=\frac{1}{R}\left(R_{,v}S_{,v}-R_{,vv}\right)\label{eq:Gvv}
\end{equation}

\begin{equation}
G_{uv}=\frac{L_{0}e^{S}}{R^{3/2}}+\frac{R_{,uv}}{R}\label{eq:guv}
\end{equation}

\begin{equation}
G_{\theta\theta}=-\frac{1}{2L_{0}}e^{-S}\sqrt{R}\left(R_{,uv}+2RS_{,uv}\right)
\end{equation}

The Einstein equation is:

\begin{equation}
G_{\mu\nu}=8\pi T_{\mu\nu}
\end{equation}

Those equations lead to the following equations for $R_{,uv}$, $S_{,uv}$:

\begin{equation}
R_{,uv}=-\frac{L_{0}e^{S}}{\sqrt{R}}+8\pi RT_{uv}\label{eq:ruv}
\end{equation}

\begin{equation}
S_{,uv}=\frac{L_{0}e^{S}}{2R^{\frac{3}{2}}}-8\pi\left(\frac{1}{2}T_{uv}+\frac{L_{0}e^{S}}{R^{\frac{3}{2}}}T_{\theta\theta}\right)\label{eq:suv}
\end{equation}

And the following equations for $T_{vv}$, $T_{uu}$:

\begin{equation}
8\pi T_{uu}=\frac{1}{R}\left(R_{,u}S_{,u}-R_{,uu}\right)\label{eq:Tuu}
\end{equation}

\begin{equation}
8\pi T_{vv}=\frac{1}{R}\left(R_{,v}S_{,v}-R_{,vv}\right)\label{eq:Tvv}
\end{equation}

The equations for $R_{,uu}$ and $R_{,vv}$, entailed in Eqs. (\ref{eq:Tuu},\ref{eq:Tvv}),
are the \emph{constraint equations}, and the equations (\ref{eq:ruv},\ref{eq:suv})
for $R_{,uv}$,$S_{,uv}$ are the \emph{evolution equations}. 

In our model, the stress-energy tensor $T_{\mu\nu}$ is composed of
two parts, the electromagnetic contribution $T_{\mu\nu}^{(EM)}$,
and the semiclassical part $T_{\mu\nu}^{(SC)}$:

\begin{equation}
T_{\mu\nu}=T_{\mu\nu}^{(EM)}+T_{\mu\nu}^{(SC)}
\end{equation}

\subsection{Solution to Maxwell's equations}

Assume that the only non-zero component of $F^{\mu\nu}$is $F^{uv}$,
as it is required by spherical symmetry. Also assume it is a function
of only $u$,$v$ as it is required by spherical symmetry.

The homogeneous Maxwell equation is trivially satisfied:

\begin{equation}
F_{\alpha\beta;\gamma}+F_{\beta\gamma;\alpha}+F_{\gamma\alpha;\beta}=0
\end{equation}
It is equivalent to 

\begin{equation}
F_{\alpha\beta,\gamma}+F_{\beta\gamma,\alpha}+F_{\gamma\alpha,\beta}=0
\end{equation}
It is easy to see the equation is trivially satisfied. 

The source-free inhomogeneous equation is:

\begin{equation}
0=F_{\quad;\beta}^{\alpha\beta}=\frac{1}{\sqrt{|g|}}\left(\sqrt{|g|}F^{\alpha\beta}\right)_{,\beta}\label{eq:Falphabeta}
\end{equation}
Where $g$ is the determinant of the metric, satisfying 

\begin{equation}
\sqrt{|g|}=r^{2}\sin\theta|g_{uv}|
\end{equation}
Taking the $\alpha=u$ component of the equation:

\begin{equation}
\left(r^{2}|g_{uv}|F^{uv}\right)_{,v}=0
\end{equation}
Taking the $\alpha=v$ component of the equation Eq. (\ref{eq:Falphabeta})
yields an analogous result:

\begin{equation}
\left(r^{2}|g_{uv}|F^{uv}\right)_{,u}=0
\end{equation}
From these two equations we get that $r^{2}|g_{uv}|F^{uv}$ is constant,
which we call $Q$. $Q$ is the electric charge. 

Therefore,

\begin{equation}
F^{uv}=\frac{Q}{r^{2}|g_{uv}|}
\end{equation}
We lower the indices to get:

\begin{equation}
F_{uv}=g_{uv}^{2}F^{vu}=-\frac{Q}{r^{2}|g_{uv}|}g_{uv}^{2}=-\frac{Q}{r^{2}}|g_{uv}|
\end{equation}
In general, we want to have currents, so we will use non-constant
$Q:$

\begin{equation}
F_{uv}=-\frac{Q(u,v)}{r^{2}}|g_{uv}|
\end{equation}

\begin{equation}
F^{uv}=\frac{Q(u,v)}{r^{2}|g_{uv}|}
\end{equation}
The inhomogeneous Maxwell's equation with sources is:

\begin{equation}
4\pi J^{\alpha}=F_{\quad;\beta}^{\alpha\beta}=\frac{1}{\sqrt{|g|}}\left(\sqrt{|g|}F^{\alpha\beta}\right)_{,\beta}
\end{equation}
Choosing $\alpha=u$:

\begin{equation}
4\pi J^{u}=\frac{1}{r^{2}\sin\theta|g_{uv}|}\left(r^{2}\sin\theta|g_{uv}|\frac{Q(u,v)}{r^{2}|g_{uv}|}\right)_{,v}=\frac{Q_{,v}}{r^{2}|g_{uv}|}
\end{equation}
Choosing $\alpha=v$ and doing a similar calculation:

\begin{equation}
4\pi J^{v}=-\frac{Q_{,u}}{r^{2}|g_{uv}|}
\end{equation}
Lowering the indices, we get:

\begin{equation}
4\pi J_{u}=g_{uv}\frac{-Q_{,u}}{r^{2}|g_{uv}|}=\frac{Q_{,u}}{r^{2}}\label{eq:J_u}
\end{equation}

\begin{equation}
4\pi J_{v}=-\frac{Q_{,v}}{r^{2}}\label{eq:J_v}
\end{equation}

\subsection{The electromagnetic stress-energy tensor}

We can now calculate the electromagnetic stress-energy tensor for
this given electromagnetic field tensor:

\begin{equation}
T_{\mu\nu}^{(EM)}=\frac{1}{4\pi}\left(F_{\mu\lambda}F_{\nu}^{\;\lambda}-\frac{1}{4}g_{\mu\nu}F_{\lambda\kappa}F^{\lambda\kappa}\right)
\end{equation}

Noting that:

\begin{equation}
F_{\lambda\kappa}F^{\lambda\kappa}=2F_{uv}F^{uv}=-2\frac{Q}{r^{2}|g_{uv}|}\frac{Q}{r^{2}}|g_{uv}|=-\frac{2Q^{2}}{r^{4}}
\end{equation}

We obtain

\begin{align}
T_{uv}^{(EM)} & =\frac{1}{4\pi}\left(F_{uv}g^{uv}F_{vu}-\frac{1}{4}g_{uv}F_{\lambda\kappa}F^{\lambda\kappa}\right)\\
 & =\frac{1}{4\pi}\left(-\frac{Q^{2}}{r^{4}}|g_{uv}|^{2}g^{uv}-\frac{1}{4}g_{uv}F_{\lambda\kappa}F^{\lambda\kappa}\right)\\
 & =\frac{1}{4\pi}\left(-\frac{Q^{2}}{r^{4}}g_{uv}+\frac{1}{4}g_{uv}\frac{2Q^{2}}{r^{4}}\right)=-\frac{1}{8\pi}\frac{Q^{2}}{r^{4}}g_{uv}
\end{align}

Recalling that 

\begin{equation}
g_{uv}=-\frac{e^{S}L_{0}}{r}\label{eq:g_uv_e^s}
\end{equation}

We get:

\begin{equation}
T_{uv}^{(EM)}=\frac{1}{8\pi}\frac{Q^{2}e^{S}L_{0}}{r^{5}}
\end{equation}

Calculating $T_{\theta\theta}^{(EM)}$, we note that the first term
$F_{\mu\lambda}F_{\nu}^{\;\lambda}$ trivially vanishes in this case,
getting:

\begin{equation}
T_{\theta\theta}^{(EM)}=\frac{1}{4\pi}\left[-\frac{1}{4}g_{\theta\theta}\left(-\frac{2Q^{2}}{r^{4}}\right)\right]=\frac{1}{8\pi}\frac{Q^{2}}{r^{2}}
\end{equation}
Because of spherical symmetry, $T_{\phi\phi}=T_{\theta\theta}\sin^{2}\theta$,
for both the electromagnetic stress-energy tensor and the semi-classical
stress energy tensor. Note that $T_{vv}^{(EM)}=T_{uu}^{(EM)}=0$,
therefore $T_{uu}=T_{uu}^{(SC)}$ and $T_{vv}=T_{vv}^{(SC)}$. We
will hereafter refer to $T_{uu}$ and $T_{vv}$ without the ``$(SC)$''
mark.

Using Einstein's equation $G_{\mu\nu}=8\pi T_{\mu\nu}$, and assuming
the only stress-energy tensor is the electromagnetic one, the evolution
equations in the presence of charge are, based on equations Eq. (\ref{eq:ruv},\ref{eq:suv})

\begin{equation}
R_{,uv}=-\frac{e^{S}L_{0}}{\sqrt{R}}+e^{S}L_{0}\frac{Q^{2}}{R^{\frac{3}{2}}}=e^{S}L_{0}F_{1}(R)\label{eq:RuvF1}
\end{equation}

\begin{equation}
S_{,uv}=\frac{e^{S}L_{0}}{2R^{\frac{3}{2}}}-\frac{3}{2}e^{S}L_{0}\frac{Q^{2}}{R^{\frac{5}{2}}}=e^{S}L_{0}F_{2}(R)\label{eq:suv_F2}
\end{equation}
where

\begin{equation}
F_{1}(R)=-\frac{1}{\sqrt{R}}+\frac{Q^{2}}{R^{\frac{3}{2}}}
\end{equation}

\begin{equation}
F_{2}(R)=\frac{1}{2R^{\frac{3}{2}}}-\frac{3}{2}\frac{Q^{2}}{R^{\frac{5}{2}}}
\end{equation}
Notice that:

\begin{equation}
F_{2}(R)=\frac{d}{dR}F_{1}(R)\label{eq:F1ddrF2}
\end{equation}
In view of this property, the PDE system (\ref{eq:RuvF1},\ref{eq:suv_F2})
belongs to the general class of PDE systems investigated in Ref. \citep{ori2007simplified}. 

Note that we here assumed that the only stress-energy tensor is the
electromagnetic one. We will later break this assumption, by taking
into account the non-zero $T_{\mu\nu}^{(SC)}$. This will lead to
new source terms in the evolution equations, see subsection \ref{subsec:Source-terms}.

Taking the constraint equation Eq. (\ref{eq:Tvv}) we get:

\begin{equation}
\widetilde{T}_{vv}\equiv8\pi r^{2}T_{vv}=R_{,v}S_{,v}-R_{,vv}\label{eq:Tvv definition}
\end{equation}
and similarly for $T_{uu}:$

\begin{equation}
\widetilde{T}_{uu}\equiv8\pi r^{2}T_{uu}=R_{,u}S_{,u}-R_{,uu}\label{eq:TuuRuSu}
\end{equation}
Here we defined the two quantities $\widetilde{T}_{vv},\widetilde{T}_{uu}.$

The quantity $\widetilde{T}_{uu}$ satisfies a conservation equation
: 

\begin{equation}
\left(\widetilde{T}_{uu}\right)_{,v}=0\label{eq:Tuu,v}
\end{equation}
To prove this, we differentiate Eq. (\ref{eq:TuuRuSu}) with respect
to $v$, and use the evolution equations Eq. (\ref{eq:ruv},\ref{eq:suv}):

\begin{widetext}

\begin{align}
\left(\widetilde{T}_{uu}\right)_{,v} & =R_{,uv}S_{,u}+R_{,u}S_{,uv}-\left(R_{,uv}\right)_{,u}\\
 & =e^{S}L_{0}F_{1}(R)S_{,u}+e^{S}L_{0}F_{2}(R)R_{,u}-\left[e^{S}L_{0}F_{1}(R)\right]_{,u}\\
 & =e^{S}L_{0}F_{1}(R)S_{,u}+e^{S}L_{0}F_{2}(R)R_{,u}-e^{S}L_{0}F_{1}(R)S_{,u}-e^{S}L_{0}\frac{dF_{1}(R)}{dR}R_{,u}=0
\end{align}
\end{widetext}In the last equation we used the relation Eq. (\ref{eq:F1ddrF2})
between $F_{1}$ and $F_{2}$. Note that we assumed $Q$ is constant
at this stage. When $Q$ changes the situation changes and we address
it in section \ref{subsec:Choice-of-source}.

In the same manner, we can derive the analogous conservation equation
for $\widetilde{T}_{vv}$ :

\begin{equation}
\left(\widetilde{T}_{vv}\right)_{,u}=0\label{eq:Tvv,u}
\end{equation}

Note, however, that when we add the semiclassical source terms Eqs.
(\ref{eq:Tuu,v}, \ref{eq:Tvv,u}) will not be valid.

\subsection{Reissner--Nordström metric}

The Reissner--Nordström metric is: \begin{widetext}

\begin{equation}
ds^{2}=-\left(1-\frac{2M}{r}+\frac{Q^{2}}{r^{2}}\right)dt^{2}+\left(1-\frac{2M}{r}+\frac{Q^{2}}{r^{2}}\right)^{-1}dr^{2}+r^{2}d\Omega^{2}
\end{equation}
\end{widetext}$r_{+},r_{-},\kappa_{+},\kappa_{-}$ are defined as
follows:

\begin{equation}
r_{+}=M+\sqrt{M^{2}-Q^{2}}
\end{equation}

\begin{equation}
r_{-}=M-\sqrt{M^{2}-Q^{2}}
\end{equation}

\begin{equation}
\kappa_{+}=\frac{r_{+}-r_{-}}{2r_{+}^{2}}=\frac{\sqrt{M^{2}-Q^{2}}}{\left(\sqrt{M^{2}-Q^{2}}+M\right)^{2}}
\end{equation}

\begin{align}
\kappa_{-} & =\frac{r_{+}-r_{-}}{2r_{-}^{2}}=\frac{\sqrt{M^{2}-Q^{2}}}{\left(M-\sqrt{M^{2}-Q^{2}}\right)^{2}}\label{eq:kappa_minus_definition}\\
 & =\frac{1}{M}\frac{\sqrt{1-\left(\frac{Q}{M}\right)^{2}}}{\left(1-\sqrt{1-\left(\frac{Q}{M}\right)^{2}}\right)^{2}}
\end{align}
Note that:

\begin{equation}
f\equiv\left(1-\frac{2M}{r}+\frac{Q^{2}}{r^{2}}\right)=\frac{\left(r-r_{-}\right)\left(r-r_{+}\right)}{r^{2}}\label{eq:r-r-*r-r+}
\end{equation}
where we defined $f(r)$. The horizons are the surfaces where $g_{uv}=0$.
They are $r=r_{+}$at the event horizon (EH) and $r=r_{-}$ at the
inner horizon (IH). 

\subsection{Ingoing Eddington coordinates}

We wish to move to different coordinates. For this purpose we define
$r_{*}$:

\begin{equation}
r_{*}=r-M+\frac{1}{2\kappa_{+}}\log\left|\frac{r-r_{+}}{M-r_{+}}\right|-\frac{1}{2\kappa_{-}}\log\left|\frac{r-r_{-}}{M-r_{-}}\right|
\end{equation}
Note that:

\begin{equation}
\frac{\partial r_{*}}{\partial r}=1+\frac{1}{2\kappa_{+}\left(r-r_{+}\right)}-\frac{1}{2\kappa_{-}\left(r-r_{-}\right)}=\left(1-\frac{2M}{r}+\frac{Q^{2}}{r^{2}}\right)^{-1}
\end{equation}
and the constant of integration was chosen such that $r_{*}(r=M)=0$.
We move from $(t,r)$ to $(v,r)$ coordinates inside the BH defined
by 

\begin{equation}
v=r_{*}(r)+t
\end{equation}
The inverse coordinate transformation is:
\begin{equation}
t=v-r_{*}(r)
\end{equation}
And its partial derivatives are:

\begin{equation}
\frac{\partial t}{\partial v}=1,\quad\frac{\partial r}{\partial v}=0
\end{equation}

\begin{equation}
\frac{\partial t}{\partial r}=-\left(1-\frac{2M}{r}+\frac{Q^{2}}{r^{2}}\right)^{-1},\quad\frac{\partial r}{\partial r}=1
\end{equation}
The metric changes according to:

\begin{equation}
g_{vv}=\frac{\partial t}{\partial v}\frac{\partial t}{\partial v}g_{tt}+\frac{\partial r}{\partial v}\frac{\partial r}{\partial v}g_{rr}=-\left(1-\frac{2M}{r}+\frac{Q^{2}}{r^{2}}\right)
\end{equation}

\begin{align}
g_{rr} & =\frac{\partial t}{\partial r}\frac{\partial t}{\partial r}g_{tt}+\frac{\partial r}{\partial r}\frac{\partial r}{\partial r}g_{rr}\\
 & =-\left(1-\frac{2M}{r}+\frac{Q^{2}}{r^{2}}\right)^{-2}\left(1-\frac{2M}{r}+\frac{Q^{2}}{r^{2}}\right)\\
 & +\left(1-\frac{2M}{r}+\frac{Q^{2}}{r^{2}}\right)^{-1}=0
\end{align}
And last metric component:

\begin{align}
g_{rv} & =\frac{\partial t}{\partial r}\frac{\partial t}{\partial v}g_{tt}+\frac{\partial r}{\partial r}\frac{\partial r}{\partial v}g_{rr}\\
 & =\left(1-\frac{2M}{r}+\frac{Q^{2}}{r^{2}}\right)^{-1}\left(1-\frac{2M}{r}+\frac{Q^{2}}{r^{2}}\right)=1
\end{align}

Thus the RN geometry in ingoing Eddington coordinates $(v,r)$ takes
the form:

\begin{equation}
ds^{2}=-\left(1-\frac{2M}{r}+\frac{Q^{2}}{r^{2}}\right)dv^{2}+2dr\;dv+r^{2}d\Omega^{2}\label{eq:ingoing eddington}
\end{equation}

\subsection{RN in double null Eddington coordinates}

We wish to move to double-null coordinates inside the black hole.
First, we'll do a coordinate transformation from $(t,r)$ to $(t,r_{*})$.
Recalling that:

\begin{equation}
\frac{\partial r_{*}}{\partial r}=\left(1-\frac{2M}{r(r_{*})}+\frac{Q^{2}}{r^{2}(r_{*})}\right)^{-1}
\end{equation}
we get:

\begin{equation}
g_{r_{*}r_{*}}=g_{rr}\frac{\partial r}{\partial r_{*}}\frac{\partial r}{\partial r_{*}}=\left(1-\frac{2M}{r(r_{*})}+\frac{Q^{2}}{r^{2}(r_{*})}\right)
\end{equation}
The metric in $(t,r_{*})$ coordinates is thus:

\begin{equation}
ds^{2}=\left(1-\frac{2M}{r(r_{*})}+\frac{Q^{2}}{r^{2}(r_{*})}\right)\left(-dt^{2}+dr_{*}^{2}\right)+r^{2}(r_{*})d\Omega^{2}\label{eq:r* metric}
\end{equation}
Now we move to double null Eddington coordinates $(u,v)$:

\begin{equation}
v=r_{*}+t
\end{equation}

\begin{equation}
u=r_{*}-t
\end{equation}
From these we get the inverse transformation:

\begin{equation}
r_{*}=\frac{u+v}{2}
\end{equation}

\begin{equation}
t=\frac{v-u}{2}
\end{equation}
We get the following relation for the differentials:

\begin{equation}
dv=dr_{*}+dt
\end{equation}

\begin{equation}
du=dr_{*}-dt
\end{equation}
Multiplying these together we get:

\begin{equation}
du\cdot dv=dr_{*}^{2}-dt^{2}
\end{equation}
Substituting this in the metric Eq. (\ref{eq:r* metric}), the Reissner--Nordström
geometry in double null coordinates is found to take the form:

\begin{equation}
ds^{2}=\left(1-\frac{2M}{r(u,v)}+\frac{Q^{2}}{r^{2}(u,v)}\right)du\;dv+r^{2}(u,v)d\Omega^{2}\label{eq:RN double null}
\end{equation}
The metric function $g_{uv}$ is equal to:

\begin{equation}
g_{uv}=\frac{1}{2}\left(1-\frac{2M}{r}+\frac{Q^{2}}{r^{2}}\right)\label{eq:g_uv_RN}
\end{equation}
The function $r(u,v)$ is given by inverting the function $r_{*}(r)$:

\begin{equation}
\frac{u+v}{2}=r_{*}=r-M+\frac{1}{2\kappa_{+}}\log\left|\frac{r-r_{+}}{M-r_{+}}\right|-\frac{1}{2\kappa_{-}}\log\left|\frac{r-r_{-}}{M-r_{-}}\right|
\end{equation}

We wish to find an expression for the derivatives of the two functions
$R(u,v)$ and $S(u,v)$ in RN near the IH. Around $r=r_{-},$we get:

\begin{equation}
\frac{u+v}{2}=r_{*}=const-\frac{1}{2\kappa_{-}}\log\left|r-r_{-}\right|
\end{equation}
Rearranging the terms, we get:

\begin{equation}
r-r_{-}=const\cdot e^{-2\kappa_{-}r_{*}}=const\cdot e^{-\kappa_{-}\left(u+v\right)}\label{eq:r_minus_r_minus}
\end{equation}
The metric component $g_{uv}$ (Eq. (\ref{eq:g_uv_RN})) around $r=r_{-}$
is (recalling also Eq. (\ref{eq:spherically symmetric metric},\ref{eq:r-r-*r-r+})):

\begin{align}
2\frac{e^{S}L_{0}}{r} & =-2g_{uv}\\
 & =-\frac{\left(r-r_{-}\right)\left(r-r_{+}\right)}{r^{2}}\approx-\frac{\left(r-r_{-}\right)\left(r_{-}-r_{+}\right)}{r_{-}^{2}}\\
 & =2\kappa_{-}\left(r-r_{-}\right)
\end{align}
Combining these equations we get:

\begin{equation}
const\cdot2\kappa_{-}e^{-\kappa_{-}\left(u+v\right)}=2\kappa_{-}\left(r-r_{-}\right)\approx2\frac{e^{s}L_{0}}{r_{-}}
\end{equation}
Taking the log of this equation,

\begin{equation}
S=-\kappa_{-}\left(u+v\right)+const
\end{equation}
So:

\begin{equation}
S_{,u}=S_{,v}=-\kappa_{-}\label{eq:svsu}
\end{equation}
near the IH. Taking the derivative of Eq. (\ref{eq:r_minus_r_minus})
with respect to $r_{*},$we get near the IH:

\begin{equation}
r_{,r_{*}}=const\cdot e^{-2\kappa_{-}r_{*}}
\end{equation}
Using this, we can calculate the derivatives of $R$ around $r=r_{-}:$

\begin{equation}
R_{,u}=R_{,r}r_{,r_{*}}r_{*,u}=2r\cdot const\cdot e^{-2\kappa_{-}r_{*}}\frac{1}{2}=const\cdot e^{-2\kappa_{-}r_{*}}\rightarrow0
\end{equation}
as $r_{*}\rightarrow\infty$, and similarly for $R_{,v}$:

\begin{equation}
R_{,v}=R_{,r}r_{,r_{*}}r_{*,v}=const\cdot e^{-2\kappa_{-}r_{*}}\rightarrow0
\end{equation}

In a similar manner, around the EH, at $r\rightarrow r_{+}$ from
below, we get:

\begin{equation}
r_{*}=const+\frac{1}{2\kappa_{+}}\log\left|r-r_{+}\right|
\end{equation}
Rearranging the terms, we get:

\begin{equation}
r-r_{+}=const\cdot e^{2\kappa_{+}r_{*}}=const\cdot e^{\kappa_{+}\left(u+v\right)}\label{eq:r_minus_r_plus}
\end{equation}
The metric component $g_{uv}$ around $r=r_{+}$is :

\begin{align}
2\frac{e^{S}L_{0}}{r} & =-2g_{uv}\\
 & =-\frac{\left(r-r_{+}\right)\left(r-r_{-}\right)}{r^{2}}\approx-\frac{\left(r-r_{+}\right)\left(r_{+}-r_{-}\right)}{r_{-}^{2}}\\
 & =-2\kappa_{+}\left(r-r_{+}\right)
\end{align}
Combining these equations we get:

\begin{equation}
const\cdot2\kappa_{+}e^{\kappa_{+}\left(u+v\right)}=2\kappa_{+}\left(r-r_{+}\right)\approx2g_{uv}\label{eq:guv_exp_kappa_plus_u+v}
\end{equation}

\subsection{Vaidya metric}

Our solution near the EH will be rooted by the ingoing charged Vaidya
metric.\citep{bonnor1970spherically}. The Vaidya metric is:

\begin{equation}
ds^{2}=-\left(1-\frac{2M(v)}{r}+\frac{Q(v)^{2}}{r^{2}}\right)dv^{2}+2dvdr+r^{2}d\Omega^{2}\label{eq:vaidya_metric}
\end{equation}
Note that this metric, when taking $M(v)=const$ and $Q(v)=const$,
is exactly equal to Eq. (\ref{eq:ingoing eddington}) ingoing Eddington
metric.

The $G_{vv}$ component of the Einstein tensor of the Vaidya metric
is:

\begin{equation}
G_{vv}=\frac{1}{r^{2}}\left(2M_{,v}-2\frac{1}{r}QQ_{,v}\right)+\left(1-\frac{2M(v)}{r}+\frac{Q(v)^{2}}{r^{2}}\right)\frac{Q^{2}}{r^{4}}
\end{equation}
The $G_{rv}$ component of the Einstein tensor of the Vaidya metric
is: 

\begin{equation}
G_{rv}=-\frac{Q^{2}}{r^{4}}
\end{equation}
We wish to know $G_{vv}$ in double null coordinates $(v,u)$, instead
of $v,r$. Double null coordinates are characterized by $g_{uu}=g_{vv}=0$.

\begin{align}
0 & =g_{vv}=\frac{\partial v}{\partial v}\frac{\partial v}{\partial v}g_{vv}+2\frac{\partial r}{\partial v}\frac{\partial v}{\partial v}g_{rv}\\
 & =-\left(1-\frac{2M}{r}+\frac{Q^{2}}{r^{2}}\right)+2\frac{\partial r}{\partial v}
\end{align}
Transformations into double null coordinates are therefore characterized
by:

\begin{equation}
\frac{\partial r}{\partial v}=\frac{1}{2}\left(1-\frac{2M}{r}+\frac{Q^{2}}{r^{2}}\right)
\end{equation}
Therefore, $G_{vv}$ is equal to

\begin{align}
G_{vv} & =\frac{\partial v}{\partial v}\frac{\partial v}{\partial v}G_{vv}+2\frac{\partial r}{\partial v}\frac{\partial v}{\partial v}G_{rv}\\
 & =\frac{1}{r^{2}}\left(2M_{,v}-2\frac{1}{r}QQ_{,v}\right)
\end{align}

Therefore, in double null coordinates,

\begin{equation}
\widetilde{T}_{vv}=r^{2}G_{vv}=2M_{,v}-2\frac{1}{r}QQ_{,v}\label{eq:tvv_region1}
\end{equation}
The inverse metric tensor is:
\begin{equation}
g^{rr}=1-\frac{2M(v)}{r}+\frac{Q(v)^{2}}{r^{2}}
\end{equation}
\begin{equation}
g^{rv}=1
\end{equation}
\begin{equation}
g^{vv}=0
\end{equation}
We raise the indices to get $G^{vv}$:
\begin{equation}
G^{vv}=g^{v\alpha}g^{v\beta}G_{\alpha\beta}=0
\end{equation}
$G^{vv}$ is the same in double null coordinates:

\begin{equation}
G^{vv}=\frac{\partial v}{\partial v}\frac{\partial v}{\partial v}G^{vv}=0
\end{equation}
We lower the indices to get $G_{UU}$:

\begin{equation}
G_{uu}=g_{uv}g_{uv}G^{vv}=0
\end{equation}
Therefore,

\begin{equation}
\widetilde{T}_{uu}=0
\end{equation}

\subsection{The source terms in the evolution equations}\label{subsec:Source-terms}

As mentioned in Eqs. (\ref{eq:Tuu,v},\ref{eq:Tvv,u}), in the absence
of $T_{uv}^{(SC)},\ensuremath{T_{\theta\theta}^{(SC)}}$ (namely in
double null fluid model) and with constant charge ($Q=const$), $\text{\ensuremath{\widetilde{T}_{uu,v}=\widetilde{T}_{vv,u}=0} }$--
that is, both $\ensuremath{\widetilde{T}_{vv}}$ and $\ensuremath{\widetilde{T}_{uu}}$
are individually conserved. Since our initial values are taken from
the Vaidya geometry near the EH, and since $\widetilde{T}_{uu}$ vanishes
there (as is evident from regularity conditions), this implies that
$\widetilde{T}_{uu}=0$ throughout. We also have $\widetilde{T}_{vv}=2M_{,v}$
, see Eq. (\ref{eq:tvv_region1}). 

When the charge is changing and there is an electric current -- which
is the general case we are actually considering in this work -- if
the current is only $J_{v}$ while $J_{u}=0$, then $\widetilde{T}_{vv}=2M_{,v}-2\frac{1}{r}QQ_{,v}$
but $\widetilde{T}_{uu}=0$ still holds.

As a matter of fact, however, the analysis of semi-classical fluxes
 at the inner horizon \citep{Zilberman_2020} (for uncharged scalar
field, yielding $J_{\mu}=0$) showed that generically $T_{uu}\neq0$
there. From the above discussion, it follows that this is only possible
because at the intermediate region between $r_{+}$ and $r_{-}$ there
are non-zero $T_{uv}^{(SC)},\ensuremath{T_{\theta\theta}^{(SC)}}$.
Thus, to allow us modeling the fact that $T_{uu}\neq0$ at the inner
horizon, we must add such non-zero $T_{uv}^{(SC)},\ensuremath{T_{\theta\theta}^{(SC)}}$
in the region between the EH and the IH. 

The evolution equations in the presence of non-vanishing $T_{uv}^{(SC)},\ensuremath{T_{\theta\theta}^{(SC)}}$
take the form:

\begin{equation}
R_{,uv}=e^{S}L_{0}F_{1}(R)+Z_{R}\label{eq:RuvF1-1-1}
\end{equation}

\begin{equation}
S_{,uv}=e^{S}L_{0}F_{2}(R)+Z_{S}\label{eq:suv_F2-1-1}
\end{equation}
where $Z_{R}$,$Z_{S}$ are the \emph{source terms} in the evolution
equation:
\begin{equation}
Z_{R}=8\pi RT_{uv}^{(SC)}\label{eq:Z_R}
\end{equation}
\begin{equation}
Z_{S}=-8\pi\left(\frac{1}{2}T_{uv}^{(SC)}+\frac{L_{0}e^{S}}{R^{\frac{3}{2}}}T_{\theta\theta}^{(SC)}\right)\label{eq:Z_S}
\end{equation}
Note that the constraint equations remain unchanged, they do not depend
at all on $T_{uv}^{(SC)}$ and $T_{\theta\theta}^{(SC)}$.

Obviously, in order to run a numerical simulation, we have to make
a concrete choice of $T_{uv}^{(SC)}$ and $T_{\theta\theta}^{(SC)}$
(as functions of $u$ and $v$). Unfortunately, these two semiclassical
functions $T_{uv}^{(SC)},\ensuremath{T_{\theta\theta}^{(SC)}}$ have
not been computed yet inside black holes. Naively one might view this
lack of knowledge of $T_{uv}^{(SC)},\ensuremath{T_{\theta\theta}^{(SC)}}$
as a crucial obstacle. This is not the case, however, as we now explain: 

The purpose of our numerical simulation is to test the validity of
an analytical approximation for the evolving semiclassical metric
(particularly near the IH), which will be described later (see Sec.
\ref{sec:Analytical-approximation}). This analytical approximation
only depends on a set of four main semiclassical input functions:
$M(v)$ (closely related to $\widetilde{T}_{vv}^{(+)}(v)$) and $Q^{v}(v)$
(closely related to $J_{v}^{(+)}(v)\propto Q_{,v}^{v}$) at the event
horizon -- and $\widetilde{T}_{uu}^{(-)}(v),$$\text{\ensuremath{\widetilde{T}_{vv}^{(-)}}}(v)$
at the inner horizon, both defined with respect to a background RN
metric with parameters $M(v)$,$Q(v)$ . This analytical approximation
does \emph{not} depend on $T_{uv}^{(SC)}$ and $T_{\theta\theta}^{(SC)}$.
{[}The actual values of $\widetilde{T}_{uu}^{(-)},$$\text{\ensuremath{\widetilde{T}_{vv}^{(-)}}}$
are affected by the presence of $T_{uv}^{(SC)},\ensuremath{T_{\theta\theta}^{(SC)}}$
between the two horizon; but nevertheless, once $\widetilde{T}_{uu}^{(-)},\text{\ensuremath{\widetilde{T}_{vv}^{(-)}}}$
are known (and indeed they have been computed, e.g. in Ref. \citep{Zilberman_2020}),
we no longer need $T_{uv}^{(SC)},\ensuremath{T_{\theta\theta}^{(SC)}}$
for the mentioned analytical approximation.{]}

The way we proceed in our numerical investigation is therefore the
following: We \emph{arbitrarily guess} the two semiclassical quantities
$T_{uv}^{(SC)},\ensuremath{T_{\theta\theta}^{(SC)}}$ (as functions
of $u,v$), and then numerically evolve the evolution equations (starting
from initial data prescribed near the EH). The choice of these two
RSET components is given in subsection \ref{subsec:Choice-of-source}.
(Overall we numerically test 7 different configurations for these
RSET components -- all of them are of order comparable to the fluxes
associated with the evaporation process.) In this process we obtain
the evolving unknown functions $R(u,v)$ and $S(u,v)$ -- and thereby
also the two near-IH flux functions $\widetilde{T}_{uu}^{(-)},$$\text{\ensuremath{\widetilde{T}_{vv}^{(-)}}}$
from the constraint equations Eqs. (\ref{eq:Tvv definition}, \ref{eq:TuuRuSu})
. We can then substitute these two flux functions in the analytical
approximation, and compare it to the actual numerically-evolving functions
$R(u,v)$ and $S(u,v)$. This method allows us to numerically test
the analytical approximation for any (quite arbitrary) choice of $T_{uv}^{(SC)}$
and $T_{\theta\theta}^{(SC)}$.

\section{Inputs from semiclassical theory}

\subsection{Hawking radiation}

We start by discussing Schwarzschild black hole. According to Hawking
\citep{hawking1975particle}, in the presence of quantum fields, the
black hole radiates away particles and will eventually evaporate.
According to Hawking, the black hole has \emph{Hawking temperature
}of $T_{H}$ (we're using units where Boltzmann constant $k_{B}=1$):
\begin{equation}
T_{H}=\frac{\hbar\kappa_{+}}{2\pi}=\frac{\hbar}{8\pi M}
\end{equation}
The most naive expression for the energy radiation rate is:

\begin{equation}
L=\sigma AT_{H}^{4}\label{eq:L_eq_A_T^4}
\end{equation}
where $\sigma=\frac{\pi^{2}}{60\hbar^{3}}$ is the Stefan-Boltzmann
constant, and $A=4\pi r_{+}^{2}=16\pi M^{2}$ is the area of the sphere
of the EH. In Schwarzschild, this luminosity is equal to:

\begin{equation}
L=\frac{\pi^{2}}{60\hbar^{3}}16\pi M^{2}\frac{\hbar^{4}}{8^{4}\pi^{4}M^{4}}=\frac{\hbar}{15360\pi M^{2}}
\end{equation}
However, black holes aren't perfect black bodies, and some waves get
reflected by the black hole. The actual luminosity is therefore equal
to:

\begin{equation}
L=\frac{\hbar}{M^{2}}P
\end{equation}
where $P$ is some dimensionless constant, which needs to be computed
numerically. According to Page \citep{page1976particle}, the constant
is equal to $P=3.366\times10^{-5}$ for photon emissions, and $P=0.3845\times10^{-5}$
for graviton emissions, and according to Elster \citep{ELSTER1983205}
$P=7.44\times10^{-5}$ for scalar field. There are two degrees of
freedom for the photon and graviton while there is a single degree
of freedom for the scalar. 

In the case of RN black hole, the luminosity depends also on the charge,
and it is of the form:

\begin{equation}
L=\frac{\hbar}{M^{2}}H_{M}\left(\frac{Q}{M}\right)
\end{equation}
for some dimensionless function $H_{M}\left(\frac{Q}{M}\right)$ of
the variable $\frac{Q}{M}$ (that may be determined numerically).

The luminosity determines the rate of change of the mass $M_{,t}$,
as observed by a distant observer:

\begin{equation}
M_{,t}^{(\infty)}=-L
\end{equation}
At the EH, the rate of change of the mass in the Eddington coordinate
$v$ is the same as the rate of change of mass seen by a distant observer:

\begin{equation}
M_{,v}^{(+)}=M_{,t}^{(\infty)}=-L=-\frac{\hbar}{M^{2}}H_{M}\left(\frac{Q}{M}\right)\label{eq:M,v}
\end{equation}
In this entire work we will assume, for simplicity, that $\frac{Q}{M}$
remains constant upon evaporation of the black hole. We will discuss
this assumption later. We can then solve this differential equation
for $M(v)$ and get:

\begin{equation}
M(v)=\left(M_{0}^{3}-3v\hbar H_{M}\left(\frac{Q}{M}\right)\right)^{\frac{1}{3}}
\end{equation}
where $M_{0}$ is the initial mass at $v=0$. We'll define the dimensionless
parameter $\varepsilon_{0}\equiv3\hbar M_{0}^{-2}H_{M}\left(\frac{Q}{M}\right)$
and obtain:

\begin{equation}
M(v)=\left(M_{0}^{3}-M_{0}^{2}\varepsilon_{0}v\right)^{\frac{1}{3}}
\end{equation}
Therefore the evaporation rate is:

\begin{equation}
M_{,v}=-\frac{1}{3}M_{0}^{2}\varepsilon_{0}\left(M_{0}^{3}-M_{0}^{2}\varepsilon_{0}v\right)^{-\frac{2}{3}}
\end{equation}
The evaporation rate at the beginning is:

\begin{equation}
M_{,v}(v=0)=-\frac{1}{3}\varepsilon_{0}\label{eq:M_,v}
\end{equation}
The evaporation time $T_{ev}$ from the initial moment (when $M=M_{0}$)
until the moment when $M(v)=0$ is:

\begin{equation}
T_{ev}=\frac{M_{0}^{3}}{3\hbar H_{M}\left(\frac{Q}{M}\right)}
\end{equation}
A convenient dimensionless quantity that characterizes the initial
mass size is $\frac{T_{ev}}{M_{0}}$:

\begin{equation}
\frac{T_{ev}}{M_{0}}=\frac{M_{0}^{2}}{3\hbar H_{M}\left(\frac{Q}{M}\right)}=\frac{1}{\varepsilon_{0}}
\end{equation}

\subsection{Choice of initial mass parameter}

Since we wish to cover significant evaporation in our numerical simulation
(we typically manage to evaporate about 53\% of the initial mass and
charge), our numerical grid must cover a grid length of order $\frac{T_{ev}}{M_{0}}\left(\frac{du}{M_{0}}\right)^{-1}$points.
($du$ is the size of a single grid step, and we usually have $\frac{du}{M_{0}}=\frac{1}{100}$.)
For astrophysical black holes of 10 solar masses, $\frac{T_{ev}}{M_{0}}\approx10^{82}$,
which is impossible to implement numerically. We can implement numerically
$\frac{T_{ev}}{M_{0}}=10^{3}$. This value definitely does not correspond
to a realistic astrophysical black hole, yet this value is still $\gg1$.
We'll therefore choose our $M_{0}$ such that $\frac{T_{ev}}{M_{0}}=10^{3}$.
The corresponding value of $\varepsilon_{0}$ is $\frac{1}{1000}$.

\subsection{Choice of units}

Recall that as mentioned above, we're using general relativistic geometrized
unit system $c=G=1$, which leaves us with a single free unit of mass
scale. We can choose this mass unit as we wish, so we'll choose it
to be our initial mass $M_{0}$, that is, $M_{0}=1$. In other words,
our mass unit is relative to the initial mass of the black hole. 

Recall the parameter $L_{0}$ from Eq. (\ref{eq:spherically symmetric metric}),
which can be chosen freely. We choose $L_{0}=M_{0}$. In our choice
of units, we therefore get $L_{0}=1$.

\subsection{Basic scaling of semiclassical $T_{\alpha}^{\;\beta}$}

Classical general relativity is scale-invariant. In particular, every
Schwarzschild black hole is similar to every other Schwarzschild black
hole via a mapping which keeps $\frac{r}{M}$ constant. Every physical
quantity whose dimensions are $[l]^{n}$ scales under this mapping
as $M^{n}$. In the case of RN, all black holes with the same $\frac{Q}{M}$
are also similar in that sense. 

In the semiclassical theory, quantities may also depend on $\hbar$.
In geometrized units ($c=G=1$) $\hbar$ has dimensions of length
squared, so $\frac{\hbar}{M^{2}}$ is dimensionless. The semiclassical
$T_{\alpha}^{\;\beta}$ has the form\footnote{We here refer to the expectation value of the renormalized stress-energy
tensor}:

\begin{equation}
T_{\alpha}^{\;\beta(SC)}=\frac{\hbar}{M^{2}}\hat{T}_{\alpha}^{\;\beta}
\end{equation}
where $\hat{T}_{\alpha}^{\;\beta}$ is independent of $\hbar$, and
therefore scales like a \emph{classical} quantity with the same dimensions
as this quantity $\hat{T}_{\alpha}^{\;\beta}$ has. But $\frac{\hbar}{M^{2}}$
is dimensionless and therefore $[\hat{T}_{\alpha}^{\;\beta}]=[T_{\alpha}^{\;\beta(SC)}]$.
$T_{\alpha}^{\;\beta(SC)}$ (like any kind of $T_{\alpha}^{\;\beta}$)
has units of energy density, that is, it has dimensions of $[T_{\alpha}^{\;\beta(SC)}]=[l]^{-2}$.
Overall, we obtain that $\hat{T}_{\alpha}^{\;\beta}$ scales like
$M^{-2}$. We therefore conclude that $T_{\alpha}^{\;\beta(SC)}$
scales as:

\begin{equation}
T_{\alpha}^{\;\beta(SC)}=\frac{\hbar}{M^{4}}H_{T_{\alpha}^{\;\beta}}\left(\frac{r}{M},\frac{Q}{M}\right)
\end{equation}
where $H_{T_{\alpha}^{\;\beta}}\left(\frac{r}{M},\frac{Q}{M}\right)$
is some dimensionless function (free of $\hbar$) that only depends
on its two dimensionless arguments. We'll denote further dimensionless
functions of this sort by $H_{...}$, where the ``$...$'' denotes
various possible subscripts which distinguish these functions from
each other. 

In particular, the scaling of $T_{vv}$ is:

\begin{equation}
T_{vv}\equiv T_{vv}^{(SC)}=\frac{\hbar}{M^{4}}H_{T_{vv}}\left(\frac{r}{M},\frac{Q}{M}\right)
\end{equation}
where, again, $H_{T_{vv}}\left(\frac{r}{M},\frac{Q}{M}\right)$ is
some dimensionless function (free of $\hbar$) of its two dimensionless
arguments. 

In Eq. (\ref{eq:Tvv definition}) we defined the quantity $\widetilde{T}_{vv}\equiv8\pi r^{2}T_{vv}$.
Correspondingly we define $\widetilde{T}_{vv}^{(+)}$ , $\widetilde{T}_{vv}^{(-)}$
, which are $\widetilde{T}_{vv}$ at the two horizons, EH and IH respectively.
They scale according to:

\begin{equation}
\widetilde{T}_{vv}^{(+)}=8\pi r_{+}^{2}\frac{\hbar}{M^{4}}H_{T_{vv}}\left(\frac{r_{+}}{M},\frac{Q}{M}\right)\equiv\frac{\hbar}{M^{2}}H_{T_{vv}}^{(+)}\left(\frac{Q}{M}\right)\label{eq:Tvv_tilde_scaling}
\end{equation}
\begin{equation}
\widetilde{T}_{vv}^{(-)}=8\pi r_{-}^{2}\frac{\hbar}{M^{4}}H_{T_{vv}}\left(\frac{r_{-}}{M},\frac{Q}{M}\right)\equiv\frac{\hbar}{M^{2}}H_{T_{vv}}^{(-)}\left(\frac{Q}{M}\right)
\end{equation}

\subsection{The charge function $Q(u,v)$}

We would like the charge of the black hole to evaporate as well as
the mass, so that the mass could evaporate completely. (If the charge
does not evaporate, the mass will stop evaporating when the black
hole approaches extremality. We are interested in the case of significantly
evaporating mass.) If there is a charged quantum field, then $Q$
will indeed evaporate, and we will have electric currents in spacetime.
We will therefore assume that there is such a charged quantum field.
Hence $Q$ is no longer a constant parameter, but a function $Q(u,v)$
of $u$ and $v$. At the horizon, we have

\begin{equation}
Q=Q^{v}(v)
\end{equation}
and in the whole spacetime we have $Q(u,v)$, which for convenience
we express as:
\begin{equation}
Q(u,v)=Q^{v}(v)+\delta Q(u,v)
\end{equation}
 Note that from Eq. (\ref{eq:J_u}) if $J_{u}=0$ throughout, then
$Q_{,u}=0$, and hence $\delta Q(u,v)=0$ because $\delta Q$ vanishes
at the EH.

We need to choose the charge evaporation function $Q^{v}(v)$. To
simplify the black hole evaporation model, we choose this function
such that:
\begin{equation}
\frac{Q^{v}(v)}{M(v)}=const
\end{equation}
Note that we choose $0<\frac{Q^{v}}{M}<1$. The derivation of the
analytical approximation below in section \ref{sec:Analytical-approximation}
is not valid in the extremal case because this approximation is based
on the assumption $\kappa_{-}>0$.

To what extent is this choice realistic? Let us consider the (arguably
hypothetical \footnote{This case is not regarded as realistic because the plasma that typically
surrounds black holes would discharge them very quickly.}) case of an astrophysically large black hole which is significantly
charged initially. Such a black hole would lose charge due to the
Schwinger pair production effect, where the strong electric field
near the EH causes electron-positron pairs to be created from the
vacuum: one of the charged particles would fall into the black hole,
reducing its charge, and the other would escape to infinity. This
effect, if possible, would quickly discharge the black hole, especially
since the charge-to-mass ratio for electrons is $\approx2\cdot10^{21}$.
A detailed analysis of schwinger pair production can be found in Ref
\citep{gibbons1975vacuum}. Schwinger pair production is effective
only if the electric field at the event horizon exceeds a critical
value $E_{cr}$. The electric field at the EH near extremality is
of order $E_{eh}=\frac{Q}{r_{+}^{2}}\approx\frac{Q}{M^{2}}=\frac{Q}{M}\frac{1}{M}$.
Therefore, if $\frac{Q}{M}\frac{1}{M}>E_{cr}$, Schwinger pair production
will quickly discharge the black hole until $\frac{Q}{M}\frac{1}{M}=E_{cr}$.
This means there is a critical mass $M_{cr}=\frac{1}{E_{cr}}$, such
that a black hole with $M>M_{cr}$ will discharge very slowly (as
$E_{eh}<E_{cr}$). In this range the black hole would lose its mass
via Hawking radiation until it reaches extremality. On the other hand,
a black hole with $M<M_{cr}$ will discharge while approximately keeping
$E_{eh}=E_{cr}$, that is, $\frac{Q}{M}\sim ME_{cr}$. This means
that in this range, $\frac{Q}{M}$ decreases roughly linearly with
$M$. This varying $\frac{Q}{M}$ would result in a more complicated
dynamics inside the evaporating black hole, which we will here avoid
for the sake of simplicity. 

One of the main reasons we are interested in charged black holes is
as a simplified toy model for spinning black holes, because the charge
$Q$ serves a similar role as the spin parameter $a$ with regards
to the formation of a Cauchy horizon inside the black hole. In the
case of a spinning black hole, Page \citep{page1976particle} found
that the angular momentum of the black hole would evaporate at a significantly
faster rate than its mass. In other words, for evaporating spinning
black hole, $\frac{a}{M}$ decreases to zero during the evaporation
process. Thus, from this perspective too, our assumption of $\frac{Q}{M}=const$
is again not realistic. Despite this assumption being unrealistic,
we still choose it here in order to simplify the evaporation model
and its analysis.

We also need to choose the function $\delta Q$. We'll choose it in
the following form:

\begin{equation}
\delta Q=\log\left(1+e^{\frac{u+v}{M(v)}}\right)M(v)^{-1}\varepsilon_{0}M_{0}^{2}\alpha\label{eq:deltaQ}
\end{equation}
This structure of $\delta Q$ contributes currents flowing in the
$r_{*}=\frac{u+v}{2}$ direction (at leading order in $\varepsilon_{0}$).
$\alpha$ is a free dimensionless parameter controlling the strength
of the semiclassical currents, which we typically choose to be of
order unity. For the dependence of $\delta Q$ on $u+v$, we chose
it to be via the function $f(x)=\log\left(1+e^{x}\right)$ as it interpolates
smoothly between $f(x)=0$ at $x\ll0$ (as desired at the EH) and
$f(x)=x$ at $x\gg0$ (corresponding to the region near the IH). The
argument of this function $f(x)$ is $x=\frac{u+v}{M(v)}$, which
is a dimensionless variable proportional to $r_{*}=\frac{u+v}{2}$.
Since $\delta Q$ is initiated from semiclassical currents, it should
be proportional to $\hbar$ ($\propto\varepsilon_{0}M_{0}^{2}$).
The later has mass squared dimensions, and the dimensions of $Q$
is $M^{1}$, therefore we also need to multiply the expression by
$M(v)^{-1}$ to get the correct mass dimensions. 

Note that we use $M(v)$ to get the correct mass dimensions, and not
$M_{0}$. Because the drift in the mass is extremely slow, at every
point we have a very good local RN approximation. We therefore may
naturally expect that the semiclassical currents -- and therefore
also $\delta Q$ -- will be well approximated by the corresponding
semiclassical quantities on RN background. From dimensional arguments,
semiclassical $\delta Q$ on RN background should take the form $\hbar M^{-1}H_{\delta Q}(\frac{r}{M},\frac{Q}{M})$.
We assume that this form also holds in the evaporation case, as explained
above, and we also substitute $\hbar=\left[3H_{M}\left(\frac{Q}{M}\right)^{-1}\right]\varepsilon_{0}M_{0}^{2}$.
For reasons explained above, we chose $H_{\delta Q}$ to be of the
form corresponding to Eq. (\ref{eq:deltaQ}). It should be emphasized
that $M_{0}$ only appears (squared) along with $\varepsilon_{0}$
as a replacement for $\hbar$.

The main motivation for introducing $\delta Q$ is to demonstrate
numerically that semiclassical drift in the charge $Q$ that may take
place and even become significant near the IH does not affect the
backreaction of the metric near the IH -- once the four input functions
$M(v)$, $Q^{v}(v)$, $\widetilde{T}_{uu}^{(-)}(v)$, $\text{\ensuremath{\widetilde{T}_{vv}^{(-)}}}(v)$
have been specified. (Note that $\delta Q$ may have an effect on
the value of $\widetilde{T}_{uu}^{(-)}(v)$, $\text{\ensuremath{\widetilde{T}_{vv}^{(-)}}}(v)$,
 similar to that of the sources $Z_{R}$,$Z_{S}$.) This fact is
already embodied in the analytical approximation -- where the charge
function required is $Q^{v}(v)$ only. Note that in the absence of
$\delta Q$, from Eq. (\ref{eq:J_u}) $J_{u}=0$. Non-vanishing $\delta Q$
allows us to explore scenarios with $J_{u}$ as well. Note that $\delta Q$
is just the simplest model we could choose which will model current
$J_{u}$. 

\subsection{Choice of source functions $Z_{R}$, $Z_{S}$}\label{subsec:Choice-of-source}

As already mentioned in subsection \ref{subsec:Source-terms}, for
the task of validating the analytical approximation, we can freely
choose the semiclassical functions which are $T_{uv}^{(SC)}$ and
$T_{\theta\theta}^{(SC)}$. This amounts to freely choosing the two
source functions $Z_{R}$, $Z_{S}$ defined in Eqs. (\ref{eq:Z_R}\ref{eq:Z_S}).

We find it convenient to choose these source functions in the following
form:

\begin{equation}
Z_{R}=\varepsilon_{0}M_{0}^{2}M^{-3}e^{S}L_{0}\left(Z_{R0}+\frac{r}{M}Z_{Rr}\right)\label{eq:RuvF1-1}
\end{equation}

\begin{equation}
Z_{S}=\varepsilon_{0}M_{0}^{2}M^{-5}e^{S}L_{0}\left(Z_{S0}+\frac{r}{M}Z_{Sr}\right)\label{eq:suv_F2-1}
\end{equation}
There are 4 parameters: $Z_{R0}$,$Z_{Rr}$,$Z_{S0}$,$Z_{Sr}$, all
dimensionless.

The reasoning behind this choice is as follows. In the semiclassical
theory these source functions vanish both at the EH and the IH. We
therefore choose these functions to be proportional to $e^{S}$, which
is close to zero near both the EH and IH\footnote{For RN background in Unruh state, it was found \citep{zilberman_unpublished}
that near the IH $Z_{R}$ is proportional to $e^{S}\sim e^{-2\kappa_{-}r_{*}}$,
whereas $Z_{S}$ decays as $r_{*}^{-5}$. In our numerical simulation,
we decided to simplify the computation scheme by assuming both to
be proportional to $e^{S}$ near the IH. We do not expect this to
change the resultant backreaction metric because $r_{*}^{-5}$ vanishes
sufficiently fast.\label{fn:footnote r*-5}}. We then multiply $e^{S}$ by the length scale $L_{0}$ similar to
the way it appears in the evolution equations, see Eqs. (\ref{eq:RuvF1},\ref{eq:suv_F2}).
The source term is semiclassical, therefore it should be proportional
to $\hbar$ ($\propto\varepsilon_{0}M_{0}^{2}$) which has mass squared
dimensions, so we want to create a dimensionless quantity from that,
leading us to the dimensionless factor $\hbar M^{-2}\propto\varepsilon_{0}M_{0}^{2}M^{-2}$.
Since $[F_{1}(R)]=\left[\frac{1}{\sqrt{R}}\right]=[l]^{-1}$, the
$Z_{R}$ source term is overall multiplied by $M^{-3}$, and since
$[F_{2}(R)]=\left[\frac{1}{R^{\frac{3}{2}}}\right]=[l]^{-3}$, the
$Z_{S}$ source term is overall multiplied by $M^{-5}$. For convenience,
we chose the spacetime dependence of these sources to be dependent
on $\frac{r}{M}$ only, and we further restricted our choice to linear
functions of this variable, determined by the four arbitrary parameters
$Z_{R0}$,$Z_{Rr}$,$Z_{S0}$,$Z_{Sr}$. When running our numerical
code, we'll choose multiple specific (arbitrary) combinations of these
parameters and the charge parameter $\alpha$.

The presence of these source terms, and / or the changing charge,
cause non-conservation of $\ensuremath{\widetilde{T}_{vv}}$ and $\ensuremath{\widetilde{T}_{uu}}$,
as already mentioned in subsection \ref{subsec:Source-terms}. We
shall now calculate the amount of this non-conservation, using the
evolution equations and the source terms in Eqs. (\ref{eq:RuvF1-1},\ref{eq:suv_F2-1}).
We start by taking the derivative of the constraint Eq. (\ref{eq:TuuRuSu}):

\begin{widetext}

\begin{align}
\left(\widetilde{T}_{uu}\right)_{,v} & =R_{,uv}S_{,u}+R_{,u}S_{,uv}-\left(R_{,uv}\right)_{,u}\\
 & =L_{0}e^{S}\left(F_{1}(R,Q)+\varepsilon_{0}M^{-3}\left(Z_{R0}+\frac{r}{M}Z_{Rr}\right)\right)S_{,u}\\
 & +\left(e^{S}L_{0}F_{2}(R,Q)+\varepsilon_{0}M^{-5}e^{S}L_{0}\left(Z_{S0}+\frac{r}{M}Z_{Sr}\right)\right)R_{,u}\\
 & -\left[L_{0}e^{S}\left(F_{1}(R,Q)+\varepsilon_{0}M^{-3}\left(Z_{R0}+\frac{r}{M}Z_{Rr}\right)\right)\right]_{,u}\\
 & =L_{0}e^{S}\left(F_{1}(R,Q)+\varepsilon_{0}M^{-3}\left(Z_{R0}+\frac{r}{M}Z_{Rr}\right)\right)S_{,u}\\
 & +\left(e^{S}L_{0}F_{2}(R,Q)+\varepsilon_{0}M^{-5}e^{S}L_{0}\left(Z_{S0}+\frac{r}{M}Z_{Sr}\right)\right)R_{,u}\\
 & -L_{0}e^{S}\left(F_{1}(R,Q)+\varepsilon_{0}M^{-3}\left(Z_{R0}+\frac{r}{M}Z_{Rr}\right)\right)S_{,u}\\
 & -e^{S}L_{0}\frac{\partial F_{1}(R,Q)}{\partial R}R_{,u}-e^{S}L_{0}\frac{\partial F_{1}(R,Q)}{\partial Q}Q_{,u}-\varepsilon_{0}M^{-3}e^{S}L_{0}\frac{r_{,u}}{M}Z_{Rr}\\
 & =e^{S}L_{0}\left[\left(\varepsilon_{0}M^{-5}\left(Z_{S0}+\frac{r}{M}Z_{Sr}\right)\right)R_{,u}-\frac{2Q}{R^{\frac{3}{2}}}Q_{,u}-\varepsilon_{0}M^{-3}\frac{r_{,u}}{M}Z_{Rr}\right]\label{eq:Tuu,v conservation}
\end{align}
And similarly for $\widetilde{T}_{vv}:$

\begin{equation}
\left(\widetilde{T}_{vv}\right)_{,u}=e^{S}L_{0}\left[\left(\varepsilon_{0}M^{-5}\left(Z_{S0}+\frac{r}{M}Z_{Sr}\right)\right)R_{,v}-\frac{2Q}{R^{\frac{3}{2}}}Q_{,v}-\varepsilon_{0}M^{-3}\frac{r_{,v}}{M}Z_{Rr}\right]\label{eq:Tvv,u conservation}
\end{equation}

\end{widetext}

From Eq. (\ref{eq:Tuu,v conservation}) we can see that if all sources
are zero and in addition $Q_{,u}=0$, then $\widetilde{T}_{uu,v}=0$.
Since $T_{uu}$ vanishes on the initial hypersurface, it follows that
in such a case, $T_{uu}$ vanishes in the entire domain of dependence.
(Also notice that the parameter $Z_{R0}$ does not affect the conservation
of either $\widetilde{T}_{uu}$ or $\widetilde{T}_{vv}$.)

As was mentioned in the Introduction, the analytical approximation
is based on a key assumption about the behavior of the RSET in the
evolving geometry. Here we shall specify this assumption. In the near-IH
RN background (in Unruh state) it was found that both $Z_{R}$ and
$Z_{S}$ \citep{zilberman_unpublished} vanish sufficiently fast in
both $u$ and $v$ and hence become negligibly small on approaching
the IH (see also footnote \ref{fn:footnote r*-5}). Our key assumption
is that in the evolving geometry too, these source functions remain
negligibly small.  This assumption leads to conservation of the fluxes
$\widetilde{T}_{uu}$ and $\widetilde{T}_{vv}$ in the near-IH region:
$\widetilde{T}_{uu,v}=0$ and $\widetilde{T}_{vv,u}=0$. To see this,
we evaluate $\widetilde{T}_{uu,v}$ by taking the $v$ derivative
of Eq. (\ref{eq:TuuRuSu}) and also substituting the evolution equations
Eqs. (\ref{eq:RuvF1-1-1},\ref{eq:suv_F2-1-1}):

\begin{align}
\left(\widetilde{T}_{uu}\right)_{,v} & =R_{,uv}S_{,u}+R_{,u}S_{,uv}-\left(R_{,uv}\right)_{,u}\\
 & =\left(e^{S}L_{0}F_{1}(R)+Z_{R}\right)S_{,u}\\
 & +\left(e^{S}L_{0}F_{2}(R)+Z_{S}\right)R_{,u}\\
 & -\left(e^{S}L_{0}F_{1}(R)+Z_{R}\right)_{,u}
\end{align}
Since $e^{S}$ decays exponentially with $u$ and $v$ (see subsection
\ref{subsec:Justifying-the-basic}), the quick vanishing of $Z_{R}$
and $Z_{S}$ leads to $\left(\widetilde{T}_{uu}\right)_{,v}=0$ in
the near-IH region. The same argument also applies to $\left(\widetilde{T}_{vv}\right)_{,u}$.

\section{Analytical approximation for the near-IH metric inside slowly evaporating
black holes}\label{sec:Analytical-approximation}

In slowly evaporating black holes, the evaporation rate

\begin{equation}
-\varepsilon=M_{,v}
\end{equation}
is small, $\varepsilon\ll1$. We can build an approximation assuming
$\varepsilon$ is small. Note that this $\varepsilon$ is distinct
from $\varepsilon_{0}$: $\varepsilon$ depends on $v$ while $\varepsilon_{0}$
is a constant associated with the substitution $v=v_{0}$. This difference
reflects the fact that at very late $v$, $\varepsilon$ is no longer
small, being proportional to $M(v)^{-2}$. 

\subsection{Regions inside black hole}

\begin{figure}
\caption{metrics inside black hole}\label{fig:g1g2g3}
\includegraphics{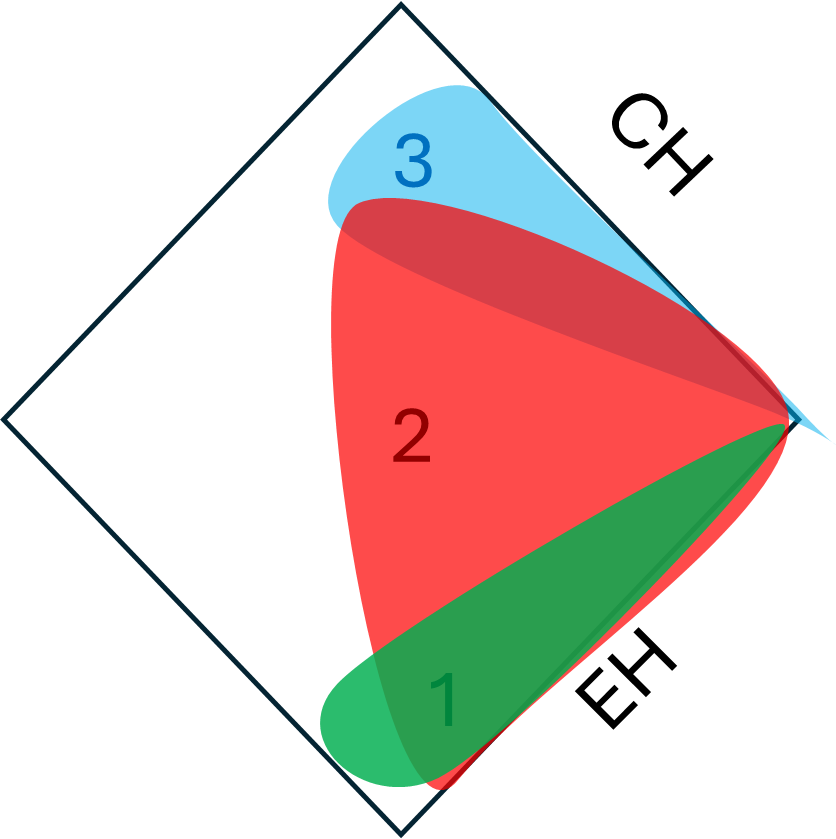}
\end{figure}

We divide the space-time domain inside the black hole to three regions,
as seen in Figure \ref{fig:g1g2g3} . Those regions, numbered 1-3,
are ordered in chronological order. Each two successive regions overlap
in their area of validity, and together they cover the entire interior
of the black hole up to the inner horizon. 

Region 1 is the neighborhood of the event horizon. This is taken here
to be ingoing charged Vaidya metric Eq. (\ref{eq:vaidya_metric}),
which we later convert into double-null coordinates.

Region 2 is between the event horizon and the inner horizon (but not
too close to the latter). This region is characterized by local RN
approximation: at each point in that region, the metric is well approximated
by the RN metric (up to deviations of order $\varepsilon$), though
with slowly varying parameters of mass $M(v)$ and charge $Q(v)$. 

It is important to note that in region 2, the local RN parameters
$M$ and $Q$ are constant along ingoing constant-$v$ rays, as these
parameters depend only on $v$ but not on $u$. Also note that region
2 completely covers region 1. 

Region 3 is the region to the future of region 2 (although there is
some overlap between these two regions) where $r_{*}\gg M$. This
means that, within the local RN approximation, we are placed very
close to the IH. Note, however, that for $u+v$ which are extremely
large -- comparable to the black hole evaporation time -- the local
RN approximation no longer hold (due to drift in $R$, see below).
For this reason, region 3 is not entirely contained in region 2.

As mentioned above, in region 2 the local RN parameter $Q$ depends
on $v$ but not on $u$. This happens because of the following reason:
at the event horizon $J_{u}=0$ as dictated by regularity reasons,
while $J_{v}$ need not vanish at the horizon. This regularity reason
is as follows: the regular Kruskal coordinate at the event horizon
is $U=e^{\kappa_{+}u}$, therefore $u=\frac{\ln U}{\kappa_{+}}$.
The quantity $J_{U}=\frac{\partial u}{\partial U}J_{u}=\frac{1}{U\kappa_{+}}J_{u}$
must be finite, however the event horizon is at $U=0$, which makes
this quantity diverge unless $J_{u}=0$. So this regularity reason
dictate that $J_{u}=0$. For similar regularity reasons, $T_{uu}=0$
at the EH.

\subsection{Basic ansatz of the analytical approximation for region 3}\label{subsec:Basic-ansatz-of}

Consider first the near-IH domain in region 2 where the geometry is
locally RN. We saw in Eq. (\ref{eq:svsu}) that in RN $S_{,u}=S_{,v}=-\kappa_{-}$
near the IH. We therefore get that $S$ is monotonically decreasing
in region 2, and becomes very negative upon increasing $u$ and $v$.
Therefore, $e^{S}$ becomes negligibly small at large $u+v$. 

For the analysis of region 3, we start by assuming that the behavior
of $S$ in region 3 is similar to region 2 near the IH in the following
sense: $S$ continues to decrease monotonically with increasing $u$
and with $v$; as a consequence, $e^{S}$ that was already negligible
in region 2 near IH, will remain negligible throughout region 3. We
justify this ansatz later. 

Using the evolution equations Eq. (\ref{eq:RuvF1-1-1},\ref{eq:suv_F2-1-1})
and neglecting the factor $e^{S}$ -- and also assuming negligibly
small sources $Z_{R}\approx Z_{S}\approx0$ (see subsection \ref{subsec:Choice-of-source})
-- we get, throughout region 3:

\begin{equation}
R_{,uv}=0,\quad S_{,uv}=0
\end{equation}
Therefore both functions $R$,$S$ in region 3 are functions of $v$
plus functions of $u$:

\begin{equation}
R(u,v)=R^{v}(v)+R^{u}(u)
\end{equation}

\begin{equation}
S(u,v)=S^{v}(v)+S^{u}(u)
\end{equation}

Because of that, $S_{,v}$ and $R_{,v}$ are functions of $v$ only:
$S_{,v}(v),R_{,v}(v).$ Similarly, $S_{,u}$ and $R_{,u}$ are functions
of $u$ only: $S_{,u}(u),R_{,u}(u)$. We still need to determine those
four functions, which is our next task. 
\begin{figure}
\caption{Diagram explaining analytical approximation}\label{fig:analytical_approx}

Region 3 is between CH and the light-blue line numbered 3. Region
2 is between EH and the red line numbered 2.

\includegraphics[scale=0.75]{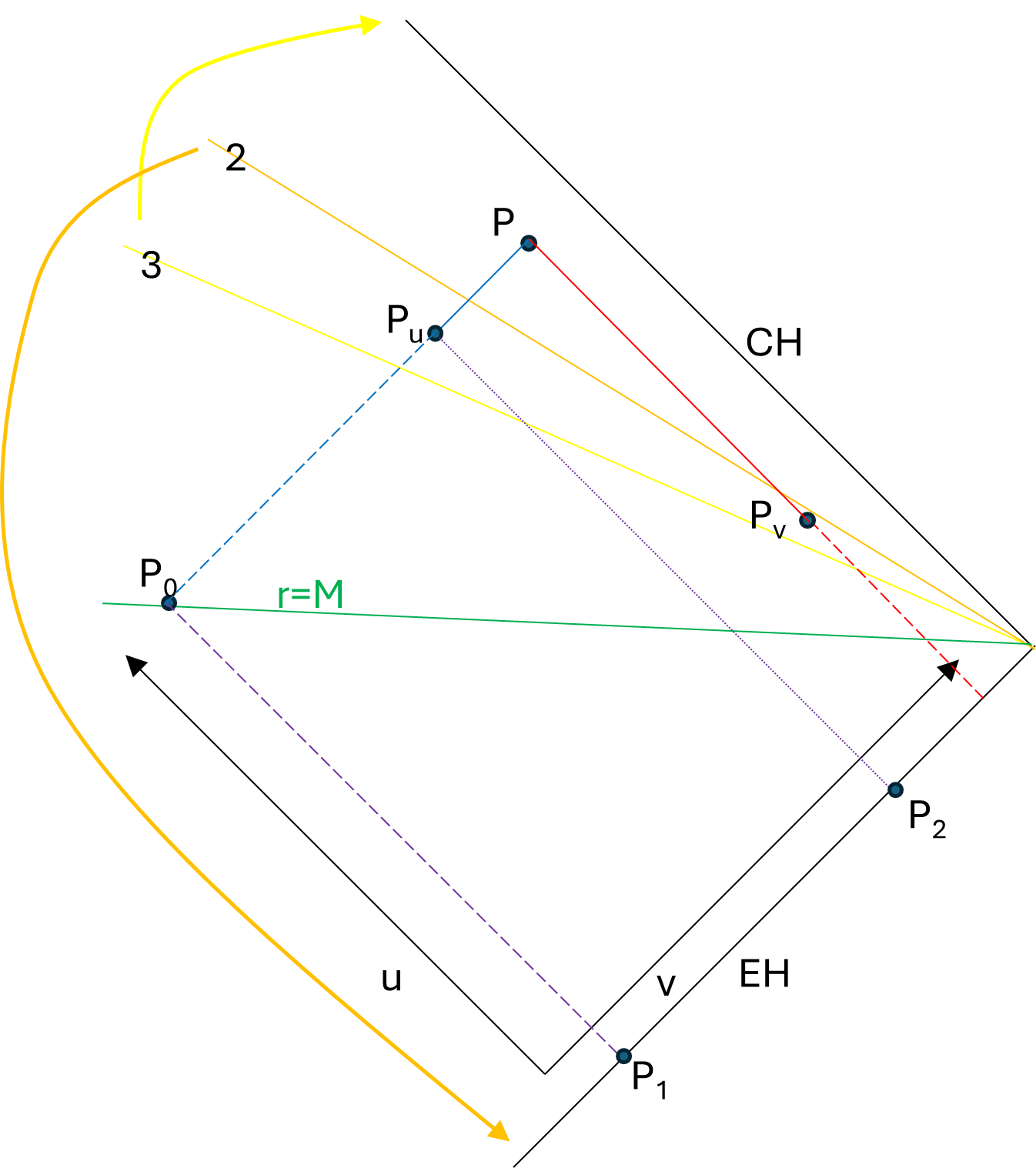}
\end{figure}

\subsection{Computing $S_{,v}$}\label{subsec:Computing-S,v}

Refer to Figure \ref{fig:analytical_approx}. Consider a point $P$
deep in region 3. We first focus on computing $S_{,v}$ at point $P$.
Since $S_{,v}(v)$ is constant along the red constant-$v$ ray, we
can read its value from point $P_{v}$, which is in the overlap of
regions 2 and 3, and therefore its value is obtained from the corresponding
expression in region 2 near IH (see Eq. (\ref{eq:svsu})): 

\begin{equation}
S_{,v}=-\kappa_{-}^{v}(v)\label{eq:S,v_kappa-}
\end{equation}
where $\kappa_{-}^{v}(v)$ refers to the value of $\kappa_{-}$ in
the local RN at point $P_{v}$:

\begin{equation}
\kappa_{-}^{v}(v)\equiv\kappa_{-}(M(v),Q^{v}(v))\label{eq:kappa_-v}
\end{equation}

\subsection{Computing $S_{,u}$}\label{subsec:s,u}

We next move to analyze $S_{,u}(u)$ at point $P$ in region 3. This
quantity is constant along the constant-$u$ blue ray, so that its
value may be obtained by carrying this blue ray until it enters the
overlap between region 3 and near-IH region 2, say until the point
marked $P_{u}$:
\begin{equation}
S_{,u}(P)=-\kappa_{-}(P_{u})\label{eq:S,u P}
\end{equation}
Here $\kappa_{-}(P_{u})$ denotes the value of $\kappa_{-}$ at the
local RN at point $P_{u}$. We pick the point $P_{u}$ in the matching
zone between region 2 and region 3. In this matching zone, both approximation
of region 2 and region 3 should apply. The approximation of region
3 requires $e^{S}$ to be negligible. The region 2 approximation requires
that the near IH drift in $R$ is still negligible, that is $\varepsilon\cdot\left(v+u\right)\ll M$.
We choose this matching zone to be at the region where $u+v$ is of
order, say, $-\frac{1}{\kappa_{-}}\log\varepsilon$. At that point,
$e^{S}$ is indeed negligible, because $S\approx-\kappa_{-}(u+v)$
so $e^{S}\approx\varepsilon\ll1$. In addition, $\varepsilon\cdot\left(v+u\right)=-\frac{1}{\kappa_{-}}\varepsilon\log\varepsilon$
is negligible compared to $M$ in that zone.

As mentioned above, the local RN at a given point in region 2 is the
same along the constant-$v$ ray that passes through that point. We
could therefore mark a constant-$v$ ray from this point $P_{u}$
up to the point $P_{2}$ on the EH.  However we prefer to pick another
point $P_{1}$ on the EH to represent the local RN at point $P_{u}$.
This point $P_{1}$ is obtained by continuing from point $P_{u}$
along the dashed blue constant-$u$ ray until we meet the hypersurface
$r=M$ (which is also the $r_{*}=0$ line) at point $P_{0}$. Then
we move from $P_{0}$ along the dashed purple constant-$v$ ray to
reach the point $P_{1}$ on the EH. We prefer to work with the representative
point $P_{1}$ rather than $P_{2}$ because this choice leads to a
significantly simpler expression for $S_{,u}$ (and $R_{,u}$) in
region 3. However, it remains to be explained why this choice, namely
the replacement of point $P_{2}$ by point $P_{1}$, is justified.

To this end, we look at the difference in the local RN parameters
$M(v)$,$Q^{v}(v)$ between points $P_{1}$and $P_{2}.$ The difference
in $v$ between $P_{1}$ and $P_{2}$ is of course equal to the difference
in $v$ between $P_{0}$ and $P_{u}$. The difference in $v$ is twice
the difference in $r_{*}$ between these two later points. But at
$P_{0}$, $r_{*}=0$, and as mentioned above, at point $P_{u}$, $v+u$
is of order, say, $-\frac{1}{\kappa_{-}}\log\varepsilon$. Since the
drift rate of $M(v)$ is of order $\varepsilon$, the relative deviation
in $M$ accumulated between $P_{1}$ and $P_{2}$ is of order $-\varepsilon\log\varepsilon\ll1$,
and can therefore be neglected here. (To appreciate the quality of
this approximation, recall that for typical astrophysical black holes
$\varepsilon<10^{-82}$, and correspondingly $-\varepsilon\log\varepsilon<2\cdot10^{-80}$.)
The same argument applies to $Q(v)$.

Since at point $P_{0}$ the coordinates $u$,$v$ satisfy \mbox{$r_{*}=\frac{u+v}{2}=0$},
we get $u_{P}=u_{P_{0}}=-v_{P_{0}}=-v_{P_{1}}$(where $u_{P}$ is
$u$ at point $P$, etc.). As explained above, in order to compute
the value of $\kappa_{-}$ at point $P_{u}$, we take the values of
$M$,$Q$ according to the $v$ coordinate of point $P_{1}$ (which
is equal to that of $P_{0}$). We are therefore interested in the
quantity:
\begin{equation}
M(v_{P_{1}})=M(v=-u_{P_{0}})=M(v=-u_{P})
\end{equation}
This motivates us to define the function $M^{u}(u)$ as follows:

\begin{equation}
M^{u}(u)\equiv M(v=-u)
\end{equation}
Similarly for $Q$:

\begin{equation}
Q^{u}(u)\equiv Q^{v}(v=-u)
\end{equation}
According to the same lines, we also define

\begin{equation}
\kappa_{-}^{u}(u)\equiv\kappa_{-}(M^{u}(u),Q^{u}(u))
\end{equation}
or in other words, 
\begin{equation}
\kappa_{-}^{u}(u)\equiv\kappa_{-}(v=-u)\label{eq:kappa u definition}
\end{equation}
For later use, we similarly define the $\kappa_{+}$ version of this
quantity:
\begin{equation}
\kappa_{+}^{u}(u)\equiv\kappa_{+}(v=-u)\label{eq:kappa plus definition}
\end{equation}
Summarizing these notational relations, we get (recalling that $u_{P}=u_{P_{u}}$):
\begin{equation}
\kappa_{-}(P_{u})=\kappa_{-}^{u}(u_{P_{u}})=\kappa_{-}^{u}(u_{P})
\end{equation}
We can therefore rewrite Eq. (\ref{eq:S,u P}) as 

\begin{equation}
S_{,u}(P)=-\kappa_{-}^{u}(u_{P})
\end{equation}
 Or in other words, for any point in region 3:

\begin{equation}
S_{,u}=-\kappa_{-}^{u}(u)\label{eq:S,u_kappa-}
\end{equation}

\subsection{Justifying the basic ansatz}\label{subsec:Justifying-the-basic}

Recalling Eqs. (\ref{eq:S,v_kappa-},\ref{eq:S,u_kappa-}) and the
fact that $\kappa_{-}>0$, we see that in region 3, $S$ continues
to decrease monotonically with both $u$ and $v$. Since $e^{S}$
was already exponentially small in the overlap between region 2 and
region 3, it is totally negligible throughout region 3 -- confirming
our former ansatz about this issue. 

Correspondingly we can neglect the factor $e^{S}$ also in (\ref{eq:Tuu,v conservation},\ref{eq:Tvv,u conservation})
and we get that:

\begin{equation}
\widetilde{T}_{uu,v}=\widetilde{T}_{vv,u}=0
\end{equation}
Therefore in region 3 $\widetilde{T}_{vv}$ is a function of $v$
and $\widetilde{T}_{uu}$ is a function of $u$.

\subsection{Computing $R_{,v}$}

We next focus on $R_{,v}$ in region 3. We recall that $\widetilde{T}_{vv}\equiv8\pi r^{2}T_{vv}=R_{,v}S_{,v}-R_{,vv}$,
and we make an ansatz that $R_{,vv}\ll R_{,v}S_{,v}$ (to be justified
later) and therefore, using Eq. (\ref{eq:S,v_kappa-}) and neglecting
$R_{,vv}$, we get:
\begin{equation}
R_{,v}=\frac{\widetilde{T}_{vv}}{S_{,v}}=-\frac{\widetilde{T}_{vv}^{(-)}}{\kappa_{-}(v)}\label{eq:R,v_Tvv_kappa}
\end{equation}
 We assume that $\widetilde{T}_{vv}\propto\varepsilon$, therefore
$R_{,v}\propto\varepsilon$.

We proceed now to show that $R_{,vv}\propto\varepsilon^{2}$ and therefore
it is indeed negligible relative to $R_{,v}S_{,v}$. We start with
a simple intuitive argument: $R_{,v}=-\frac{\widetilde{T}_{vv}^{(-)}}{\kappa_{-}(v)}$
is already of order $\varepsilon$. Therefore $R_{,vv}$ would exclusively
be made of contributions from the $v$ derivative of either $\kappa_{-}(v)$
or $\widetilde{T}_{vv}^{(-)}$ . But since both these quantities are
uniquely determined by $M$ and $Q^{v}$, their $v$ derivative only
originates from the $v$ derivatives of $M$ and $Q^{v}$. These $v$
derivatives are obviously proportional to $\varepsilon$, and therefore
$R_{,vv}\propto\varepsilon^{2}$.

We shall also give here a more systematic derivation of this property.
Let $F$ be a quantity that can be expressed as:

\begin{equation}
F=M^{m}\varepsilon^{n}H_{F}\left(\frac{Q^{v}(v)}{M(v)}\right)\label{eq:F_M_epsilon_H}
\end{equation}
for some dimensionless function $H_{F}$ that doesn't depend on $\hbar$.
It's important to note that $H_{F}$ doesn't depend on the position
in any way other than via the dependence of $Q^{v}$ and $M$ on $v$.
(All the functions $H_{i}\left(\frac{Q}{M}\right)$ that will appear
below will all be of this class: dimensionless and won't depend on
$\hbar$, and won't depend on the position except via $Q^{v}(v)$,$M(v)$)
Then we'll prove that:
\begin{equation}
F_{,v}=M^{m-1}\varepsilon^{n+1}H_{F'}\left(\frac{Q^{v}(v)}{M(v)}\right)\label{eq:F,v_M_epsilon}
\end{equation}
for some new dimensionless function $H_{F'}$. For that purpose, we
calculate $\left(\frac{Q^{v}}{M}\right)_{,v}:$
\begin{equation}
\left(\frac{Q^{v}}{M}\right)_{,v}=-\frac{Q^{v}M_{,v}}{M^{2}}+\frac{Q_{,v}^{v}}{M}
\end{equation}
From semiclassical theory we know that Eq. (\ref{eq:M,v}) (setting
$M^{(+)}\rightarrow M$ and $Q\rightarrow Q^{v}$):
\begin{equation}
M_{,v}\equiv-\varepsilon=-\frac{\hbar}{M^{2}}H_{M}\left(\frac{Q^{v}}{M}\right)\label{eq:M,v_epsilon}
\end{equation}
and similarly for $Q_{,v}^{v}$:
\begin{equation}
Q_{,v}^{v}=-\frac{\hbar}{M^{2}}H_{Q}\left(\frac{Q^{v}}{M}\right)=-\varepsilon\frac{H_{Q}\left(\frac{Q^{v}}{M}\right)}{H_{M}\left(\frac{Q^{v}}{M}\right)}\equiv-\varepsilon H_{q}\left(\frac{Q^{v}}{M}\right)
\end{equation}
Therefore,
\begin{equation}
\left(\frac{Q^{v}}{M}\right)_{,v}=\frac{\varepsilon Q^{v}}{M^{2}}-\frac{\varepsilon H_{q}\left(\frac{Q^{v}}{M}\right)}{M}\equiv\frac{\varepsilon}{M}H_{v}\left(\frac{Q^{v}}{M}\right)
\end{equation}

We can now use this result to calculate $\frac{d}{dv}H\left(\frac{Q^{v}}{M}\right)$:

\begin{equation}
\frac{d}{dv}H\left(\frac{Q^{v}}{M}\right)=H'\left(\frac{Q^{v}}{M}\right)\frac{d}{dv}\left(\frac{Q^{v}}{M}\right)=H'\left(\frac{Q^{v}}{M}\right)\frac{\varepsilon}{M}H_{v}\left(\frac{Q^{v}}{M}\right)
\end{equation}
where $H'$ denotes the derivative of the function $H$ with respect
to its argument $\frac{Q^{v}}{M}$. We next calculate $\varepsilon_{,v}$
by differentiating Eq. (\ref{eq:M,v_epsilon}):
\begin{align}
\varepsilon_{,v} & =-2\frac{\hbar M_{,v}}{M^{3}}H_{M}\left(\frac{Q^{v}}{M}\right)+\frac{\hbar}{M^{2}}\frac{d}{dv}H_{M}\left(\frac{Q^{v}}{M}\right)\\
 & =-2\frac{\hbar M_{,v}}{M^{3}}H_{M}\left(\frac{Q^{v}}{M}\right)+\frac{\hbar}{M^{2}}H'_{M}\left(\frac{Q^{v}}{M}\right)\frac{\varepsilon}{M}H_{v}\left(\frac{Q^{v}}{M}\right)\\
 & =\frac{\hbar}{M^{2}}\frac{\varepsilon}{M}\left[2H_{M}\left(\frac{Q^{v}}{M}\right)+H'_{M}\left(\frac{Q^{v}}{M}\right)H_{v}\left(\frac{Q^{v}}{M}\right)\right]\\
 & \equiv\frac{\varepsilon^{2}}{M}H_{\varepsilon}\left(\frac{Q^{v}}{M}\right)
\end{align}
where $H'_{M}$ denotes the derivative of the function $H_{M}$ with
respect to its argument $\frac{Q^{v}}{M}$. Finally, we take the derivative
of $F$ in Eq. (\ref{eq:F_M_epsilon_H}) to get:

\begin{widetext}

\begin{align}
F_{,v} & =mM_{,v}M^{m-1}\varepsilon^{n}H_{F}\left(\frac{Q^{v}}{M}\right)+n\varepsilon_{,v}M^{m}\varepsilon^{n-1}H_{F}\left(\frac{Q^{v}}{M}\right)+M^{m}\varepsilon^{n}H'_{F}\left(\frac{Q^{v}}{M}\right)\frac{\varepsilon}{M}H_{v}\left(\frac{Q^{v}}{M}\right)\\
 & =-M^{m-1}\varepsilon^{n+1}mH_{F}\left(\frac{Q^{v}}{M}\right)+M^{m-1}\varepsilon^{n+1}nH_{\varepsilon}\left(\frac{Q^{v}}{M}\right)H_{F}\left(\frac{Q^{v}}{M}\right)\\
 & +M^{m-1}\varepsilon^{n+1}H'_{F}\left(\frac{Q^{v}}{M}\right)H_{v}\left(\frac{Q^{v}}{M}\right)\\
 & \equiv M^{m-1}\varepsilon^{n+1}H_{F'}\left(\frac{Q^{v}}{M}\right)
\end{align}
\end{widetext}We proved Eq. (\ref{eq:F,v_M_epsilon}). Now we will
use this result for Eq. (\ref{eq:R,v_Tvv_kappa}). First, recall from
Eq. (\ref{eq:Tvv_tilde_scaling}) that $\widetilde{T}_{vv}^{(-)}$
can be expressed as:

\begin{equation}
\widetilde{T}_{vv}^{(-)}=\varepsilon H_{T_{vv}}^{(-)}\left(\frac{Q^{v}}{M}\right)
\end{equation}
Secondly, recall that for $\kappa_{-},$Eq. (\ref{eq:kappa_minus_definition})
can be written as:
\begin{equation}
\kappa_{-}=\frac{1}{M}H_{\kappa}\left(\frac{Q^{v}}{M}\right)
\end{equation}
Therefore, 
\begin{equation}
R_{,v}=-\frac{\widetilde{T}_{vv}^{(-)}}{\kappa_{-}(v)}=-\varepsilon M\frac{H_{T_{vv}}^{(-)}\left(\frac{Q^{v}}{M}\right)}{H_{\kappa}\left(\frac{Q^{v}}{M}\right)}\equiv\varepsilon MH_{R}\left(\frac{Q^{v}}{M}\right)
\end{equation}
We can now apply Eq. (\ref{eq:F,v_M_epsilon}) to $F=R_{,v}$ to get
$R_{,vv}$:

\begin{equation}
R_{,vv}=\varepsilon^{2}H_{vv}\left(\frac{Q^{v}}{M}\right)
\end{equation}
We also express $R_{,v}S_{,v}$ in a similar format:
\begin{align}
R_{,v}S_{,v} & =-R_{,v}\kappa_{-}=\left[\varepsilon MH_{R}\left(\frac{Q^{v}}{M}\right)\right]\left[\frac{1}{M}H_{\kappa}\left(\frac{Q^{v}}{M}\right)\right]\\
 & =\varepsilon H_{RS}\left(\frac{Q^{v}}{M}\right)
\end{align}
This confirms our ansatz that $R_{,vv}\ll R_{,v}S_{,v}$, as $R_{,v}\propto\varepsilon$,
while $R_{,vv}\propto\varepsilon^{2}$. Our final result is therefore,
throughout region 3,

\begin{equation}
R_{,v}(u,v)=-\frac{\widetilde{T}_{vv}^{(-)}(v)}{\kappa_{-}^{v}(v)}\label{eq:RvTvvK}
\end{equation}

\subsection{Computing $R_{,u}$}

We next focus on $R_{,u}$. We recall that \mbox{$\widetilde{T}_{uu}\equiv8\pi r^{2}T_{uu}=R_{,u}S_{,u}-R_{,uu}$},
and we make a similar ansatz that $R_{,uu}\ll R_{,u}S_{,u}$ because
$R_{,uu}\propto\varepsilon^{2}$ while $R_{,u}\propto\varepsilon$
(justifying it is similar to the $R_{,v}$ case discussed above).
Therefore, using Eq. (\ref{eq:S,u_kappa-}), and recalling that in
region 3, $\widetilde{T}_{uu}=\widetilde{T}_{uu}^{(-)}(v=-u)$, we
get:

\begin{equation}
R_{,u}(u,v)=-\frac{\widetilde{T}_{uu}^{(-)}(v=-u)}{\kappa_{-}^{u}(u)}\label{eq:RuTuuK}
\end{equation}

To avoid confusion we should note the following: $\widetilde{T}_{uu}^{(-)}$
was defined as a function of $v$ (as the near-IH $\widetilde{T}_{uu}$
value in RN with parameters $M(v)$ and $Q(v)$). We also saw that
in region 3, $\widetilde{T}_{uu}$ depends only on $u$ (see subsection
\ref{subsec:Justifying-the-basic}). The expression we used above
$\widetilde{T}_{uu}=\widetilde{T}_{uu}^{(-)}(v=-u)$ is indeed consistent
with both of these requirements.

\subsection{Summary}\label{subsec:Summary-analytical-approx}

Eqs. (\ref{eq:S,v_kappa-},\ref{eq:S,u_kappa-},\ref{eq:RvTvvK},\ref{eq:RuTuuK})
describe the analytical approximation for the quantities $S_{,v}$,
$S_{,u}$, $R_{,v}$, $R_{,u}$ in region 3. As can be seen in these
equations, this analytical approximation only depends on a set of
four semiclassical input functions: $M(v)$ and $Q^{v}(v)$ (or alternatively
$\widetilde{T}_{vv}^{(+)}(v)$ and $J_{v}^{(+)}(v)$) at the event
horizon, and $\widetilde{T}_{uu}^{(-)}(v)$, $\text{\ensuremath{\widetilde{T}_{vv}^{(-)}}}(v)$
at the inner horizon. (Recall, these two last quantities are defined,
via the semiclassical theory, as the near-IH fluxes in RN with parameters
$M(v)$ and $Q(v)$, and are therefore viewed as functions of $v$.)

\section{Numerical algorithm}\label{sec:Numerical-algorithm}

We would like to integrate the evolution equations Eq. (\ref{eq:RuvF1},\ref{eq:suv_F2})
numerically. We do that by putting both $R$ and $S$ on a discrete
grid. We denote the distance between grid points in the $u$ direction
as $du$ and the distance between grid points in the $v$ direction
as $dv$. The grid points are accessed by two integer indices $i,j$.
For simplicity we chose $du=dv$ throughout this work. 

The basic numerical algorithm computes $R$ at the point $(i,j)$
(marked red in Figure \ref{fig:numerical_points}) as a combination
of 3 previous points (marked blue), and of the derivative $R_{,uv}$
evaluated at the middle point $(i-\sfrac{1}{2},j-\sfrac{1}{2})$ (marked
green), as illustrated in Figure \ref{fig:numerical_points}, as follows:

\begin{figure}
\caption{Points in numerical algorithm}\label{fig:numerical_points}

\includegraphics[scale=0.5]{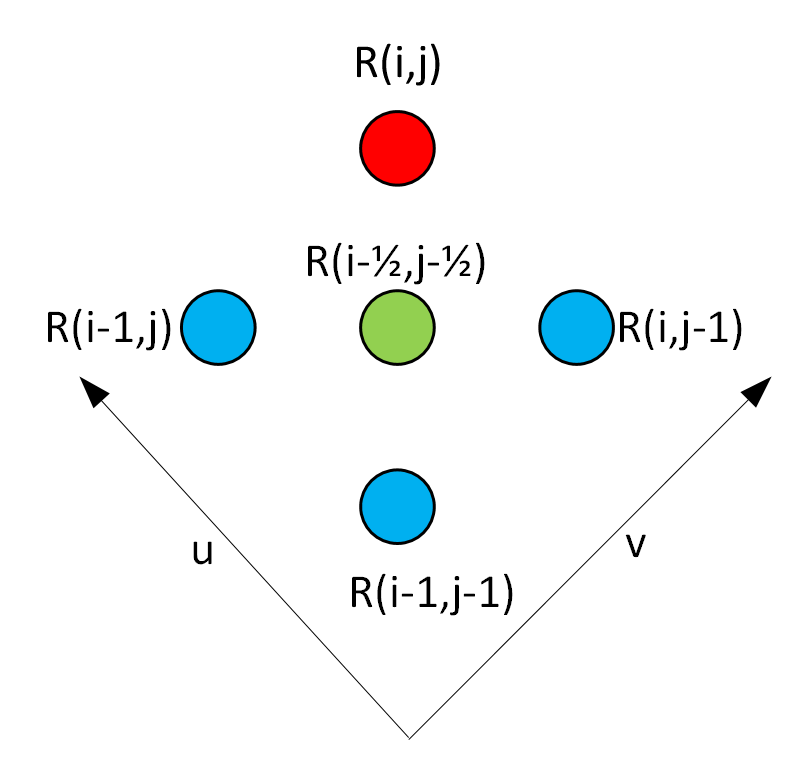}
\end{figure}

\begin{widetext}

\begin{equation}
R(i,j)=R(i-1,j)+R(i,j-1)-R(i-1,j-1)+du\;dvR_{,uv}(i-\sfrac{1}{2},j-\sfrac{1}{2})
\end{equation}

A similar calculation is done for $S$:

\begin{equation}
S(i,j)=S(i-1,j)+S(i,j-1)-S(i-1,j-1)+du\;dvS_{,uv}(i-\sfrac{1}{2},j-\sfrac{1}{2})
\end{equation}

\end{widetext}

Note that the green point $(i-\sfrac{1}{2},j-\sfrac{1}{2})$ is not
a grid point, as indicated by the half integer indices. To evaluate
quantities at that point, we take the average of the relevant quantities
in the two neighboring points $(i-1,j)$ and $(i,j-1)$. 

Such a discrete numerical algorithm necessarily introduces a numerical
error, called truncation error, which in this specific algorithm should
be proportional to $du\cdot dv=du^{2}$. We can measure this numerical
error by running the algorithm with two different values of $du$
and comparing the results obtained in these two runs, see subsection
\ref{subsec:Finite-du-error}.

\section{Initial values}

In the vicinity of the EH, we presumably have, to a good approximation,
a charged Vaidya solution, which evaporates at the rate of $\varepsilon$.
Recall Eq. (\ref{eq:vaidya_metric}):

\begin{equation}
ds^{2}=-\left(1-\frac{2M(v)}{r}+\frac{Q(v)^{2}}{r^{2}}\right)dv^{2}+2dvdr+r^{2}d\Omega^{2}\label{eq:vaidya_metric-1}
\end{equation}
In order to connect to the formalism developed above in subsection
\ref{subsec:Spherically-symmetric-Einsteins} we have to transform
it into double null coordinates. Unfortunately, we cannot implement
this transformation analytically. Nevertheless, since $\varepsilon\ll1$,
at each point in this Vaidya geometry there is a local approximately-RN
solution. Our strategy will be to use this fact to construct the (approximate)
metric for the Vaidya geometry in double null coordinates.

\subsection{``naive $g_{1}$ metric''}\label{subsec:naive metric}

We choose a certain $v$ value which we denote $v_{0}$. Due to the
smallness of $\varepsilon$ there is a range of $v$ in the vicinity
of $v_{0}$ in which the geometry is RN to a good approximation. The
Vaidya metric there reduces to the RN geometry in ingoing Eddington
coordinates $v,r$. We now transform it to double null coordinates
$u,v$. Taking Eq. (\ref{eq:RN double null}) and applying with local
$M(v_{0})$ and $Q(v_{0})$ parameters:

\begin{equation}
ds^{2}=\left(1-\frac{2M(v_{0})}{r(u,v)}+\frac{Q(v_{0})^{2}}{r(u,v)^{2}}\right)du\;dv+r(u,v)^{2}d\Omega^{2}\label{eq:v0 naive metric}
\end{equation}
where $r(u,v)$ is a function of $\frac{u+v}{2}=r_{*}$ given implicitly
by the relation:\begin{widetext}

\begin{equation}
r_{*}=r-M(v_{0})+\frac{1}{2\kappa_{+}(v_{0})}\log\left|\frac{r-r_{+}(v_{0})}{M(v_{0})-r_{+}(v_{0})}\right|-\frac{1}{2\kappa_{-}(v_{0})}\log\left|\frac{r-r_{-}(v_{0})}{M(v_{0})-r_{-}(v_{0})}\right|\label{eq:rstar naive metric v0}
\end{equation}

\end{widetext}

We will now seek to construct a global metric that applies in the
entire range of $v$. To achieve this we will simply replace the parameter
$v_{0}$ with $v$, resulting in the following metric which we denote
``naive $g_{1}$ metric'' (for reasons that will be clarified soon):
\begin{equation}
ds^{2}=\left(1-\frac{2M(v)}{r(u,v)}+\frac{Q(v)^{2}}{r(u,v)^{2}}\right)du\;dv+r(u,v)^{2}d\Omega^{2}\label{eq:naive metric}
\end{equation}
The equation defining $r(u,v)$ then becomes:\begin{widetext}
\begin{equation}
\frac{u+v}{2}\equiv r_{*}=r-M(v)+\frac{1}{2\kappa_{+}(v)}\log\left|\frac{r-r_{+}(v)}{M(v)-r_{+}(v)}\right|-\frac{1}{2\kappa_{-}(v)}\log\left|\frac{r-r_{-}(v)}{M(v)-r_{-}(v)}\right|\label{eq:rstar naive metric}
\end{equation}
\end{widetext}For later use, we define $f$ which is a function of
$v$,$r$:

\begin{equation}
f=1-\frac{2M(v)}{r}+\frac{Q(v)^{2}}{r^{2}}
\end{equation}
Note that 
\begin{equation}
\left(\frac{\partial r}{\partial r_{*}}\right)_{v}=f\label{eq:dr_drstar}
\end{equation}

The metric in Eq. (\ref{eq:naive metric}) provides a description
of the Vaidya metric Eq. (\ref{eq:vaidya_metric-1}) in double null
coordinates, up to deviation of order $\varepsilon\ll1$. We shall
now discuss the regularity of this metric at the EH.

Already in RN, the metric in $u$,$v$ Eddington coordinates isn't
regular at the EH, as $g_{uv}=0$ and hence the metric has zero determinant
there. Nevertheless, in RN we can define double null Kruskal coordinate
$U(u)$ and $V(v)$ with the transformation 
\begin{equation}
U=e^{\kappa_{+}u},\hfill V=e^{\kappa_{+}v}\label{eq:kruskal uv}
\end{equation}
Attempting to generalize these Kruskal coordinates to our case, since
$\kappa_{+}$ is a function of $v$ (unlike the RN case) $U=e^{\kappa_{+}(v)u}$
is not a function of only $u$, and therefore it is not a null coordinate.
Even worse, it turns out that the naive metric (\ref{eq:naive metric})
geometry is not regular at the EH: we will show in Appendix A that
the Ricci scalar actually diverges there. To solve this problem we
will move to the metric $\widetilde{g_{1}}$ (see next subsection). 

Note that this problem arose when we moved from $v_{0}$ in the local
RN metric Eq. (\ref{eq:v0 naive metric}) to the global $v$-dependent
metric (\ref{eq:naive metric}). We need to implement this transition
in a different manner to avoid this pitfall, as we will now describe. 

\subsection{The $\widetilde{g_{1}}$ metric}\label{subsec:The-g1_tilde-metric}

First, we will take the local Reissner--Nordström metric (at $v=v_{0}$)
in double null coordinates Eq. (\ref{eq:v0 naive metric}), and transform
it to different double null coordinates $\widetilde{u}(u),\widetilde{v}(v)$
defined by:

\begin{equation}
\widetilde{v}\equiv\stackrel[0]{v}{\int}\kappa_{+}(v_{0})dv,\qquad\widetilde{u}\equiv\stackrel[0]{u}{\int}\kappa_{+}(v_{0})du\label{eq:vtildeutilde}
\end{equation}
Here $\kappa_{+}$is still a constant, but that will change soon once
we make the parameters vary with $v$ . Then the new metric is:\begin{widetext}

\begin{equation}
ds^{2}=\frac{1}{\kappa_{+}(v_{0})^{2}}\left(1-\frac{2M(v_{0})}{r}+\frac{Q(v_{0})^{2}}{r^{2}}\right)d\widetilde{u}\;d\widetilde{v}+r^{2}\left(d\theta^{2}+\sin^{2}\theta d\phi^{2}\right)
\end{equation}
\end{widetext}We define $\widetilde{r_{*}}$ correspondingly as:
\begin{equation}
\widetilde{r_{*}}=\frac{\widetilde{v}+\widetilde{u}}{2}
\end{equation}
which satisfies:

\begin{equation}
\widetilde{r_{*}}=\kappa_{+}(v_{0})r_{*}
\end{equation}

Now we take this metric and make its parameters $M$,$Q$ (and hence,
also $\kappa_{+}$) vary with $\widetilde{v}$. Our metric is then:

\begin{equation}
ds^{2}=\frac{1}{\kappa_{+}(\widetilde{v})^{2}}\left(1-\frac{2M(\widetilde{v})}{r}+\frac{Q(\widetilde{v})^{2}}{r^{2}}\right)d\widetilde{u}\;d\widetilde{v}+r^{2}\left(d\theta^{2}+\sin^{2}\theta d\phi^{2}\right)
\end{equation}
The definition of $r(\widetilde{u},\widetilde{v})$ is through the
relation:\begin{widetext}
\begin{equation}
\frac{\widetilde{u}+\widetilde{v}}{2}\equiv\widetilde{r_{*}}=\kappa_{+}(v)\left(r-M(v)+\frac{1}{2\kappa_{+}(v)}\log\left|\frac{r-r_{+}(v)}{M(v)-r_{+}(v)}\right|-\frac{1}{2\kappa_{-}(v)}\log\left|\frac{r-r_{-}(v)}{M(v)-r_{-}(v)}\right|\right)
\end{equation}
\end{widetext}Note that when $\widetilde{u}+\widetilde{v}=0$, $r=M(v)$.

We shall call this metric ``the $\widetilde{g_{1}}$ metric''.

Next we discuss the regularity of this geometry at the EH. This metric,
as is, still isn't regular, because $g_{\widetilde{u}\widetilde{v}}=0$
at the EH. However, like in RN, we can transform it to Kruskal coordinates
(identical to those defined in Eq. (\ref{eq:kruskal uv})) which now
take the form:

\begin{equation}
U=e^{\widetilde{u}},\qquad V=e^{\widetilde{v}}
\end{equation}
Recalling that our metric is just like RN along a line of constant
$v$, and recalling Eq. (\ref{eq:guv_exp_kappa_plus_u+v}) , the metric
function $g_{\widetilde{u}\widetilde{v}}$ around $r_{+}$ will be
\begin{equation}
g_{\widetilde{u}\widetilde{v}}=\frac{g_{uv}}{\kappa_{+}^{2}(v)}=-W(v)\cdot e^{\kappa_{+}\left(u+v\right)}=-W(v)\cdot e^{\widetilde{u}+\widetilde{v}}
\end{equation}
where $W(v)$ is a non-zero function which depends only on $v$. Transforming
into $U,V$ we get:

\begin{equation}
g_{UV}=-W(v)
\end{equation}
Our new metric is therefore regular at the EH.

Next we wish to construct initial values for the metric and its derivatives.
More specifically, we need the expressions for $R$, $\widetilde{S}$,
$R_{,\widetilde{u}},$ $\widetilde{S}_{,\widetilde{u}}$, where $\widetilde{S}$
denotes the value of $S$ in the coordinates $\widetilde{u},\widetilde{v}$,
that is, $\frac{e^{\widetilde{S}}}{r}=-g_{\widetilde{u}\widetilde{v}}$.
(The derivatives $R_{,\widetilde{u}},$ $\widetilde{S}_{,\widetilde{u}}$
allow us to obtain the initial time derivatives of $R$ and $\widetilde{S}$,
see appendix B.) Recall that 
\begin{equation}
R=r^{2}
\end{equation}
We define:
\begin{equation}
r_{*}\equiv\frac{\widetilde{r_{*}}}{\kappa_{+}^{v}(v)}=\frac{\widetilde{u}+\widetilde{v}}{2\kappa_{+}^{v}(v)}\label{eq:r star g1}
\end{equation}
Therefore, using Eq. (\ref{eq:dr_drstar}), the derivative of $r$
with respect to $\widetilde{u}$ is:

\begin{equation}
\frac{\partial r}{\partial\widetilde{u}}=\left(\frac{\partial r}{\partial r_{*}}\right)_{v}\left(\frac{\partial r_{*}}{\partial\widetilde{u}}\right)_{v}=\frac{f}{2\kappa_{+}^{v}(v)}\label{eq:r,utilde}
\end{equation}
The derivative of $R$ on the initial values is then:

\begin{equation}
R_{,\widetilde{u}}=2rr_{,\widetilde{u}}=\frac{rf}{\kappa_{+}(v)}\label{eq:R,utilde}
\end{equation}
To construct $\widetilde{S}$ and its derivative, we first recall
that $g_{\widetilde{u}\widetilde{v}}$ on the initial values is:

\begin{equation}
\frac{e^{\widetilde{S}}}{r}=-g_{\widetilde{u}\widetilde{v}}=-\frac{f}{2\kappa_{+}(v)^{2}}
\end{equation}
namely,

\begin{equation}
\widetilde{S}=\log\left(-\frac{rf}{2\kappa_{+}^{2}(v)}\right)
\end{equation}
The derivative of $\widetilde{S}$ with respect to $\widetilde{u}$
is:

\begin{equation}
\widetilde{S}_{,\widetilde{u}}=\left(\frac{\partial}{\partial r}\log\left(-rf\right)\right)r_{,\widetilde{u}}=\left(\frac{\partial}{\partial r}\log\left(-rf\right)\right)\frac{f}{2\kappa_{+}^{2}}
\end{equation}

We got a metric which has a regular geometry, and we found the needed
initial values for this metric. However, this metric uses $\widetilde{v},\widetilde{u}$
coordinates which are not convenient to work with. For the numerical
simulation, it will be more convenient to use null coordinates similar
to Eddington $u,v$ coordinates. In the following subsection, we will
move to such more convenient coordinates.

\subsection{$g_{1}$ metric}

We wish to do coordinate transformation from $\widetilde{g_{1}}$
with $\widetilde{v},\widetilde{u}$ coordinates to new double null
coordinates whose scaling (with regards to $M$) is more similar to
the Eddington $v,u$ coordinates in RN. (Because it is just coordinate
transformation, it will keep the geometry regular at the EH.) For
simplicity, we'll call these new coordinates $v,u$ (however, this
new metric should \emph{not} be confused with the ``naive metric''
described in subsection \ref{subsec:naive metric}). We shall call
the new resulting metric simply ``the $g_{1}$ metric''. 

The desired coordinate transformation is:

\begin{equation}
\widetilde{v}(v)\equiv\stackrel[0]{v}{\int}\kappa_{+}^{v}(v)dv,\hfill\widetilde{u}(u)\equiv\stackrel[0]{u}{\int}\kappa_{+}^{u}(u)du\label{eq:vtildeutilde-1}
\end{equation}
where $\kappa_{+}^{u}(u)$ was defined in Eq. (\ref{eq:kappa plus definition}).
These coordinates satisfy the following relations:

\begin{equation}
\frac{d\widetilde{v}}{dv}=\kappa_{+}^{v}(v),\hfill\frac{d\widetilde{u}}{du}=\kappa_{+}^{u}(u)
\end{equation}
Notice that the definition of $\widetilde{u}(u)$ uses $\kappa_{+}^{u}(u)$
rather than $\kappa_{+}^{v}(v)$ because it must be a function of
only $u$ (not $v$), so that $u$ will be a null coordinate. The
metric function $g_{\widetilde{u}\widetilde{v}}$ now transforms to:

\begin{align}
g_{uv} & =g_{\widetilde{u}\widetilde{v}}\frac{\partial\widetilde{u}}{\partial u}\frac{\partial\widetilde{v}}{\partial v}\\
 & =\left(1-\frac{2M(v)}{r}+\frac{Q(v)^{2}}{r^{2}}\right)\frac{1}{2\kappa_{+}^{v}(v)^{2}}\cdot\left(\kappa_{+}^{v}(v)\kappa_{+}^{u}(u)\right)
\end{align}
The new line element is therefore:

\begin{equation}
ds^{2}=\frac{\kappa_{+}^{u}(u)}{\kappa_{+}^{v}(v)}fdu\;dv+r^{2}d\Omega^{2}\label{eq:g2_metric}
\end{equation}
where, recall,

\begin{equation}
f=1-\frac{2M(v)}{r}+\frac{Q(v)^{2}}{r^{2}}\label{eq:f g2}
\end{equation}

We'll now construct the four functions which will be needed for initial
values: $R$, $S$, $R_{,u}$, $S_{,u}$. See appendix B for why we
require these $u$ derivatives. The function $S$ is:

\begin{equation}
S=\log\left(-rg_{uv}\right)=\log\left(-\frac{1}{2}rf\right)+\log\left(\kappa_{+}^{u}(u)\right)-\log\left(\kappa_{+}^{v}(v)\right)
\end{equation}
To get the function $R=r^{2}$, we recall the definition of $r_{*}$
given in Eq. (\ref{eq:r star g1})

\begin{equation}
r_{*}=\frac{\widetilde{u}+\widetilde{v}}{2\kappa_{+}^{v}(v)}
\end{equation}
where now $\widetilde{u},\text{\ensuremath{\widetilde{v}} are given by Eq. }$(\ref{eq:vtildeutilde-1})
as functions of our new coordinates $u$,$v$ respectively. Then we
find $r(u,v)$ as the inverse function of the relation:\begin{widetext}

\begin{equation}
\frac{\widetilde{u}(u)+\widetilde{v}(v)}{2\kappa_{+}^{v}(v)}=r_{*}=r-M+\frac{1}{2\kappa_{+}}\log\left|\frac{r-r_{+}}{M-r_{+}}\right|-\frac{1}{2\kappa_{-}}\log\left|\frac{r-r_{-}}{M-r_{-}}\right|
\end{equation}
\end{widetext}The $u$ derivative of $r$ is then (using Eq. (\ref{eq:r,utilde})):

\begin{equation}
r_{,u}=r_{,\widetilde{u}}\widetilde{u}_{,u}=f\frac{\kappa_{+}^{u}(u)}{2\kappa_{+}^{v}(v)}
\end{equation}
and therefore:

\begin{equation}
R_{,u}=2rr_{,u}=rf\frac{\kappa_{+}^{u}(u)}{\kappa_{+}^{v}(v)}
\end{equation}
The derivative of $S$ with respect to $u$ is:
\begin{equation}
S_{,u}=\left(\frac{\partial}{\partial r}\log\left(-rf\right)\right)r_{,u}+\frac{1}{\kappa_{+}^{u}(u)}\frac{d}{du}\kappa_{+}^{u}(u)
\end{equation}
where in $\frac{\partial}{\partial r}\log\left(-rf\right)$ the derivative
is taken with $M$,$Q$ being constant:

\begin{equation}
\frac{\partial}{\partial r}\log\left(-rf\right)=\frac{1-\frac{Q^{2}}{r^{2}}}{rf}
\end{equation}
Also note that:

\begin{equation}
\frac{d}{du}\left(\kappa_{+}^{u}(u)\right)=\frac{\partial}{\partial M}\kappa_{+}^{u}(u)\frac{d}{du}M^{u}(u)+\frac{\partial}{\partial Q}\kappa_{+}^{u}(u)\frac{d}{du}Q^{u}(u)
\end{equation}

Finally, we show that if $u+v=0,$then $\widetilde{u}+\widetilde{v}=0$
and therefore $r=M(v)$. Choosing a specific value $v=v_{0}$, from
$u+v=0$ we obtain that $u=-v_{0}.$ We then get from Eq. (\ref{eq:vtildeutilde-1})
that:

\begin{equation}
\widetilde{v}+\widetilde{u}=\stackrel[0]{v_{0}}{\int}\kappa_{+}^{v}(v)dv+\stackrel[0]{-v_{0}}{\int}\kappa_{+}^{u}(u)du
\end{equation}
Using the definition of $\kappa_{+}^{u}(u)$:

\begin{equation}
\kappa_{+}^{u}(u)=\kappa_{+}^{v}(v=-u)
\end{equation}
we get that:

\begin{align}
\widetilde{v}+\widetilde{u} & =\stackrel[0]{v_{0}}{\int}\kappa_{+}^{v}(v)dv+\stackrel[0]{-v_{0}}{\int}\kappa_{+}^{v}(-u)du\\
 & =\stackrel[0]{v_{0}}{\int}\kappa_{+}^{v}(v)dv-\stackrel[0]{v_{0}}{\int}\kappa_{+}^{v}(w)dw=0
\end{align}
(In the last equality we changed the integration variable $w=-u$.)
We therefore conclude that when $u+v=0$, $\widetilde{u}+\widetilde{v}=0$
and therefore (as seen in subsection \ref{subsec:The-g1_tilde-metric})
$r=M(v)$.

\subsection{Initial hypersurface and the range of the $v$ and $u$ coordinates}

Basically, we have freedom to add constant to the coordinate $v$.
Once we've chosen this additive constant for $v$, the other coordinate
$u$ is uniquely fixed by the condition $r_{*}(r=M)=0$, recalling
that $r_{*}\equiv\frac{u+v}{2}$. \footnote{Note that this is the case in the metric $g_{1}$ from which we pick
the initial values near the EH. In the semiclassical metric that evolves
later on, the relation $r_{*}(r=M)=0$ holds only up to deviations
of order $\varepsilon$.}

We mark $T\equiv\frac{u+v}{2}$. We choose the initial-value surface
to be a surface of constant $T$, which we denote by $T_{0}$. $T_{0}$
is one of the parameters of the numerical simulation, which we set
to $T_{0}=-30$ unless stated otherwise. Note that $r-r_{+}=const\cdot e^{2\kappa_{+}T}$,
and therefore, the deviation of $r$ at the initial hypersurface from
its EH value $r_{+}$ is exponential in $T_{0}$. In all our simulations,
$r-r_{+}<4\cdot10^{-7}$ on the initial hypersurface.

In our numerical simulation, $v$ runs along the initial hypersurface
from $0$ up to $\Delta v$, where $\Delta v$ is a parameter specifying
how large is the numerical domain. We will set $\Delta v=900$ in
most cases. Recall that we use units where $M_{0}=1$, where $M_{0}$
is the initial BH mass.

\begin{figure*}
\caption{v and u coordinates}\label{fig:uv_figure}
\includegraphics{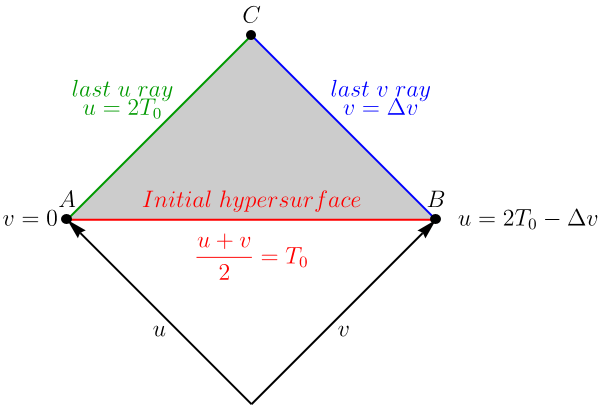}
\end{figure*}

At the point $v=0$ on the initial hypersurface, namely the point
marked A in Figure \ref{fig:uv_figure}, $\frac{u+v}{2}=\frac{u}{2}=T_{0}$
and therefore $u=2T_{0}$ there. At the point $v=\Delta v$ on the
initial hypersurface, namely the point marked B, \mbox{$\frac{u+v}{2}=\frac{u+\Delta v}{2}=T_{0}$}
and therefore $u=2T_{0}-\Delta v$. Therefore, the choice of the value
of $T_{0}$ determines the range of $u$: $u$ runs from $2T_{0}-\Delta v$
to $2T_{0}$. To summarize, the point A is located at $v=0,u=2T_{0}$,
the point B at $v=\Delta v,u=2T_{0}-\Delta v$; therefore the point
C is at $v=\Delta v,u=2T_{0}-\Delta v$.

The domain of the numerical integration is the grey equilateral right
triangle shown in Figure \ref{fig:uv_figure}. The hypotenuse of the
triangle is the initial hypersurface, and the two sides are defined
by the two null rays $u=2T_{0}$ and $v=\Delta v$.

We will usually show plots of the various quantities along the two
final rays: $u=2T_{0}$ and $v=\Delta v$. The $u=2T_{0}$ ray is
called the \emph{last u ray}, and the $v=\Delta v$ ray is called
the \emph{last v ray}. When we plot quantities along the last $u$
ray $u=2T_{0}$, the plot will be a function of v, and when we plot
quantities along the last $v$ ray $v=\Delta v$, the plot will be
a function of u. Usually in our simulation, as mentioned above, $\Delta v=900$
and $T_{0}=-30$, so the mentioned two rays are $u=-60$ and $v=900$.

\part{Numerical results}

We now present the results of running the numerical code. Recall that
we set our units such that $M_{0}=L_{0}=1$. As a default, we set
$du=dv=\frac{1}{100}$, meaning there are hundred grid points per
1 unit of $u$ or $v$. We set the initial-moment parameter $T_{0}=-30$
as a default (unless stated otherwise). Recall that the range of $v$
is usually from $v=0$ to $v=\Delta v=900$, and correspondingly the
range of $u$ is from $u=2T_{0}-\Delta v=-960$ to $u=2T_{0}=-60$.
Recall that we set $\varepsilon_{0}=\frac{1}{1000}.$

First we will present the results of running with no source terms
(which mean that $T_{uv}=T_{\theta\theta}=0$) in section \ref{sec:Numerical-results-run0}.
We will present and explore the resulting quantities $R,S$, $\widetilde{T}_{vv}$
and $\widetilde{T}_{uu}$. In region 2 we will compare the evolving
metric versus that of the corresponding local RN metric. Then we turn
to explore region 3, namely the near-IH region. We will graphically
display both the numerical result and the corresponding analytical
approximation from section \ref{sec:Analytical-approximation}. Then
we shall also display the deviation from the analytical approximation
caused primarily by finite $\varepsilon_{0}$.

Then we will present the result of running with all sources set to
1 in section \ref{sec:Numerical-results-run1}. We shall again display
the quantities of the analytical approximation and their corresponding
numerical results, as well as the deviations from the analytical approximation. 

Then in the next stage in section \ref{sec:Finite-du-and-T0} we will
analyze the consequences of running with finite $T_{0}$ (the value
of $T=\frac{u+v}{2}$ on the initial surface) and finite $du$ --
by running with several different values of these parameters and comparing
the results. 

Afterwards in section \ref{sec:Independence-from-the-detail-of-sources}
we will display that the detail of sources doesn't matter, only the
values of $\widetilde{T}_{vv}^{(-)}$ and $\widetilde{T}_{uu}^{(-)}$.
We will do it by running with 3 different combinations of the parameters
$Z_{R0}$,$Z_{Rr}$,$Z_{S0}$,$Z_{Sr}$ chosen such that they all
lead to the same $\widetilde{T}_{vv}^{(-)}$ and $\widetilde{T}_{uu}^{(-)}$.
We display the metric functions $R$ and $S$ in these configurations,
showing that they are the same in those 3 configurations.

Lastly we display the results of running with three other different
combinations of source parameters and show the validity of the analytical
approximation in all those runs in section \ref{sec:runs2-to-run4}

Note that because we are trying to validate the analytical approximation,
the numerical simulations do not incorporate the analytical approximation
in their assumptions.

\section{Numerical results - run with no sources}\label{sec:Numerical-results-run0}

We start by presenting the results of running with zero source terms,
$Z_{R0}=$$Z_{Rr}$$=Z_{S0}$$=Z_{Sr}$$=0$, as well as $\delta Q=0$,
for $\frac{Q}{M}=0.8$. In this case, $T_{uu}=0$ throughout the whole
spacetime, as already mentioned in subsection \ref{subsec:Source-terms}
above. We will first look at the entire domain of integration (including
all three regions). Then we shall have a more focused look: first
on region 2, and then on region 3. In region 2 we will compare the
results with the local RN metric. In region 3 we will compare the
metric functions to the analytical approximation from section \ref{sec:Analytical-approximation}.
We will then analyze the deviation from the analytical approximation.

\subsection{Overview}

\begin{figure*}
\caption{R}

\subfloat[$R(v)$ on last u ray\label{fig:R(v)_allzeros}]{\includegraphics{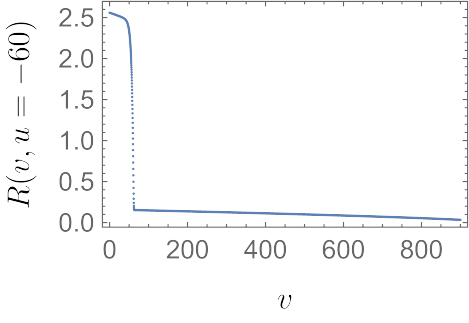}}\subfloat[$R(u)$ on last v ray\label{fig:R(u)_allzeros}]{

\includegraphics{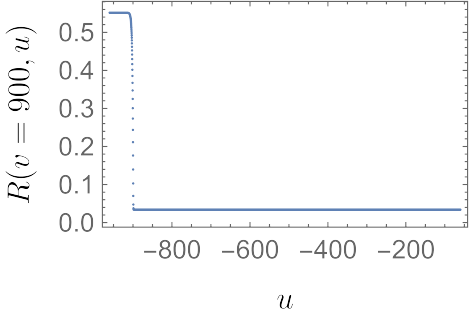}}
\end{figure*}

As one can observe in Figure \ref{fig:R(v)_allzeros}, showing $R(v)$,
$R$ starts at a value of $R_{+}\equiv r_{+}^{2}=2.56$ in region
1, drifts down because of the negative $M_{,v}$ (still in region
1); then it ``falls down'' during the active region (region 2)
-- as it does in RN -- and ends near $R_{-}\equiv r_{-}^{2}=0.16$
near IH, at the beginning of region 3. Note that this apparently rapid
decrease in $R$ is actually smooth and regular. This decrease in
$R$ from $r_{+}$ to $r_{-}$ happens in a timescale of order a few
times $M$ (just like in RN). It only appears sharp because of the
horizontal scale which spans a very large range of $v$. Then $R(v)$
drifts slightly further downwards with a small negative $R_{,v}$
according to Eq. (\ref{eq:RvTvvK}).

The second panel \ref{fig:R(u)_allzeros}, displaying $R(u)$, shows
a slightly different behavior. Because $T_{uu}=0$ (unlike $T_{vv}$
which is not zero), $R$ stays almost constant in region 1, \mbox{$R\approx r_{+}^{2}\left(M(v=900),Q(v=900)\right)=0.551535$},
and also in region 3 where \mbox{$R\approx r_{-}^{2}\left(M(v=900),Q(v=900)\right)=0.034471$}.
It only varies significantly in region 2, similar to the RN case.
In this special case where $J_{u}=0$ and $T_{uu}=0$ the geometry
is the charged Vaidya solution, therefore along an incoming null ray
with $v=const$ all quantities behave exactly like in the corresponding
RN.

\begin{figure*}
\caption{S}\label{fig:S_allzeroes}

\subfloat[S(v) on last u ray\label{fig:S(v)}]{

\includegraphics{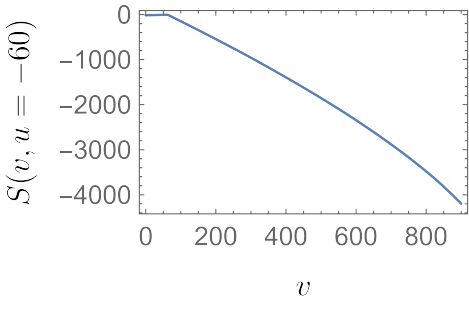}}\subfloat[S(u) on last v ray\label{fig:S(u)}]{

\includegraphics{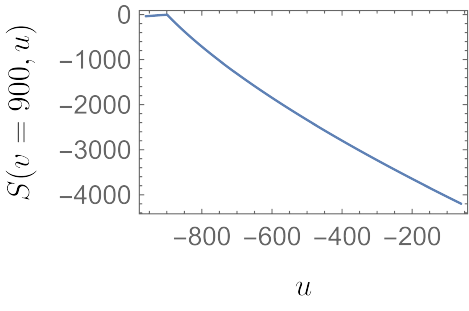}}
\end{figure*}

The following figure \ref{fig:S_allzeroes} shows the behavior of
$S$. We first focus on $S(v)$ in panel \ref{fig:S(v)}. In region
1, $S_{,v}=\kappa_{+}(v)$, and hence $S$ increases towards region
2 (but this is hard to notice in the figure because of the vertical
scale, which mostly shows region 3. The behavior in regions 1 and
2 is better seen in Figure \ref{fig:S(v)RN} below). In region 2,
$S$ peaks and then goes down as $S_{,v}=-\kappa_{-}(v)$ on approaching
the near-IH region. In region 3, $S$ keeps drifting down over the
long stretch from $v=100$ until $v=900$ for the same reason. $S$
eventually reaches very negative values of -4000 and even below. Therefore,
\mbox{$e^{S}\approx0$} to a very good approximation. 

Panel \ref{fig:S(u)} is very similar to panel \ref{fig:S(v)}, and
shows $S$ as a function of $u$, along the other ray $v=900$.

\begin{figure*}
\caption{$\widetilde{T}_{uu}$ on last v,u rays}\label{fig:Tuu_allzeros}
\subfloat[$\widetilde{T}_{uu}$ on last u ray]{\includegraphics{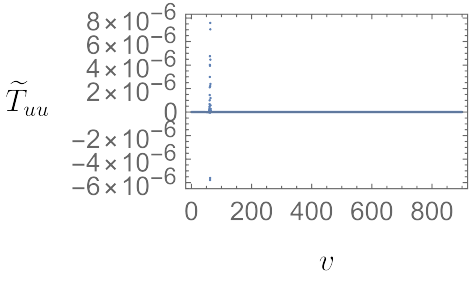}}\subfloat[$\widetilde{T}_{uu}$ on last v ray]{

\includegraphics{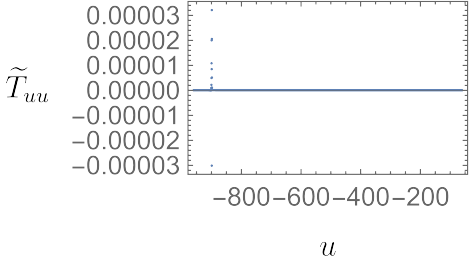}}
\end{figure*}

Next, we show in Figure \ref{fig:Tuu_allzeros} that $\widetilde{T}_{uu}=0$
in the entire domain of integration. The small deviation from zero
(primarily seen in region 2) is due to numerical truncation error
(we verified this by comparing different $du$ values).

\begin{figure*}
\caption{$\widetilde{T}_{vv}$ on last u ray}\label{fig:Tvv(v)}

Blue curve is barely seen as it is underneath the orange curve.

\includegraphics{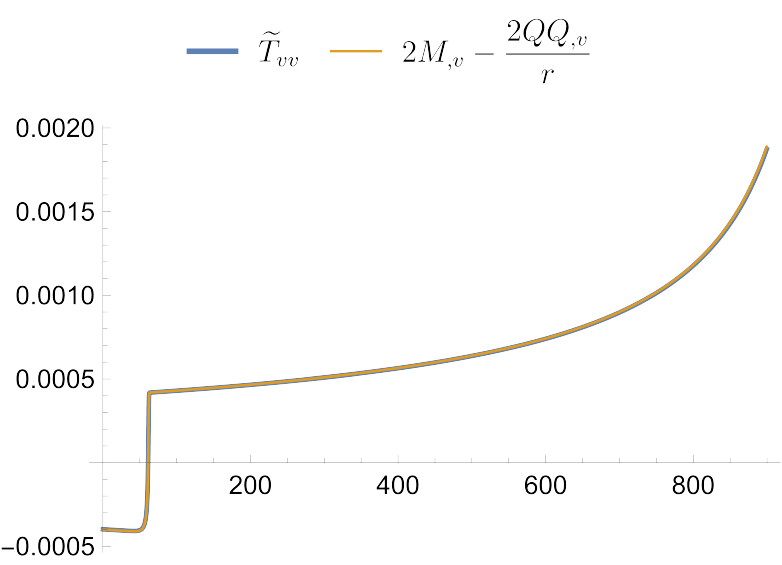}
\end{figure*}

We continue to Figure \ref{fig:Tvv(v)} which shows both $\widetilde{T}_{vv}$
and the quantity

\begin{equation}
2M_{,v}-2\frac{1}{r}QQ_{,v}\label{eq:M,v-1/rQQ,v}
\end{equation}
which is the expected value of $\widetilde{T}_{vv}$ according to
Eq. (\ref{eq:tvv_region1}). The two curves overlap, demonstrating
that $\widetilde{T}_{vv}$ matches its expected value. In region 2,
$\widetilde{T}_{vv}(v)$ goes up from -0.0004 to +0.0004 (due to the
rapid change in $r$ in the above expression $2\frac{1}{r}QQ_{,v}$).
It then continues to drift in region 3 (though more slowly) as all
factors $M_{,v}$ , $r$,$Q$, $Q_{,v}$ appearing in \ref{eq:M,v-1/rQQ,v}
slowly drift there with increasing $v$.

\begin{figure}
\caption{$\widetilde{T}_{vv}(u)$ on last v ray}\label{fig:Tvv(u)}

\includegraphics{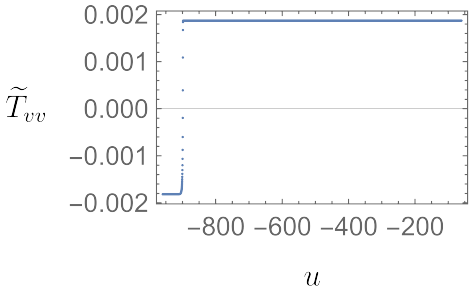}
\end{figure}

Next we see in Figure \ref{fig:Tvv(u)} that $\widetilde{T}_{vv}(u)$
is constant in regions 1 and 3, and it only varies in region 2. This
is because $M_{,v}$ and $Q_{,v}$ and $Q$ depend only on $v$, hence
these quantities are constant along the last $v$ ray. Therefore the
only change in the expression \ref{eq:M,v-1/rQQ,v} comes from changing
$r$. But $r$ doesn't change in regions 1 and 3, it only changes
in region 2. 

\begin{figure*}
\caption{$\widetilde{T}_{vv}$ and $\widetilde{T}_{uu}$ on initial hypersurface}\label{fig:Initial-values}
\subfloat[$\widetilde{T}_{vv}$ on initial hypersurface\label{fig:Tvv-on-initial-values}]{

\includegraphics{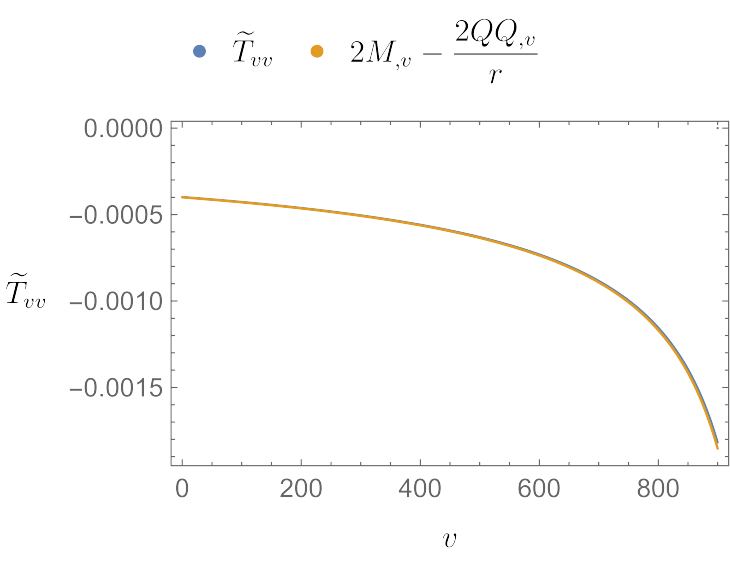}}

\subfloat[$\widetilde{T}_{vv}-\left(2M_{,v}-2\frac{1}{r}QQ_{,v}\right)$ on
the initial hypersurface\label{fig:Tvv-on-initial-values-expected}]{

\includegraphics{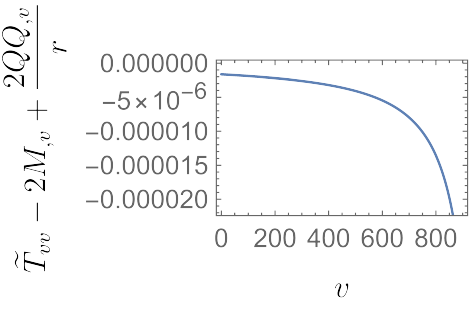}}\subfloat[$\widetilde{T}_{uu}$ on initial hypersurface\label{fig:Tuu-on-initial-values}]{

\includegraphics{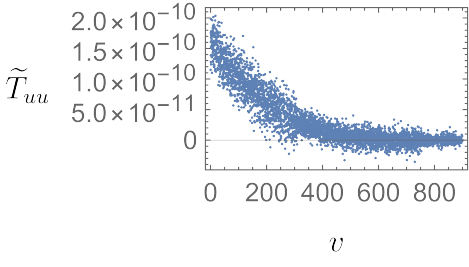}}
\end{figure*}

In Figure \ref{fig:Initial-values} we elaborate on the fluxes $\widetilde{T}_{vv}$
and $\widetilde{T}_{uu}$ on the initial hypersurface $T=T_{0}$ (the
red line in Figure \ref{fig:uv_figure}). Panel \ref{fig:Tvv-on-initial-values}
compares $\widetilde{T}_{vv}$ along with its theoretically expected
value $2M_{,v}-2\frac{1}{r}QQ_{,v}$. $\widetilde{T}_{vv}$ is computed
according to \ref{eq:Tvv definition}. We display the small difference
between these two quantities in Figure \ref{fig:Tvv-on-initial-values-expected},
which shows the error reaching at most $2\cdot10^{-5}$. We see that
the computed value matches the expected value well. Figure \ref{fig:Tuu-on-initial-values},
shows that up to numerical errors ($<2\cdot10^{-10}$, mainly roundoff),
$T_{uu}=0$ on the initial hypersurface line, as it should.

\subsection{Region 2: comparison with the local RN approximation}

Now we will show several figures which test the approximation of region
2 as a drifting RN metric.

\begin{figure*}
\caption{computed R vs. RN value of R}
\subfloat[R computed vs. drifting RN on last u ray\label{fig:R-computed-vs-RN-drifting}]{

\includegraphics{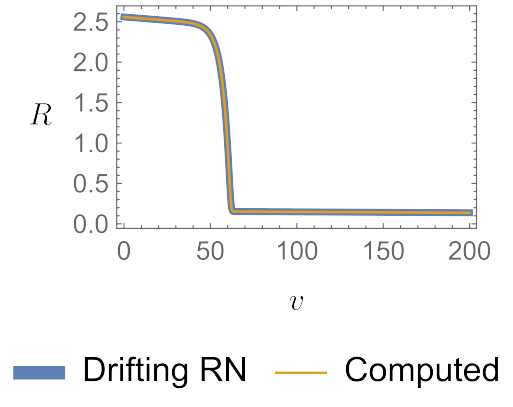}}\subfloat[R(u) computed vs. RN on last v ray\label{fig:R(u)-computed-vs-RN}]{

\includegraphics{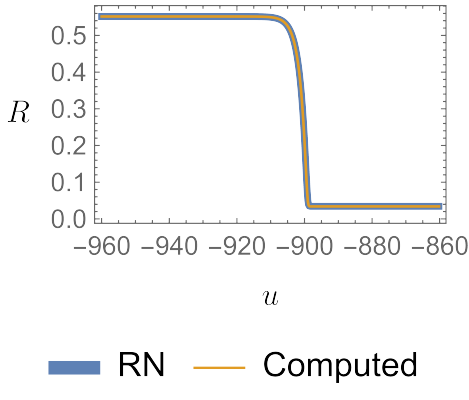}}
\end{figure*}

In Figure \ref{fig:R-computed-vs-RN-drifting} we compare the numerically-computed
$R$ with that corresponding to the RN metric (in double null Eddington
coordinates) but with its $M$ and $Q$ parameters replaced with $M(v)$
and $Q^{v}(v)$. Notice the complete overlap of the two curves .

Subsequently, Figure \ref{fig:R(u)-computed-vs-RN} shows $R(u)$
along the last $v$ ray, versus RN with constant $M,Q$ corresponding
to $M(v=900)$ and $Q^{v}(v=900)$. Notice that because there is no
drift of $M$ and $Q$ parameters in the $u$ direction, the two curves
nicely overlap.

\begin{figure*}
\caption{S computed vs. RN}\label{fig:S(u/v)RN}
\subfloat[S(v) computed vs. RN on last u ray\label{fig:S(v)RN}]{

\includegraphics{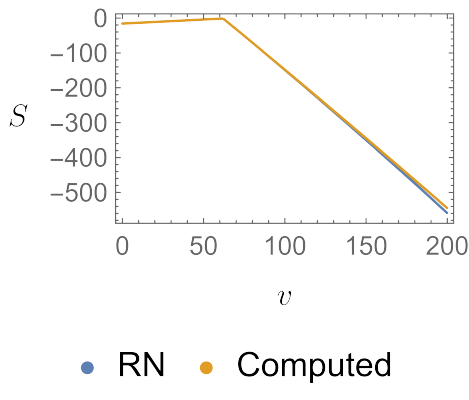}}\subfloat[S(u) computed vs. RN on last v ray\label{fig:S(u)RN}]{

\includegraphics{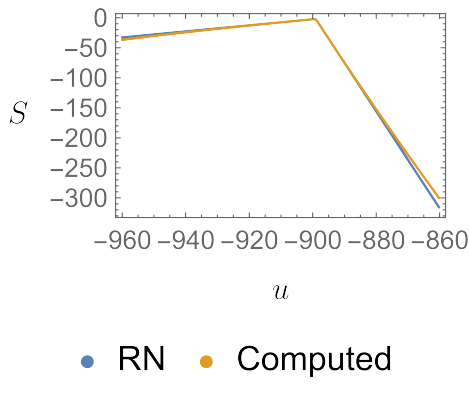}}
\end{figure*}

We move on to show $S$ in Figure \ref{fig:S(u/v)RN} in region 2,
versus the value of $S$ in the corresponding RN metric, determined
by 

\begin{equation}
\frac{e^{S}}{r}=\frac{1}{2}\left(1-\frac{2M}{r}+\frac{Q^{2}}{r^{2}}\right)
\end{equation}
In Panel \ref{fig:S(v)RN} we plot $S(v)$ along the $u=-60$ ray.
Then in panel \ref{fig:S(u)RN} we plot $S(u)$ along the last ray
$v=900$. In both cases, the two curves are in good agreement. The
visible deviation at late times is expected because this range is
further from the domain of validity of the local RN approximation
which is our focus here.

\subsection{Region 3: comparison with the analytical approximation }\label{subsec:Region-3-analytical_allzeroes}

In this subsection, we will compare the numerically computed quantities
$R_{,u},$$R_{,v},$$S_{,u}$,$S_{,v}$ in region 3 with their predicted
values according to Eqs. (\ref{eq:S,v_kappa-},\ref{eq:S,u_kappa-},\ref{eq:RvTvvK},\ref{eq:RuTuuK}).

\begin{figure*}
\caption{Numerical result and analytical approximation for $S_{,v}$, $S_{,u}$}
\subfloat[Comparison of $S_{,v}$ and $-\kappa_{-}^{v}$ on last u ray\label{fig:s,v_kappa-}]{

\includegraphics{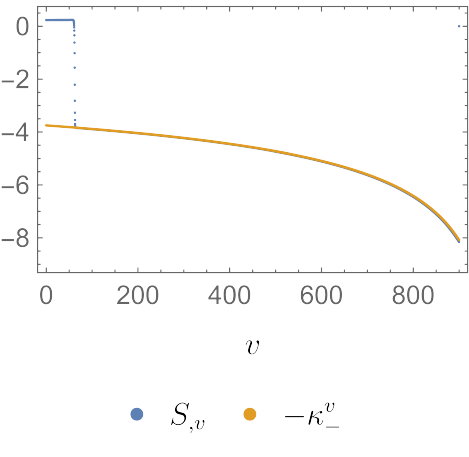}}\subfloat[Comparison of $S_{,u}$ and $-\kappa_{-}^{u}$ on last v ray\label{fig:s,u_kappa-}]{

\includegraphics{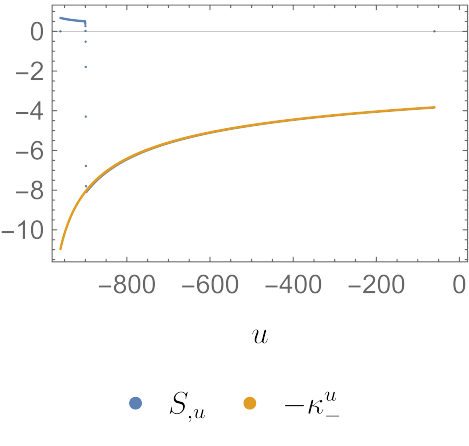}}
\end{figure*}

We start with Figure \ref{fig:s,v_kappa-}, which compares $S_{,v}$
and $-\kappa_{-}(v)$, both displayed as a function of $v$. They
are indeed very close, as predicted by the analytical approximation.
Then we see in Figure \ref{fig:s,u_kappa-} that the approximation
is good for $S_{,u}$ as well, when comparing it to $-\kappa_{-}^{u}$.

\begin{figure*}
\caption{Analytical approximation of $R_{,v}$, $R_{,u}$ compared with their
numerical values}
\subfloat[$R_{,v}$ and $-\frac{\widetilde{T}_{vv}^{(-)}}{\kappa_{-}}$ on last
u ray\label{fig:R,v_all_zeroes}]{

\includegraphics{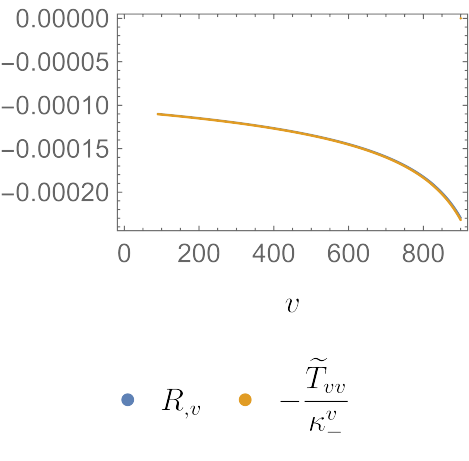}}\subfloat[$R_{,u}$ and $-\frac{\widetilde{T}_{uu}^{(-)}}{\kappa_{-}}$ on last
v ray\label{fig:R,u_all_zeroes}]{

\includegraphics{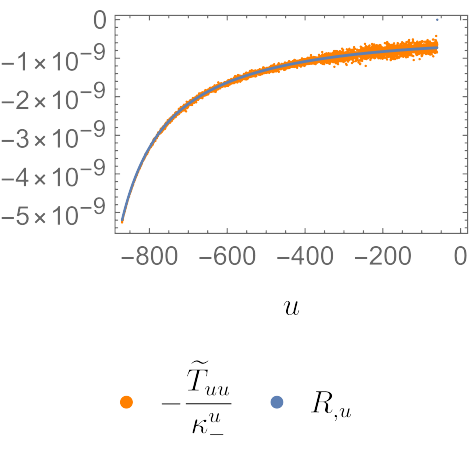}}
\end{figure*}

Moving on to the derivatives of $R$, we see in Figure \ref{fig:R,v_all_zeroes}
that the analytical approximation Eq. (\ref{eq:RvTvvK}) works, comparing
$R_{,v}$ with $-\frac{\widetilde{T}_{vv}^{(-)}}{\kappa_{-}}$ as
they vary as functions of $v$.

Finally, we show in Figure \ref{fig:R,u_all_zeroes} the quantities
$R_{,u}$ and $-\frac{\widetilde{T}_{uu}^{(-)}}{\kappa_{-}^{u}}$
in region 3. These quantities should both vanish, as mentioned in
subsection \ref{subsec:Source-terms}, because we have no sources
in the present run. In the figure we see that both quantities are
very small. It is remarkable that their erroneous non-vanishing value
is still consistent with Eq. (\ref{eq:RuTuuK}). Our interpretation
of this situation is as follows: these erroneous non-vanishing values
happens because of finite-$du$ error. This causes a drift in $\widetilde{T}_{uu}$
leading to non-vanishing $\widetilde{T}_{uu}^{(-)}$-- which in turn
leads to non-vanishing $R_{,u}$ in region 3, in accordance with Eq.
(\ref{eq:RuTuuK}).

To check this interpretation, we show in Figure \ref{fig:R,u_different_dus}
the magnitude of $R_{,u}$ for two different values of $du$: $\frac{1}{100}$
and $\frac{1}{50}$. The third curve shows the $du=\frac{1}{50}$
case multiplied by $\frac{1}{4}$. We can see that it perfectly overlaps
with the curve of $du=\frac{1}{100}$. This shows that the erroneous
non-vanishing value of $R_{,u}$ indeed scales with $du^{2}$, implying
that it is associated with a finite-$du$ truncation error.

\begin{figure*}
\caption{$R_{,u}$ with different $du$ values on last v ray}\label{fig:R,u_different_dus}

Note that the blue curve completely overlap with the green curve.

\includegraphics[scale=0.8]{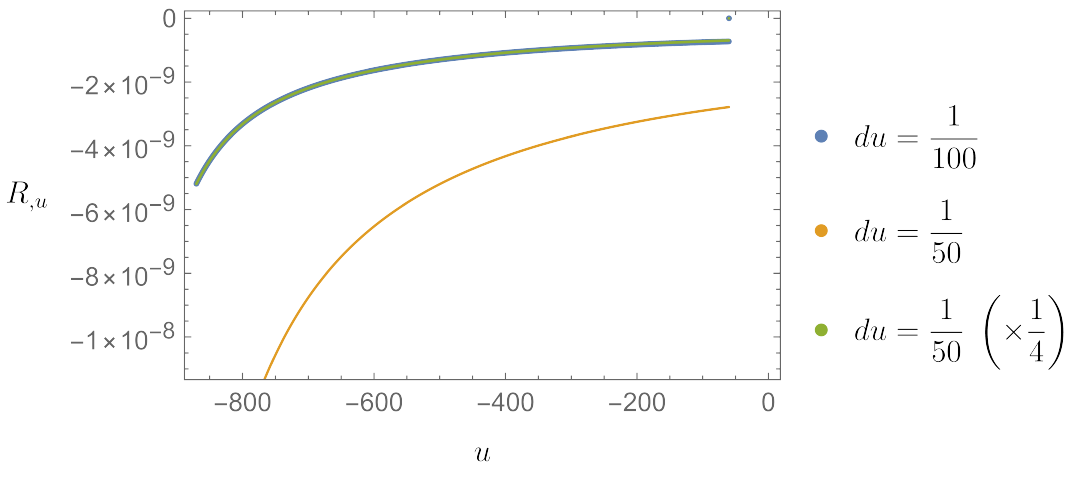}
\end{figure*}

\subsection{Deviation from the analytical approximation (region 3)}\label{subsec:Deviation-analytical-approx-allzeroes}

We will now explore and discuss the deviation of the computed evolving
$R$ and $S$ from the analytical approximation. We expect these deviations
to be caused by finite $\varepsilon_{0}$, as we shall test this expectation
below.

\subsubsection{Deviation in $S_{,v}$}

First, we plot the deviation $S_{,v}+\kappa_{-}$ in Figure \ref{fig:S,v-kappa_all_zeroes}.
(To appreciate the magnitude of this deviation, note that it is roughly
about 1\% of the value of $S_{,v}$ as seen in Figure \ref{fig:s,v_kappa-}.)

\begin{figure*}
\caption{Deviation from analytical approximation of $S_{,v}$, $R_{,v}$}

\subfloat[$S_{,v}+\kappa_{-}$ on last u ray\label{fig:S,v-kappa_all_zeroes}]{

\includegraphics{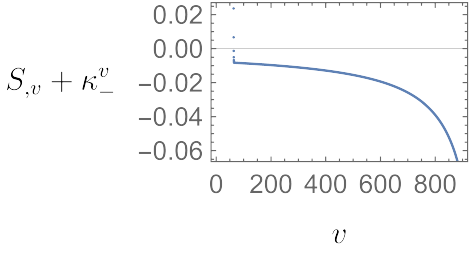}}\subfloat[$R_{,v}+\frac{\widetilde{T}_{vv}}{\kappa_{-}}$ on last u ray\label{fig:Tvv-approx}]{

\includegraphics{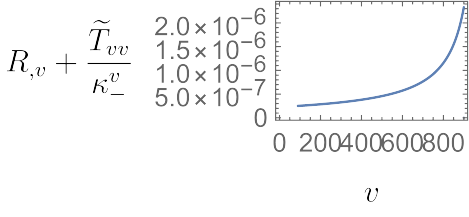}}
\end{figure*}
We also show the deviation of $R_{,v}$ in Figure \ref{fig:Tvv-approx}

To analyze whether this deviation scales with $\varepsilon_{0},$we
do two runs: one with $\varepsilon_{0}=\frac{1}{1000}$ as usual,
and another one with $\varepsilon_{0}=\frac{1}{500}.$ To compare
the two different runs, we plot the mentioned deviations as a function
of $M^{v}(v)$ instead of as a function of $v$, to account for the
fact that the evaporation rate is different in these two cases. Figure
\ref{fig:s,v epsilon comparison x2} compares two different $\varepsilon_{0}$
runs. (Note that in this plot, the low $v$ values correspond to high
$M$, with $v=0$ at $M=1$.) If the deviation is proportional to
$\varepsilon_{0}$, we would expect it to be twice as big in $\varepsilon_{0}=\frac{1}{500}$
compared to $\varepsilon_{0}=\frac{1}{1000}$. 
\begin{figure*}
\caption{$S_{,v}+\kappa_{-}$ with two $\varepsilon_{0}$ values, scaled by
a factor of $\frac{1}{2}$, on the last u ray}\label{fig:s,v epsilon comparison x2}

\includegraphics[scale=0.7]{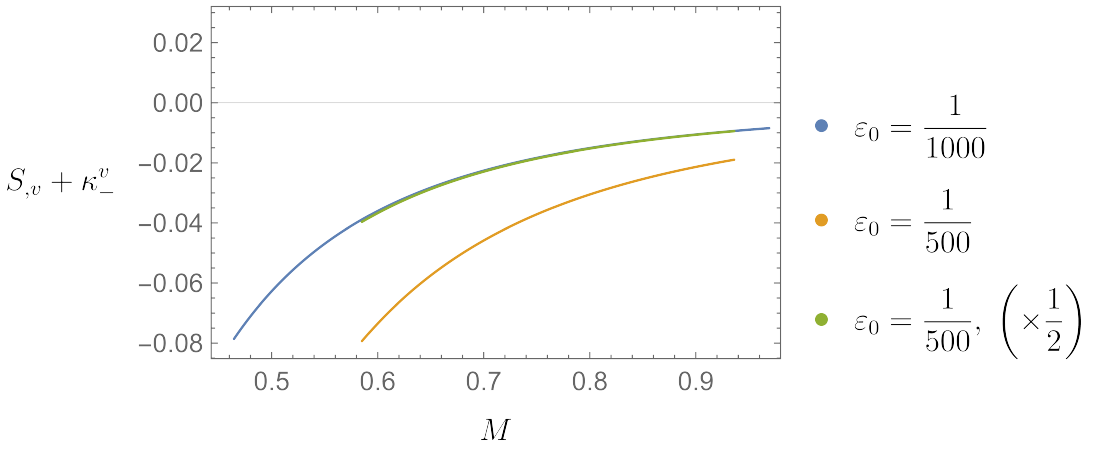}
\end{figure*}
To test this assumption, Figure \ref{fig:s,v epsilon comparison x2}
scales the curve of $\varepsilon_{0}=\frac{1}{500}$ by $\frac{1}{2}$
(green curve), and now we can see the two curves (green and blue)
overlap, confirming that the deviation indeed scales with $\varepsilon_{0}$--meaning
that it is a finite-$\varepsilon_{0}$ effect. We also display the
unscaled $\varepsilon_{0}=\frac{1}{500}$ curve, as is, in orange.
Note that the graph does not start from $M=1$, because here we focus
on the analytical approximation of region 3, therefore we restricted
the curves to the range of $v$ for which the last $u$ ray is within
region 3. In the next figure we see that in earlier $M$ values there
is a large deviation from the analytical approximation of region 3.

We also wish to know how this finite-$\varepsilon_{0}$ deviation
scales with $M$. From basic semiclassical scaling, one would expect
this deviation to scale as $\propto M^{-3}$. This is because the
dimensions of $S_{,v}$ is $[S_{,v}]=[l]^{-1}$ (as $S$ is dimensionless),
and we need to multiply this by the dimensionless factor $\frac{\hbar}{M^{2}}$
(as this deviation in $S_{,v}$ is a semiclassical effect), so the
overall scaling goes as $M^{-3}$. We therefore check it by plotting
this deviation multiplied by $M^{3},$and check that we get relatively
constant line. This approximately constant line is seen in Figure
\ref{fig:s,v epsilon scaling} (note the narrow range of the vertical
scale).

\begin{figure*}
\caption{$\left(S_{,v}+\kappa_{-}\right)M^{3}$ on last u ray}\label{fig:s,v epsilon scaling}

\includegraphics[scale=0.7]{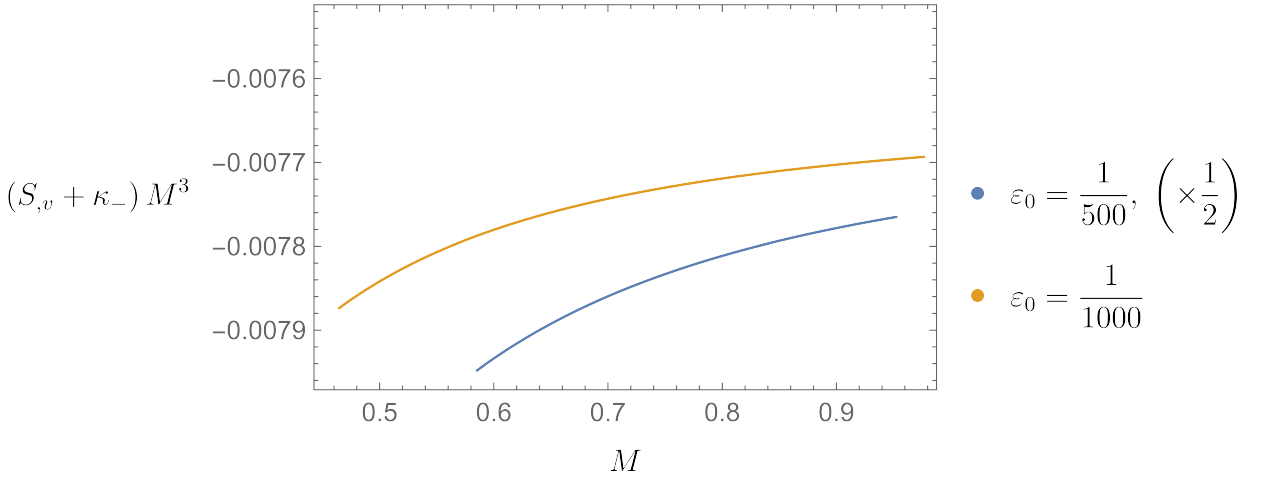}
\end{figure*}

\subsubsection{Deviation in $R_{,v}$}\label{subsec:Deviation-in-R,v}

\begin{figure*}
\caption{$R_{,v}+\frac{\widetilde{T}_{vv}^{(-)}}{\kappa_{-}}$ with two $\varepsilon_{0}$
values, one of them scaled by $\frac{1}{4}$, on the last u ray}\label{fig:Tvv-approx epsilon comparison x4}

\includegraphics[scale=0.7]{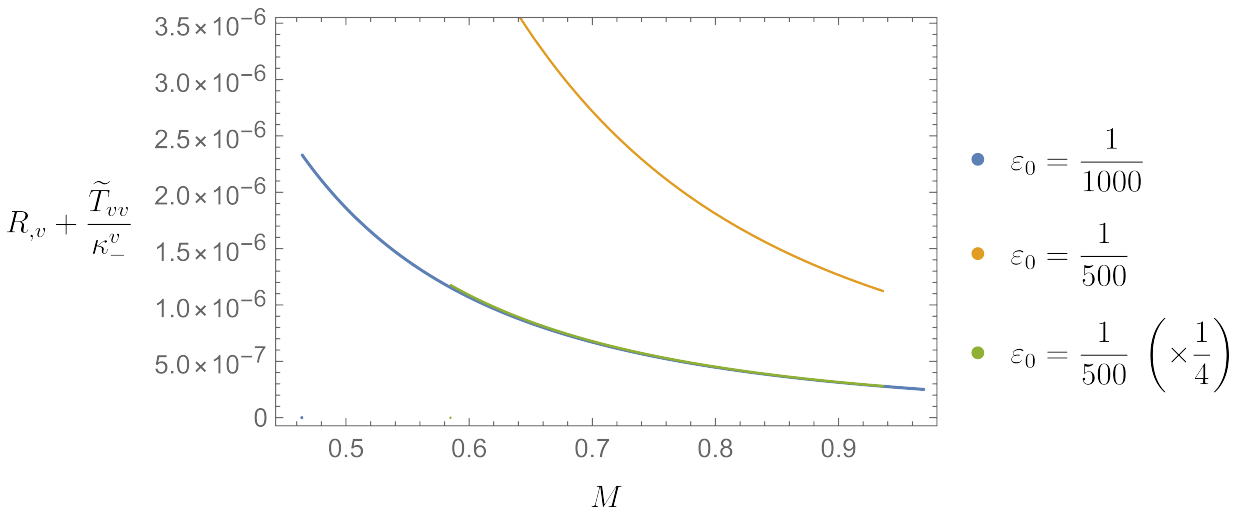}
\end{figure*}

Subsequently, we show in Figure \ref{fig:Tvv-approx epsilon comparison x4}
the deviation from the analytical approximation for $R_{,v}$ for
two different $\varepsilon_{0}$ values. This deviation is equal to
$R_{,v}+\frac{\widetilde{T}_{vv}^{(-)}(v)}{\kappa_{-}(v)}$, see Eq.
(\ref{eq:RvTvvK}) .

This time the deviation is expected to scale as $\varepsilon_{0}^{2}$.
This is because $R_{,v}$ is proportional to $\varepsilon_{0}$ already
at the leading order, so that the finite-$\varepsilon_{0}$ deviation
should be proportional to $\varepsilon_{0}^{2}$. We test this assumption
by checking that the curve of $\varepsilon_{0}=\frac{1}{500}$ is
four times as big as the curve for $\varepsilon_{0}=\frac{1}{1000}$
-- by multiplying the curve of $\varepsilon_{0}=\frac{1}{500}$ by
a factor of $\frac{1}{4}$.

We can see in the figure that the two curves (green and blue) overlap,
confirming the scaling by $\varepsilon_{0}^{2}$ as expected.

\begin{figure*}
\caption{$\left(R_{,v}+\frac{\widetilde{T}_{vv}^{(-)}}{\kappa_{-}}\right)M^{3}$
scaling of the deviation on last u ray}\label{fig:Tvv-approx epsilon scaling x4}

\includegraphics[scale=0.8]{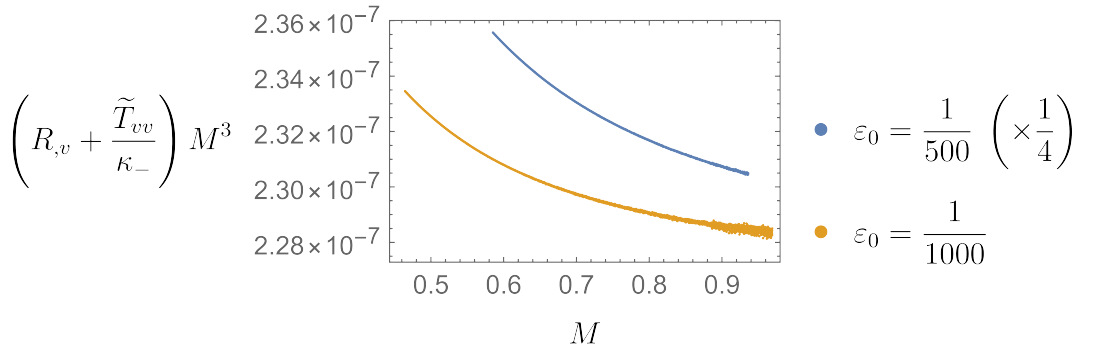}
\end{figure*}

We also wish to know how the mentioned deviation scales with $M$.
This deviation scales as $\propto M^{-3}$, and we show it by plotting
the deviation times $M^{3},$and check that we get a relatively constant
curve. The reason that the deviation scales as $M^{-3}$ is because
the dimensions of $[R_{,v}]=[l]$ (as $R$ has dimensions of length
squared), and we need to further multiply it by a factor of $\varepsilon_{0}^{2}\propto\frac{\hbar^{2}}{M^{4}}$,
so that the total power of $M$ becomes $M^{-3}$. This is indeed
seen in Figure \ref{fig:Tvv-approx epsilon scaling x4} -- note the
small scale of the vertical axis.

\subsubsection{Deviation of $S_{,u}$}

The following Figure \ref{fig:s,u epsilon comparison 2x} shows the
deviation of $S_{,u}$ from $\kappa_{-}^{u}(u)$. Note that we're
now using $M^{u}(u)$ as the horizontal axis, as we need the mass
function that depends on $u$. Like the case of $S_{,v}$, we expect
the deviation of $\varepsilon_{0}=\frac{1}{500}$ to be twice as big
as the deviation for $\varepsilon_{0}=\frac{1}{1000}$. 

\begin{figure*}
\caption{$S_{,u}+\kappa_{-}$ with two $\varepsilon_{0}$ values with one scaled
by $\frac{1}{2}$, on the last v ray}\label{fig:s,u epsilon comparison 2x}

\includegraphics[scale=0.7]{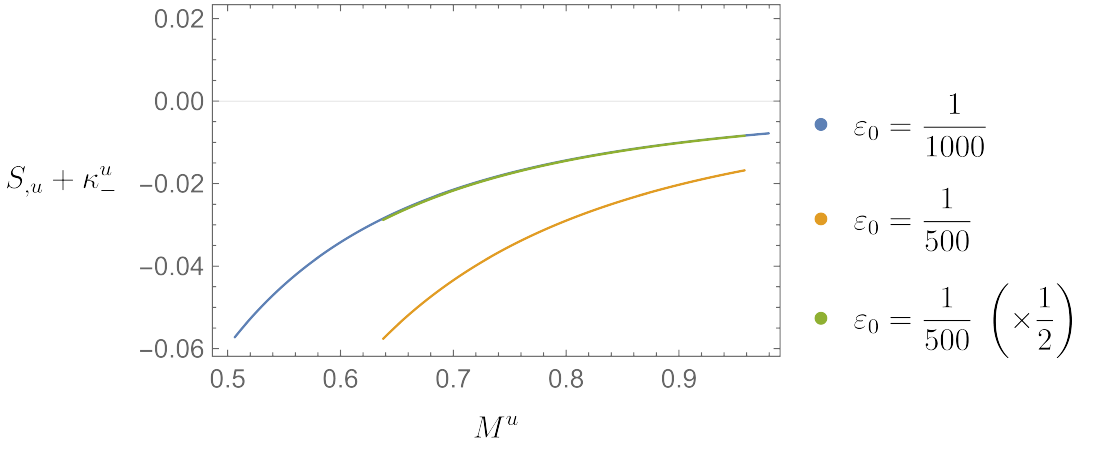}
\end{figure*}
We therefore scale the curve of $\varepsilon_{0}=\frac{1}{500}$ by
a factor of $\frac{1}{2}$ in Figure \ref{fig:s,u epsilon comparison 2x}
in green. This figure also shows the unscaled version in orange. The
green and blue curves overlap, confirming the scaling by $\varepsilon_{0}$.

In the current run, because of the lack of sources,$\widetilde{T}_{uu}=0$
and $R_{,u}=0$ (in region 3), therefore we do not attempt to explore
the finite-$\varepsilon_{0}$ deviation in this case.

While it might seem that the $\varepsilon_{0}$ deviation is not that
small in the present case, note that in astrophysical black holes
$\varepsilon_{0}$ is proportional to $10^{-82}$ and so it is a very
good approximation for these black holes.

\section{Numerical results -- run with all sources set to 1 }\label{sec:Numerical-results-run1}

We will now present a numerical run where all the sources are turned
on; specifically we shall here choose: $Z_{R0}=$$Z_{Rr}$$=Z_{S0}$$=Z_{Sr}$$=1$,
and $\alpha=1$ for $\delta Q$ , with $\frac{Q}{M}=0.8$.

\subsection{Overview}

\begin{figure*}
\caption{R on last u,v rays}\label{fig:R_run1}
\subfloat[R(v) on last u ray\label{fig:R(v)_run1}]{\includegraphics{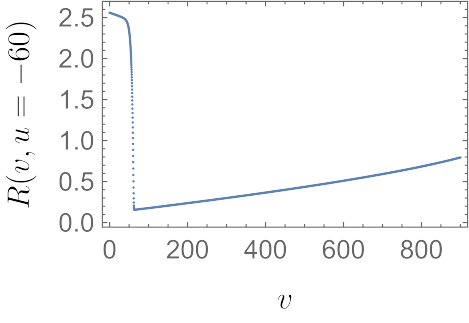}}\subfloat[R(u) on last v ray\label{fig:R(u)_run1}]{\includegraphics{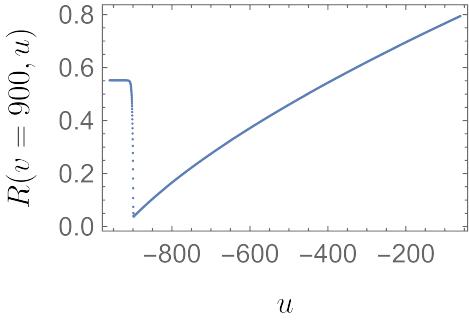}}
\end{figure*}

As one can observe in Figure \ref{fig:R(v)_run1}, showing $R(v)$
along the last-$u$ ray, in region 1 $R$ starts at $r_{+}^{2}(M=1,Q=0.8)=2.56$,
and slowly drifts down because of non-vanishing $M_{,v}$. Afterwards
it quickly falls down during the active region (region 2) -- just
like in RN -- and ends up near $r_{-}^{2}(M=1,Q=0.8)=0.16$ at the
beginning of region 3. Up until region 3, this looks pretty much the
same as the case of zero sources shown in Figure \ref{fig:R(v)_allzeros}.

However, in region 3, the two figures look differently: the difference
in the value of the sources in these two cases leads to a difference
in the corresponding values of $\widetilde{T}_{vv}^{(-)}$. In particular,
the two cases have different signs for $\widetilde{T}_{vv}^{(-)}$.
Thus, in Figure \ref{fig:R(v)_run1} (run with non-vanishing sources)
$R(v)$ goes up in region 3 because of negative $\widetilde{T}_{vv}^{(-)}$,
whereas in Figure \ref{fig:R(v)_allzeros} (run with vanishing sources)
$R$ goes down because of positive $\widetilde{T}_{vv}^{(-)}$. (Recall
that $R_{,v}=-\frac{\widetilde{T}_{vv}^{(-)}}{\kappa_{-}}$ and so
$R_{,v}$ is of the opposite sign to $\widetilde{T}_{vv}^{(-)}$.)

In the next Figure \ref{fig:R(u)_run1}, we look at $R(u)$ along
the last $v$ ray. We can see that $R$ starts at $r_{+}^{2}=0.551535$
(corresponding to the final values of $M$ and $Q$ along the EH)
and it is almost exactly constant throughout region 1, then it falls
down in region 2, similar to Figure \ref{fig:R(u)_allzeros}. Because
of the non-vanishing sources, $\widetilde{T}_{uu}$ no longer vanishes.
In the present case, $\widetilde{T}_{uu}^{(-)}$ is negative and so
$R(u)$ drifts upwards in region 3. This is in contrast to the case
of Figure \ref{fig:R(u)_allzeros} where, due to vanishing sources,
$\widetilde{T}_{uu}^{(-)}=0$ and $R(u)$ stays constant.

\begin{figure*}
\caption{$\widetilde{T}_{uu}(v)$ and $\widetilde{T}_{uu}(u)$ on last u,v
rays}\label{fig:Tuu(v)(u)}
\subfloat[$\widetilde{T}_{uu}(v)$ on last u ray\label{fig:Tuu(v)_run1}]{\includegraphics{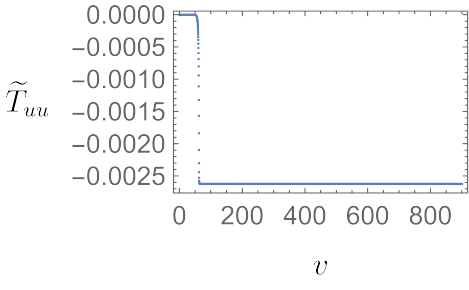}}\subfloat[$\widetilde{T}_{uu}(u)$ on last v ray\label{fig:Tuu(u)_run1}]{\includegraphics{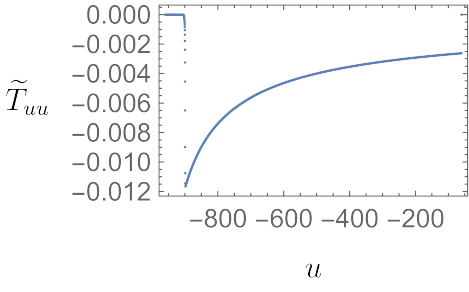}}
\end{figure*}

The following Figure \ref{fig:Tuu(v)(u)} shows $\widetilde{T}_{uu}$
on the two rays, namely last-$u$ (Figure \ref{fig:Tuu(v)_run1})
and last-$v$ (Figure \ref{fig:Tuu(u)_run1}) rays. As can be seen
in both panels, in region 1 near the initial hypersurface, $\widetilde{T}_{uu}=0$.
Then the non-vanishing sources in region 2 cause a non-vanishing,
negative $\widetilde{T}_{uu}$. 

\begin{figure*}
\caption{$\widetilde{T}_{vv}(u,v)$}\label{fig:Tvv(u,v)_run1}
\subfloat[$\widetilde{T}_{vv}(v)$ on last u ray\label{fig:Tvv(v)-run1}]{\includegraphics{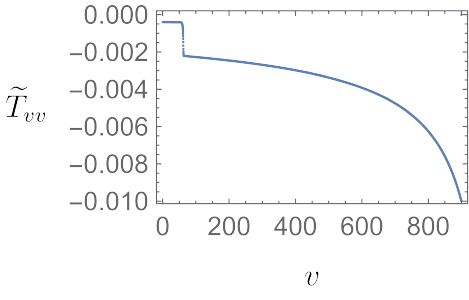}}\subfloat[$\widetilde{T}_{vv}(u)$ on last v ray\label{fig:Tvv(u)-run1}]{\includegraphics{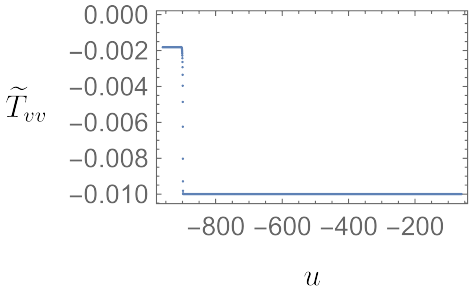}}
\end{figure*}

In Figure \ref{fig:Tvv(u,v)_run1} we display $\widetilde{T}_{vv}$
on both rays, namely last-$u$ (Figure \ref{fig:Tvv(v)-run1}) and
last-$v$ (Figure \ref{fig:Tvv(u)-run1}) rays. In Figure \ref{fig:Tvv(v)-run1}
we can see that the chosen source values caused a negative $\widetilde{T}_{vv}$
in region 3, in contrast to Figure \ref{fig:Tvv(v)} where $\widetilde{T}_{vv}$
is positive there.

Plotting $\widetilde{T}_{vv}$ as a function of $u$ on the last-$v$
ray in Figure \ref{fig:Tvv(u)-run1}, we see that it is almost exactly
constant in region 1 and 3, it only changes in region 2 as a result
of the non-vanishing sources there.

\begin{figure*}
\caption{Scaling of $\widetilde{T}_{uu}$ and $\widetilde{T}_{vv}$}
\subfloat[$\widetilde{T}_{vv}(v)M^{2}(v)$ on last u ray\label{fig:Tvv(v)M^2-run1}]{\includegraphics{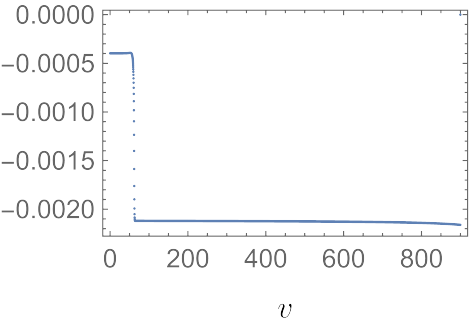}}\subfloat[$\widetilde{T}_{uu}M^{u}(u)^{2}$ on last v ray \label{fig:Tuu(u)Mu^2}]{\includegraphics{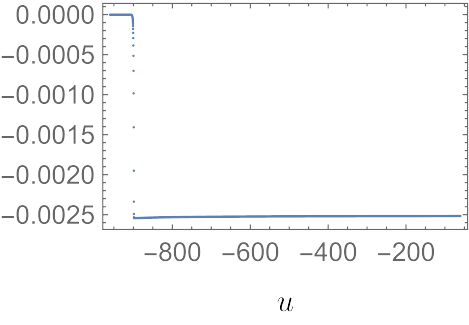}}
\end{figure*}

The scaling of both $\widetilde{T}_{vv}^{(+)}$ and $\widetilde{T}_{vv}^{(-)}$
should\footnote{This simple scaling holds in our case because $\frac{Q}{M}$ is constant
along the EH, and in addition all source terms were constructed with
the appropriate $M$ scaling.} be $\propto M^{-2}$ and to illustrate that, we plot $\widetilde{T}_{vv}M^{2}$
in Figure \ref{fig:Tvv(v)M^2-run1}. This quantity is approximately
constant in both region 1 and region 3, confirming that $\widetilde{T}_{vv}\propto M^{-2}$
there.

Next we do a similar check but for $\widetilde{T}_{uu}(u)$ rather
than $\widetilde{T}_{vv}(v)$. To this end, we now plot $\widetilde{T}_{uu}M^{u}(u)^{2}$
as a function of $u$ in Figure \ref{fig:Tuu(u)Mu^2}. This quantity
is again approximately constant in region 1 and region 3, confirming
that $\widetilde{T}_{uu}\propto M^{u}(u)^{-2}$ in both these regions. 

Because the sources only affect the metric during and after region
2, in region 1 the results are (almost) exactly the same as the no-sources
run. In particular the behavior of $\widetilde{T}_{vv}(v)$ on the
initial hypersurface in the present case is exactly the same as seen
in Figure \ref{fig:Tvv-on-initial-values}.

\subsection{Region 3: comparison with the analytical approximation}

Like in the corresponding subsection \ref{subsec:Region-3-analytical_allzeroes},
we will now compare $R_{,u},$$R_{,v},$$S_{,u}$,$S_{,v}$ with their
expected values according to Eqs. (\ref{eq:S,v_kappa-},\ref{eq:S,u_kappa-},\ref{eq:RvTvvK},\ref{eq:RuTuuK}).

\begin{figure*}
\caption{Analytical approximation of $S_{,v}$ and $S_{,u}$}
\subfloat[$S_{,v}$ and $-\kappa_{-}$ on last u ray\label{fig:s,v_kappa-run1}]{

\includegraphics{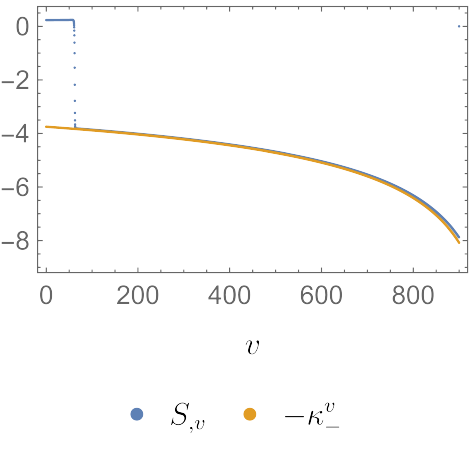}}\subfloat[$S_{,u}$ and $-\kappa_{-}^{u}$ on last v ray\label{fig:s,u_kappa-run1}]{

\includegraphics{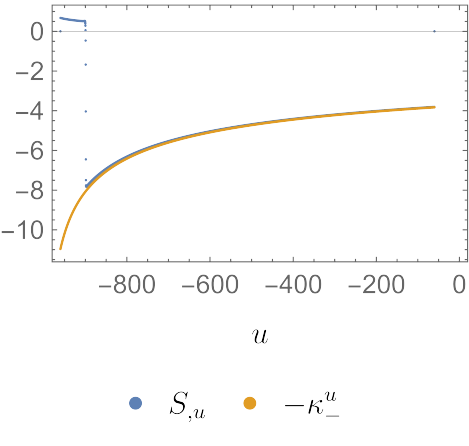}}
\end{figure*}

The first Figure \ref{fig:s,v_kappa-run1}, show $S_{,v}(v)$ and
$-\kappa_{-}(v)$ along the last-$u$ ray. They are indeed very close,
as predicted by the analytical approximation. However, there is a
small visible gap near the end at around $v\sim900$ (which is almost
invisible in the corresponding Figure \ref{fig:s,v_kappa-}). We will
show in the next subsection that this is a finite-$\varepsilon_{0}$
deviation.

We see in the following Figure \ref{fig:s,u_kappa-run1} that the
approximation is good for $S_{,u}$ as well. Similar to the $S_{,v}$case,
there is a small visible gap near the end at $u=-950$ (which doesn't
appear at the corresponding Figure \ref{fig:s,u_kappa-}). This is
again a finite-$\varepsilon_{0}$ deviation.

\begin{figure*}
\caption{Analytical approximation of $R_{,v}$ and $R_{,u}$}
\subfloat[$R_{,v}$ and $-\frac{\widetilde{T}_{vv}}{\kappa_{-}}$ on last u
ray\label{fig:R,v_all_zeroes-run1}]{

\includegraphics{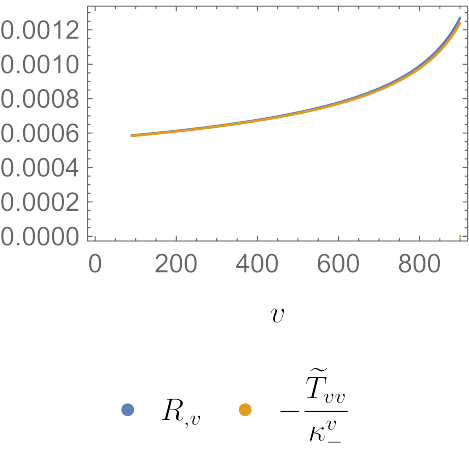}}\subfloat[$R_{,u}$ and $-\frac{\widetilde{T}_{uu}}{\kappa_{-}^{u}}$ on last
v ray\label{fig:R,u_all_zeroes-run1}]{

\includegraphics{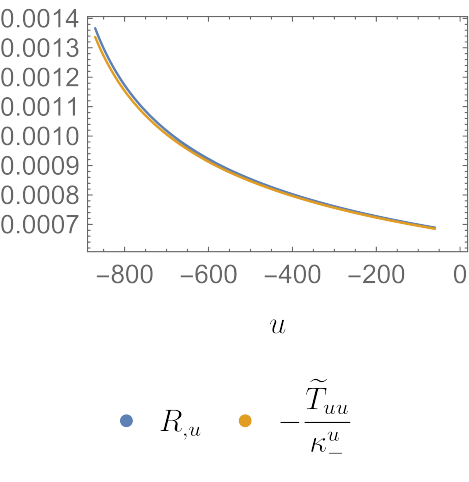}}
\end{figure*}

We move on to explore the analytical approximation for $R$. In Figure
\ref{fig:R,v_all_zeroes-run1}, we see that the analytical approximation
for $R_{,v}$ Eq. (\ref{eq:RvTvvK}) works well. In this case, in
contrast to Figure \ref{fig:R,v_all_zeroes}, $R_{,v}>0$ because
in the present case the chosen sources lead to $\widetilde{T}_{vv}<0$.

Finally, we show in Figure \ref{fig:R,u_all_zeroes-run1} that the
analytical approximation for $R_{,u}$ Eq. (\ref{eq:RuTuuK}) works
as well. This time, unlike Figure \ref{fig:R,u_all_zeroes} (the case
of vanishing sources), $\widetilde{T}_{uu}\neq0$ because of the non-vanishing
sources. 

\subsection{Impact of $\delta Q$}

\begin{figure*}
\caption{$Q^{v}$ and $Q=Q^{v}+\delta Q$ on the last u ray}\label{fig:Q_plot}

\includegraphics{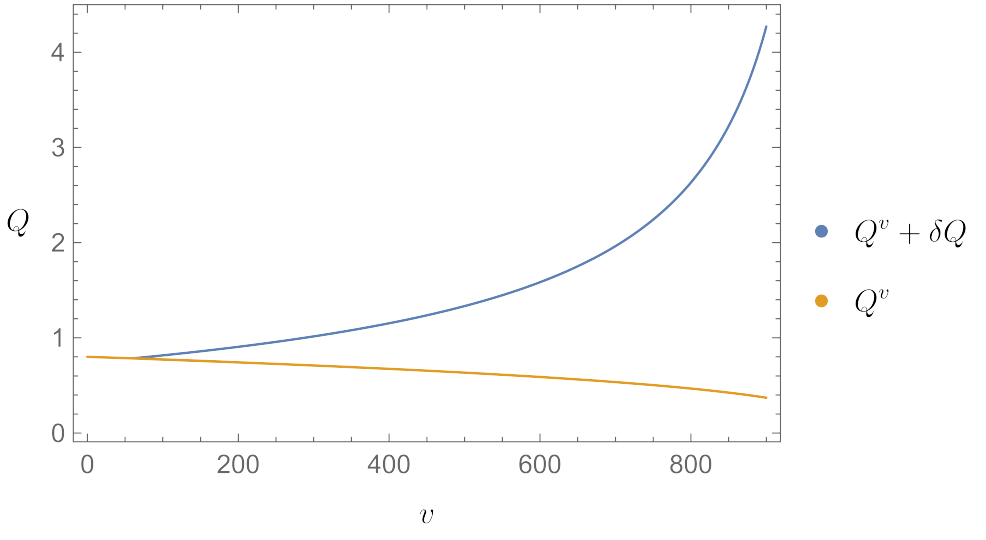}
\end{figure*}

In this run we used $\alpha=1$ parameter for $\delta Q$ (unlike
the previous run which had $\alpha=0$). First we will show in Figure
\ref{fig:Q_plot} the value of $Q(u,v)$ along the last-$u$ ray (including
the $\delta Q$ term) and compare it with $Q^{v}(v)$ which is the
value of $Q$ at the EH. As seen in the figure, the value of $Q$
drifts by a lot and reaches 4 at maximal $v$, meaning that $Q(u,v)$
is greater than $M(v)$ -- which might naively be regarded as ``overcharging''.

Recall that the analytical approximation for $S_{,v}$ is $-\kappa_{-}^{v}$,
and that $\kappa_{-}^{v}(v)$ was defined to be $\kappa_{-}^{v}(M(v),Q^{v}(v))$,
see Eq. (\ref{eq:kappa_-v}). One might be tempted to think that perhaps
the more relevant charge parameter, to be used in $\kappa_{-}^{v}(v)$,
would be $Q(u,v)=Q^{v}(v)+\delta Q(u,v)$ rather than just $Q^{v}(v)$.
The purpose of the next figure is to check this issue.  

In Figure \ref{fig:S,v with different Q} we show $S_{,v}$ along
with two ``candidate'' $-\kappa_{-}^{v}$ curves: the green with
the correct choice $Q^{v}(v)$, and red with the incorrect choice
$Q(u,v)=\delta Q+Q^{v}$. As can be clearly seen in the figure, the
green curve matches the blue curve, while the red curve quickly deviates
from them in region 3. (The red curve is cut short because $M$ becomes
bigger than $Q(u,v)$, meaning the corresponding $\kappa_{-}$ would
have a negative square root.) This figure indicates that using $Q^{v}(v)$
in the analytical approximation is indeed the correct choice. This
is not surprising, of course, because our construction of the analytical
approximation for $S_{,v}$ in subsection \ref{subsec:Computing-S,v}
provided the $S_{,v}=-\kappa_{-}^{v}$ specifically with the charge
parameter $Q^{v}(v)$, see Eq. (\ref{eq:kappa_-v}).

\begin{figure*}
\caption{$S_{,v}$ and $-\kappa_{-}$ on last u ray, using both $Q^{v}$ and
$Q=\delta Q+Q^{v}$ for $-\kappa_{-}$}\label{fig:S,v with different Q}

\includegraphics[scale=0.7]{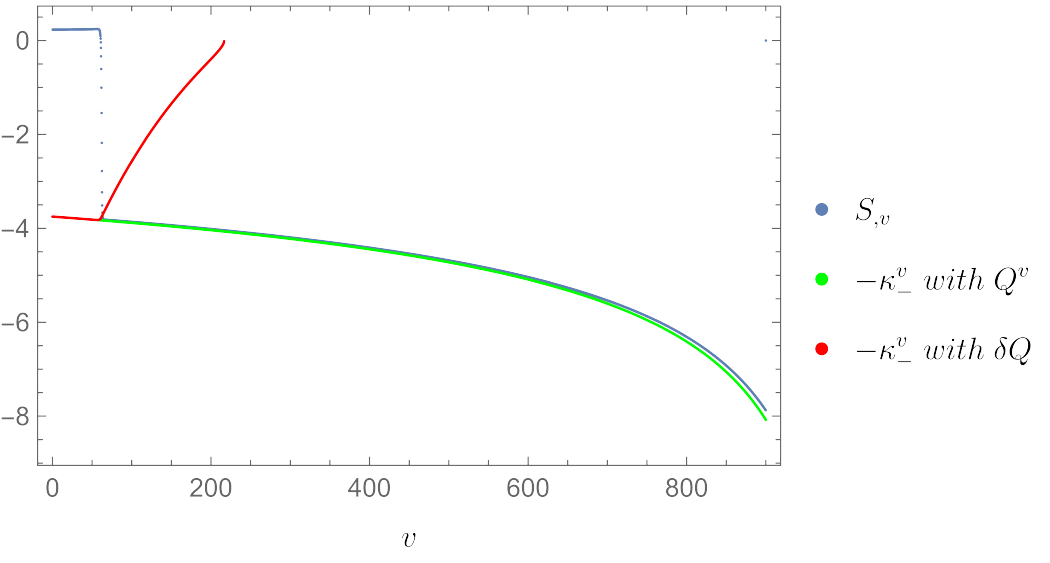}
\end{figure*}

\subsection{Deviations from the analytical approximation}

In this section we intend to analyze the deviation from the analytical
approximation in order to test whether it is a finite-$\varepsilon_{0}$
effect. To this end, similar to subsection \ref{subsec:Deviation-analytical-approx-allzeroes}
above, we'll compare the deviation from the analytical approximation
for two different $\varepsilon_{0}$ values: $\frac{1}{500}$ and
$\frac{1}{1000}$. We will again scale the $\varepsilon_{0}=\frac{1}{500}$
curve by a factor of $\frac{1}{2}$ for $S$ and $\frac{1}{4}$ for
$R$. As before, we will plot these scaled deviations as a function
of the mass $M(v)$. We will see that the two curves overlap, confirming
that the deviation scales as $\varepsilon_{0}$ for $S$ and as $\varepsilon_{0}^{2}$
for $R$. We shall also show the unscaled $\varepsilon_{0}=\frac{1}{500}$
curve in orange.

\begin{figure*}
\caption{$S_{,v}+\kappa_{-}$ with two $\varepsilon_{0}$, scaled by $\frac{1}{2}$
on last u ray}\label{fig:s,v epsilon comparison x2-run1}

\includegraphics[scale=0.8]{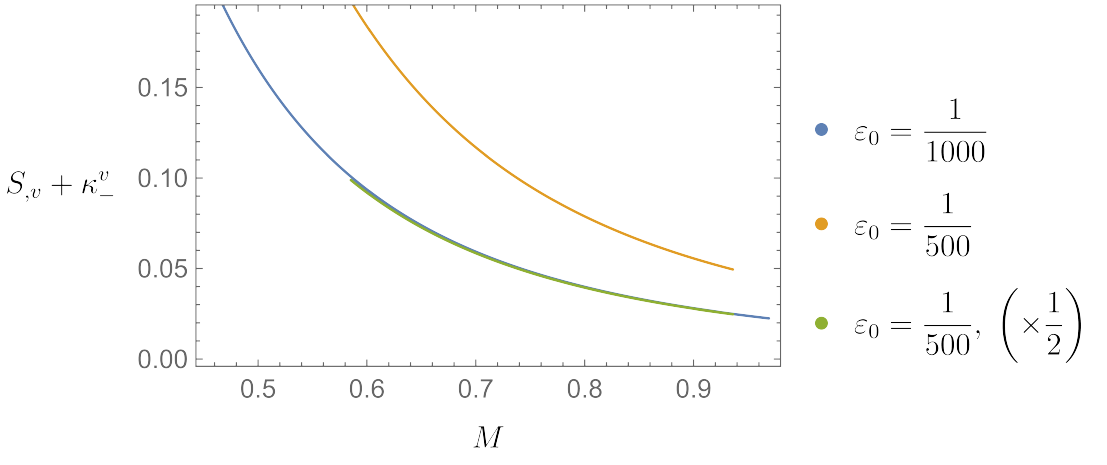}
\end{figure*}

The first Figure \ref{fig:s,v epsilon comparison x2-run1} shows the
deviation of $S_{,v}$ for the two mentioned values of $\varepsilon_{0}$,
with the $\varepsilon_{0}=\frac{1}{500}$ deviations scaled by $\frac{1}{2}$.
The two curves (green and blue) indeed overlap, confirming the scaling
of the deviation as $\varepsilon_{0}.$ Comparing to Figure \ref{fig:s,v epsilon comparison x2}
(vanishing sources run), we see that the deviation is larger by around
a factor of $\sim2$ in the current case, but this change is of no
importance.

\begin{figure*}
\caption{$S_{,u}+\kappa_{-}^{u}$ with two $\varepsilon_{0}$, scaled by $\frac{1}{2}$
on last v ray}\label{fig:s,u epsilon comparison x2-run1}

\includegraphics[scale=0.8]{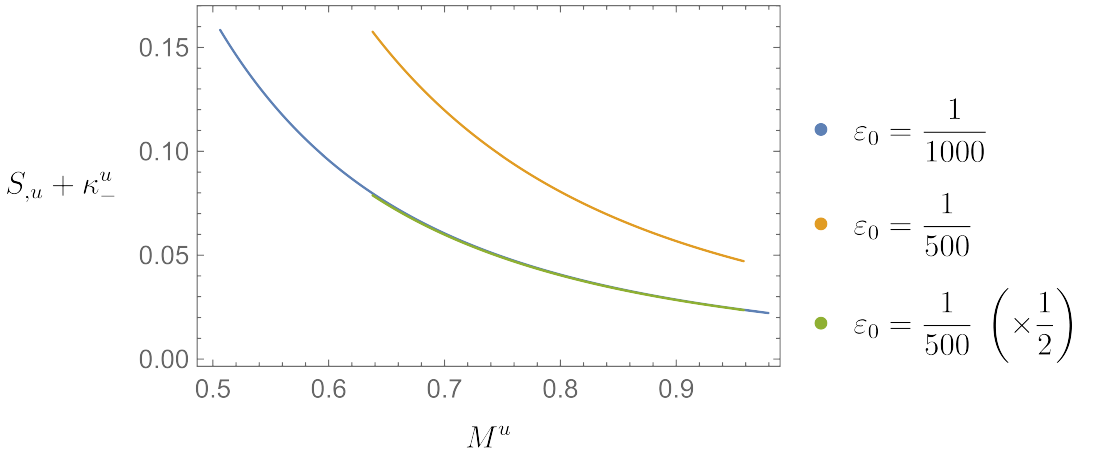}
\end{figure*}

The second Figure \ref{fig:s,u epsilon comparison x2-run1} shows
the deviation of $S_{,u}$ for the same two values of $\varepsilon_{0}$,
with the $\varepsilon_{0}=\frac{1}{500}$ deviations scaled by $\frac{1}{2}$.
The two curves (blue and green) again overlap, confirming a scaling
of the deviation in $S_{,u}$ as $\varepsilon_{0}$.

\begin{figure*}
\caption{$R_{,v}+\frac{\widetilde{T}_{vv}}{\kappa_{-}}$ with two $\varepsilon_{0}$
and scaled by $\frac{1}{4}$ on last u ray}\label{fig:Tvv-approx epsilon comparison x4-run1}

\includegraphics[scale=0.8]{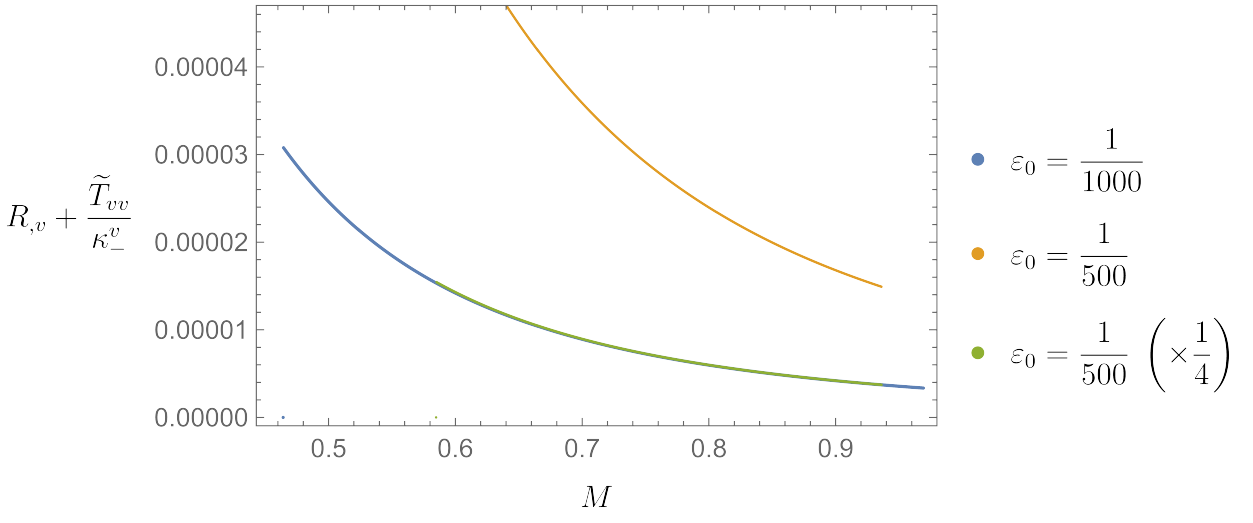}
\end{figure*}

The following Figure \ref{fig:Tvv-approx epsilon comparison x4-run1}
shows the deviation of $R_{,v}$ for the two values of $\varepsilon_{0}$,
with $\varepsilon_{0}=\frac{1}{500}$ curve scaled by $\frac{1}{4}$.
The two curves overlap, confirming scaling with $\text{\ensuremath{\varepsilon_{0}^{2}}}$
-- as expected from $R$ derivatives in region 3, see explanation
in subsection \ref{subsec:Deviation-in-R,v}.

\begin{figure*}
\caption{$R_{,u}+\frac{\widetilde{T}_{uu}}{\kappa_{-}^{u}}$ with two $\varepsilon_{0}$
and scaled by $\frac{1}{4}$ on last v ray}\label{fig:Tuu-approx epsilon comparison x4-run1}

\includegraphics[scale=0.7]{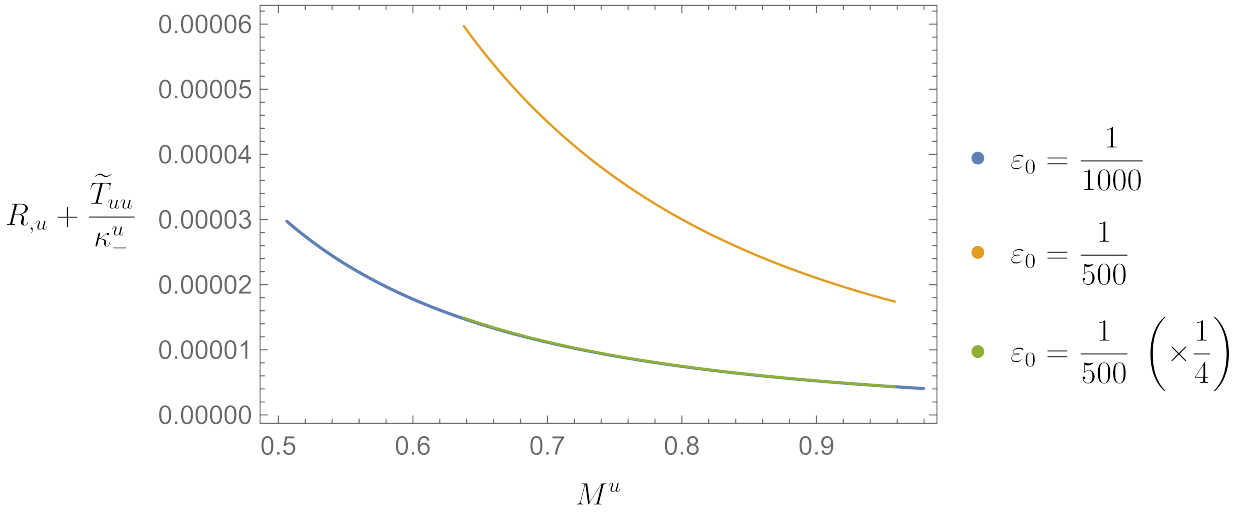}
\end{figure*}

Finally we explore the deviation of $R_{,u}$ from its analytical
approximation. Unlike the no-sources run considered in subsection
\ref{subsec:Deviation-analytical-approx-allzeroes}, this time there
is non-zero $\widetilde{T}_{uu}$ and hence non-vanishing $R_{,u}$
in region 3, and Figure \ref{fig:Tuu-approx epsilon comparison x4-run1}
shows the deviation of $R_{,u}$, with the $\varepsilon_{0}=\frac{1}{500}$
curve scaled by $\frac{1}{4}$. The two curves (green and blue) again
overlap, confirming the scaling of the deviation as $\text{\ensuremath{\varepsilon_{0}^{2}}}.$
(We expect the same scaling of the deviation for $R_{,u}$ as for
$R_{,v}$.)

\section{Finite-$du$ and finite-$T_{0}$ errors}\label{sec:Finite-du-and-T0}

In this section, we shall explore two kinds of errors: 
\begin{enumerate}
\item The numerical error associated with finite $du$ (namely the truncation
error),
\item The error associated with setting the initial values on a hypersurface
of finite $T_{0}$ value -- rather than at the event horizon itself,
which would correspond to $T_{0}\to-\infty$.
\end{enumerate}
Specifically, we will work here with the set of parameters of section
\ref{sec:Numerical-results-run1}:

$Z_{R0}=$$Z_{Rr}$$=Z_{S0}$$=Z_{Sr}$$=Z_{*}=1$, and $\alpha=1$
for $\delta Q$, and with $\frac{Q}{M}=0.8$. We shall focus primarily
on the magnitude of these errors in region 3.

\subsection{Finite-$du$ error}\label{subsec:Finite-du-error}

Our algorithm discretizes spacetime into a two-dimensional grid with
separation $du$ (we use $du=dv$) between two adjacent grid locations.
Because the distance between grid points is finite, we have an error
in our numerical calculation, which increases with the size of $du$.
We would like to estimate this error which results from having finite
$du$.

To estimate the magnitude of this error, we run the numerical code
with three different $du$ values: $du=\frac{1}{50}$, $du=\frac{1}{100}$,
and $du=\frac{1}{200}$, and compare the results. The difference between
the runs $du=\frac{1}{50}$ and $du=\frac{1}{100}$ provides an estimate
for the truncation error in the $du=\frac{1}{50}$ run (up to a factor
$\frac{4}{3}$ which we explain below), and similarly the difference
between the $du=\frac{1}{100}$ and $du=\frac{1}{200}$ runs provides
an estimate for the truncation error in the $du=\frac{1}{100}$ run,
as we will now explain. 

Let $f\left(du\right)$ denote some numerically computed quantity
(that necessarily depends on $du$), and let $e\left(du\right)$ denote
the numerical error in the computation of that quantity. We expect
the error to scale as $du^{2}$, as already mentioned above in section
\ref{sec:Numerical-algorithm} (and as will be confirmed below):
\begin{equation}
e\left(du\right)=\beta\cdot du^{2}
\end{equation}
where $\beta$ is a certain parameter. Under this expectation, we
evaluate the difference between the value of the quantity $f$ computed
with two different $du$ values:
\begin{align}
f\left(du\right)-f\left(\frac{1}{2}du\right) & =e\left(du\right)-e\left(\frac{1}{2}du\right)\\
 & =\beta\left(du^{2}-\left(\frac{1}{2}du\right)^{2}\right)=\frac{3}{4}\beta du^{2}
\end{align}
Therefore, we can compute the parameter $\beta$:

\begin{equation}
\beta=\frac{4}{3}\frac{f\left(du\right)-f\left(\frac{1}{2}du\right)}{du^{2}}
\end{equation}
Thus, the error in $du$ is equal to:

\begin{equation}
e\left(du\right)=\frac{4}{3}\left(f\left(du\right)-f\left(\frac{1}{2}du\right)\right)
\end{equation}
In particular, if we wish to estimate the numerical error $e\left(\frac{1}{100}\right)$
given the values of a certain quantity $f$ at $du=\frac{1}{100}$
and $du=\frac{1}{200}$, we get:

\begin{equation}
e\left(\frac{1}{100}\right)=\frac{4}{3}\left[f\left(\frac{1}{100}\right)-f\left(\frac{1}{200}\right)\right]
\end{equation}
(and similarly for $du=\frac{1}{50}$).

\begin{figure}
\caption{Truncation error in $R$ on last u ray}\label{fig:R error diff}

\includegraphics[width=1\linewidth]{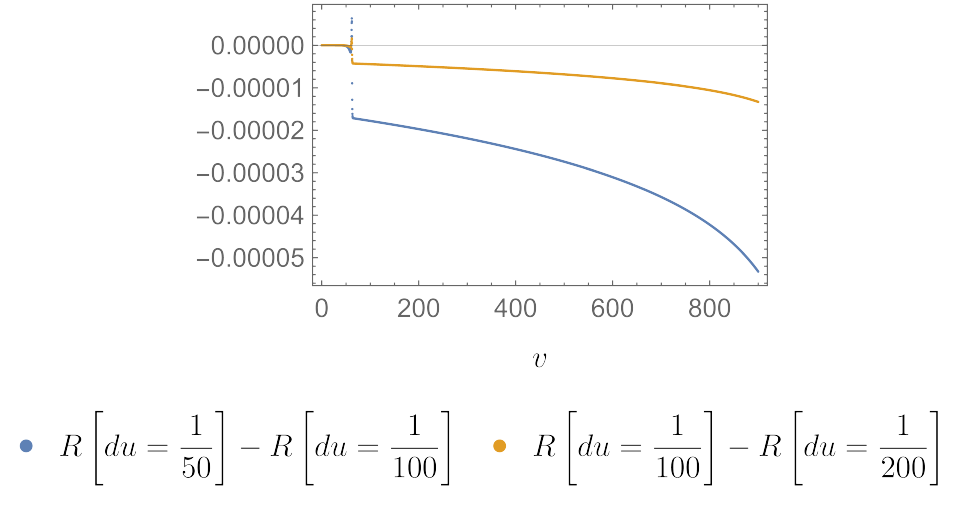}
\end{figure}

We first present the results for the truncation error in $R$ in Figure
\ref{fig:R error diff}, which shows the two differences between running
with different $du$ values. Note how in region 1 the error is very
close to zero. The error for $du=\frac{1}{100}$ is of order $\sim10^{-5}$
 as indicated by the orange curve.

\begin{figure}
\caption{Ratio of truncation errors $\frac{R\left[du=\frac{1}{50}\right]-R\left[du=\frac{1}{100}\right]}{R\left[du=\frac{1}{100}\right]-R\left[du=\frac{1}{200}\right]}$
on last u ray}\label{fig:R error ratio}

\includegraphics[width=1\linewidth]{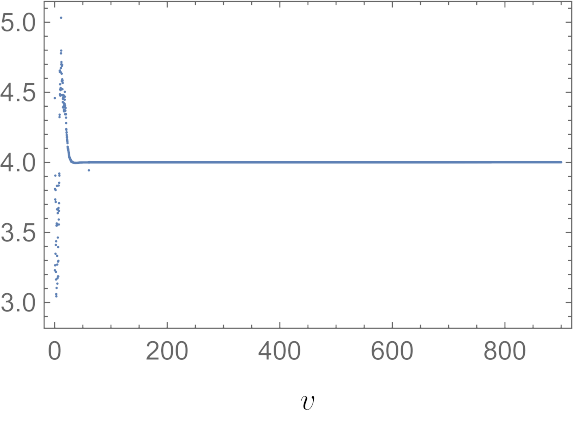}
\end{figure}

We next present the ratio of these two differences in Figure \ref{fig:R error ratio}.
As expected, the ratio is equal to 4, confirming error scaling as
$du^{2}$. We point out that the error in region 1 (not seen in this
figure) is exceptional: it is very small (see Figure \ref{fig:R error diff})
and it doesn't scale as $du^{2}$. In this region, the truncation
error is negligible and other errors seem to dominate.

We shall now present similar findings for other quantities $S$, $S_{,v}$,$R_{,v}$:

\begin{figure}
\caption{Truncation error in S on last u ray}\label{fig:S error diff}

\includegraphics[width=1\linewidth]{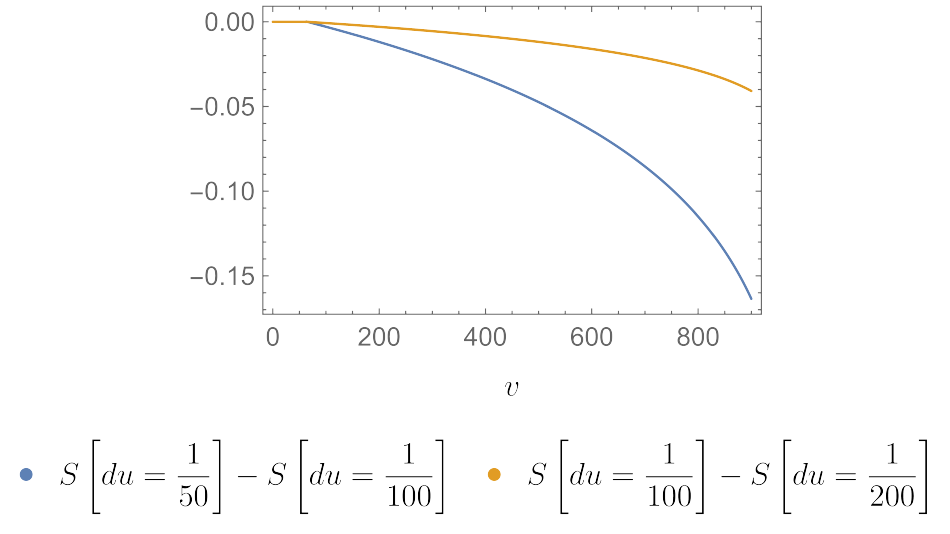}
\end{figure}

Figure \ref{fig:S error diff} shows the truncation error in $S$.
Note that while the error reaches a relatively high value, 0.04, $S$
itself reaches $\sim4000$ so the relative error is again of order
$10^{-5}$.

\begin{figure}
\caption{Ratio of truncation errors $\frac{S\left[du=\frac{1}{50}\right]-S\left[du=\frac{1}{100}\right]}{S\left[du=\frac{1}{100}\right]-S\left[du=\frac{1}{200}\right]}$
on last u ray}\label{fig:S error ratio}

\includegraphics[width=1\linewidth]{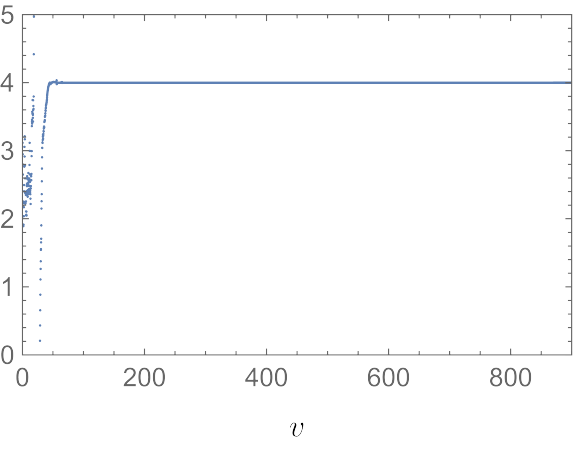}
\end{figure}
We display the ratio of these two truncation errors in Figure \ref{fig:S error ratio},
and it can be seen that in region 2 and 3 the ratio is 4, confirming
error scaling with $du^{2},$as before. In region 1, the error is
very small and doesn't scale with $du$ for the same reason explained
above.

\begin{figure}
\caption{Truncation error in $R_{,v}$ on last u ray}\label{fig:R,v error diff}

\includegraphics[width=1\linewidth]{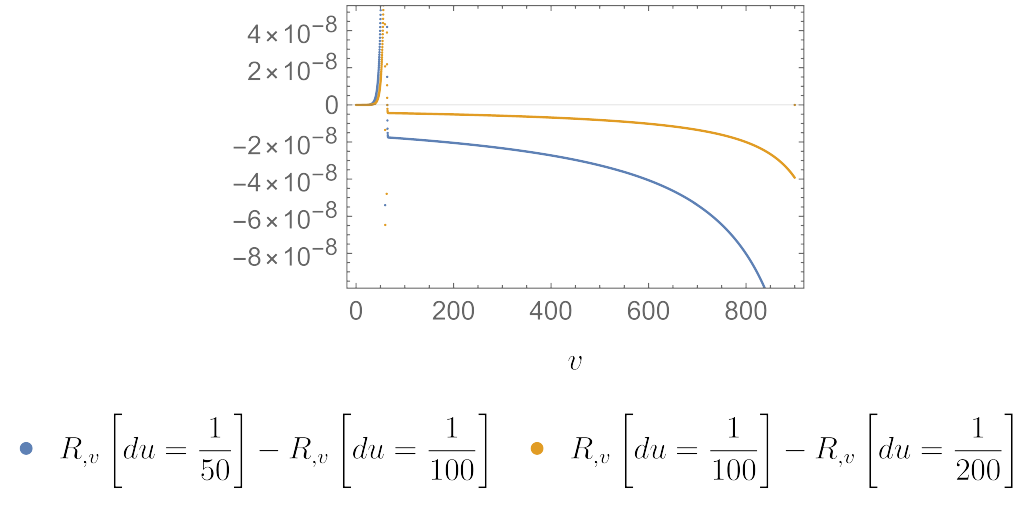}
\end{figure}

Next we show the errors in $R_{,v}$ in Figure \ref{fig:R,v error diff}.
In region 1, the errors are roughly equal because truncation error
isn't the leading cause of errors. In region 3 they seem to scale
as expected. We computed the ratio in Figure \ref{fig:R,v error ratio},
and it appears concentrated around 4, as expected from $du^{2}$ error
scaling. It appears more jittery because of round-off error.

\begin{figure}
\caption{Ratio of truncation errors $\frac{R_{,v}\left[du=\frac{1}{50}\right]-R_{,v}\left[du=\frac{1}{100}\right]}{R_{,v}\left[du=\frac{1}{100}\right]-R_{,v}\left[du=\frac{1}{200}\right]}$
on last u ray}\label{fig:R,v error ratio}

\includegraphics[width=1\linewidth]{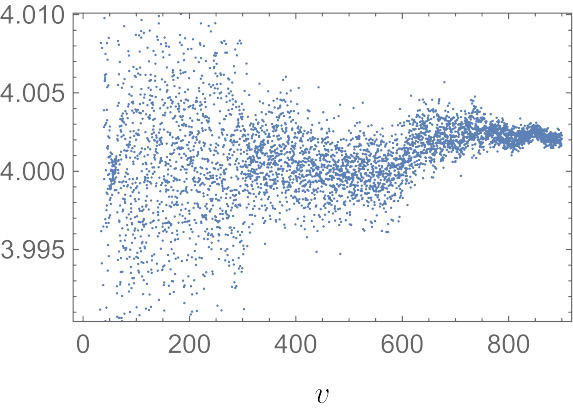}
\end{figure}

\begin{figure}
\caption{Truncation error in $S_{,v}$ on last u ray}\label{fig:S,v error diff}

\includegraphics[width=1\linewidth]{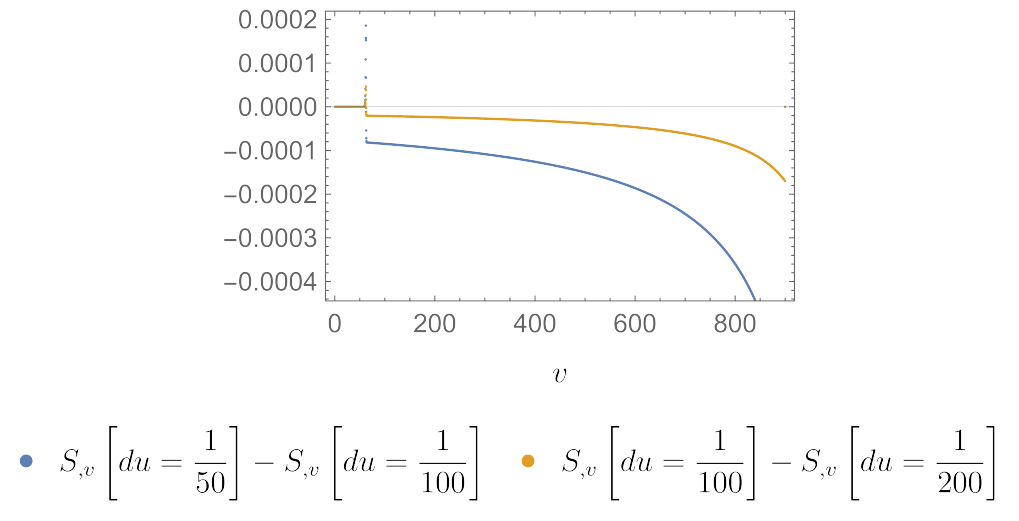}
\end{figure}

In Figure \ref{fig:S,v error diff} we show the error in $S_{,v}$.
It is of order $10^{-4}.$ The error in region 1 is again close to
zero.

\begin{figure}
\caption{Ratio of truncation errors $\frac{S_{,v}\left[du=\frac{1}{50}\right]-S_{,v}\left[du=\frac{1}{100}\right]}{S_{,v}\left[du=\frac{1}{100}\right]-S_{,v}\left[du=\frac{1}{200}\right]}$
on last u ray}\label{fig:S,v error ratio}

\includegraphics[width=1\linewidth]{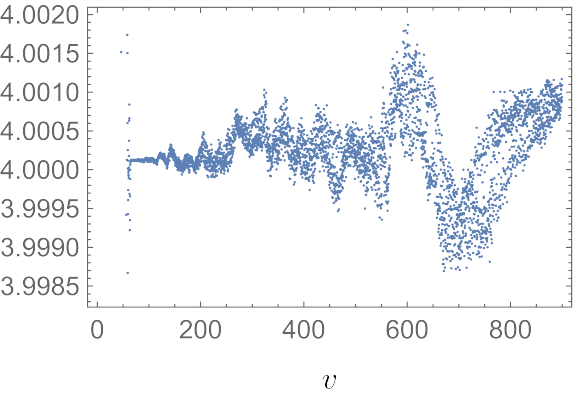}
\end{figure}
We see in Figure \ref{fig:S,v error ratio} that the ratio is also
around 4 in region 3, as expected from $du^{2}$ scaling.

We shall not show the truncation errors in $R_{,u}$ and $S_{,u}$
because they are similar to the corresponding truncation errors in
$R_{,v}$ and $S_{,v}$.

Based on all these tests, we conclude that the truncation error in
our numerical code indeed scales as $du^{2}$ (as it should). 

\subsection{Finite $T_{0}$ errors}

Ideally, we would choose $T_{0}=-\infty$ and start the evolution
at the event horizon. However, we can't do that because that would
require infinite precision (as the difference of $R$ from $r_{+}^{2}$
is exponentially small in $T_{0}$); and furthermore, at a certain
point choosing too negative $T_{0}$ value would result in increasing
errors. Our goal here is to quantify the error resulting from choosing
a finite value for $T_{0}$. To do that, we run the code with four
different values of $T_{0}$, and analyze the results, as explained
below. We compare the results of the different runs on a \emph{common}
$u=const$ ray -- namely on a certain $u=const$ ray that appears
in the numerical grids of all 4 runs. (Note that such a common $u=const$
ray corresponds to different grid indices in the different runs.) 

Specifically, we run the code with the following values of $T_{0}$:
$-20,-25,-30,-35$. In principle we expect the finite-$T_{0}$ error
to be exponentially small in $T_{0}$. Therefore we expect this error
to be very small for all 4 cases. Yet the differences are measurable
(see below) and we shall analyze these differences. We chose the common
ray to be $u=-80$ (which is close to the maximal $u$ values in all
these grids). 

\begin{figure*}
\caption{Difference in R between different $T_{0}$ values on last u ray}\label{fig:RDiff T0 comparison}

\includegraphics[scale=0.8]{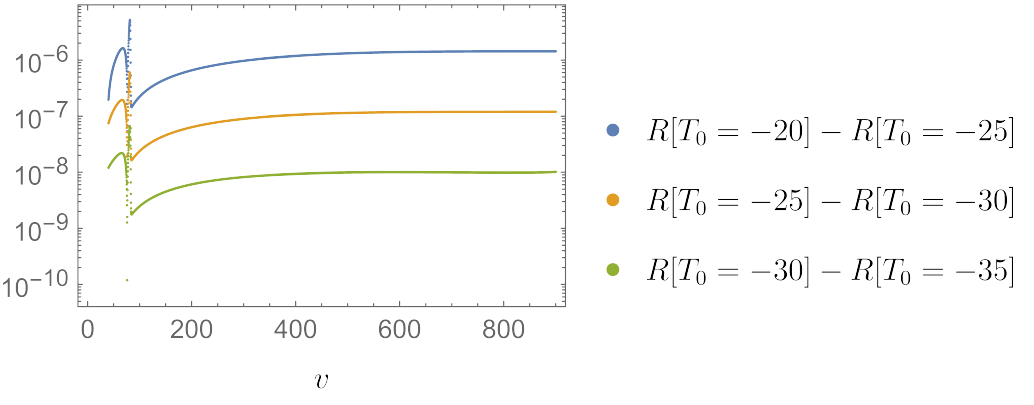}
\end{figure*}

In figure \ref{fig:RDiff T0 comparison}, we show the difference in
$R$ between the three pairs of successive $T_{0}$ runs. We see
in the graph that this difference decreases by a factor of approximately
10 between two such successive pairs. Notice that $e^{2\kappa_{+}\Delta T_{0}}\approx10.4$
where $\Delta T_{0}=5$ is the difference in $T_{0}$ between two
successive runs, which nicely agrees with this factor of $\sim10$. 

Throughout this work we use, as default, $T_{0}=-30$, and the corresponding
finite $T_{0}$ error is estimated to be approximately of order $10^{-8}$
in region 3 (as seen in the green curve).

\begin{figure*}
\caption{Difference in S between different $T_{0}$ values on last u ray}\label{fig:SDiff T0 comparison}

\includegraphics[scale=0.8]{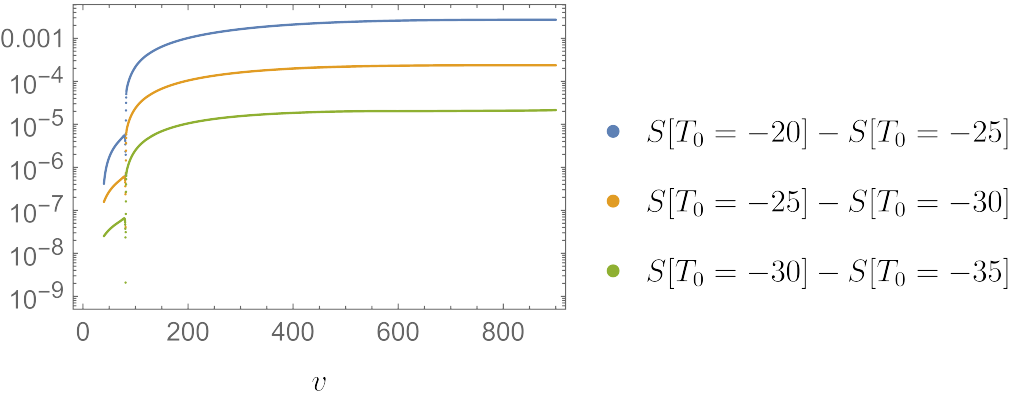}
\end{figure*}

In figure \ref{fig:SDiff T0 comparison}, we show the difference in
$S$ between the three pairs of successive $T_{0}$ runs. We can see
a similar behavior for $S$ as it was for $R$: the gaps here again
correspond to the error being reduced by a factor of approximately
10. Here the error in $T_{0}=-30$ is approximately of order $10^{-5}$.

\begin{figure*}
\caption{Difference in $R_{,v}$ between different $T_{0}$ values on last
u ray}\label{fig:RvDiff T0 comparison}

\includegraphics[scale=0.8]{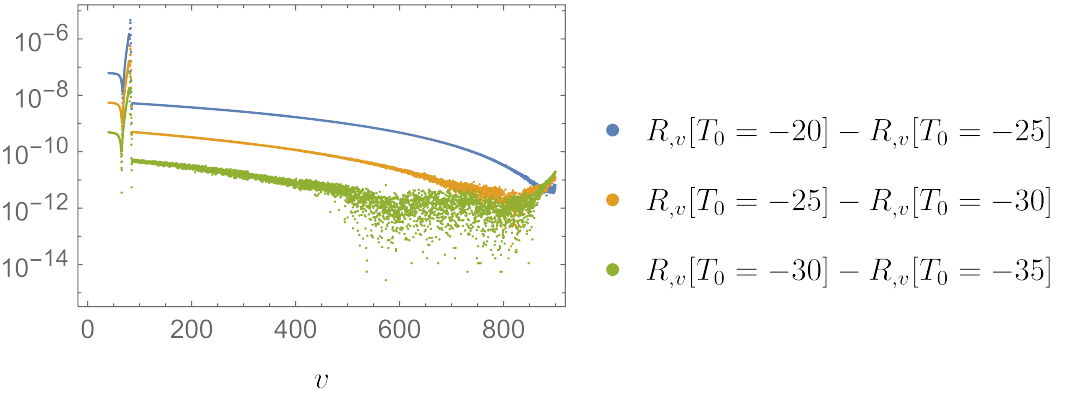}
\end{figure*}

In the following Figure \ref{fig:RvDiff T0 comparison}, we see the
differences in $R_{,v}$ between pairs of successive $T_{0}$ runs.
In the region $v>600$ we see that the round-off error starts to dominate
over the finite-$T_{0}$ error. The error in $T_{0}=-30$ is found
to be of order $10^{-10}$ (or smaller) in region 3.

\begin{figure*}
\caption{Difference in $S_{,v}$ between different $T_{0}$ values on last
u ray}\label{fig:SvDiff T0 comparison}

\includegraphics[scale=0.8]{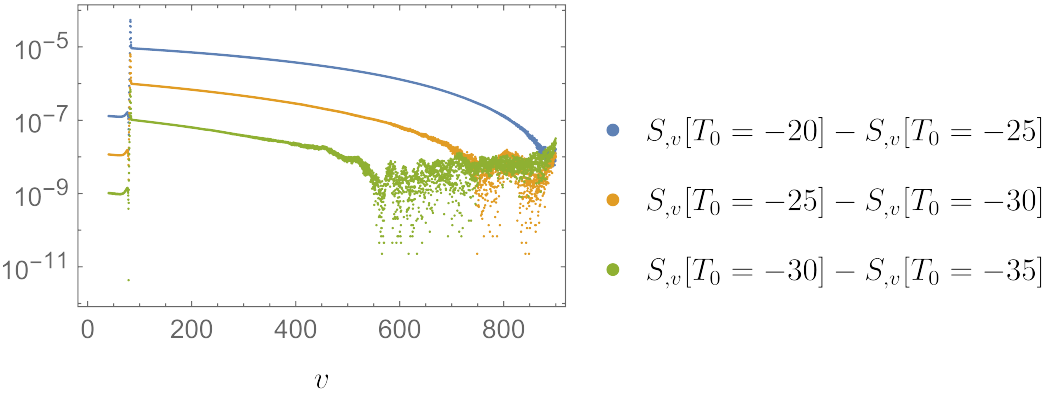}
\end{figure*}

Moving on to Figure \ref{fig:SvDiff T0 comparison}, which shows the
corresponding differences in the values of $S_{,v}$, we again see
that the roundoff error dominates at $v>600$. The finite-$T_{0}$
error in $T_{0}=-30$ is of order $10^{-7}$.

$S_{,u}$ and $R_{,u}$ display similar features to $S_{,v}$ and
$R_{,v}$ respectively: the finite-$T_{0}$ error is reduced by a
factor of $\sim10$ when $T_{0}$ decreases by 5. We shall not display
them here.

In conclusion, for our standard choice $T_{0}=-30$, all the above
quantities show that the finite-$T_{0}$ error is negligibly small.

\section{Independence from the detail of sources}\label{sec:Independence-from-the-detail-of-sources}

As was already mentioned in subsection \ref{subsec:Summary-analytical-approx},
our analytical approximation for region 3 only depends on the four
input functions of $M(v)$, $Q(v)$, $\widetilde{T}_{vv}^{(-)}(v)$
and $\widetilde{T}_{uu}^{(-)}(v)$. Once $M(v)$ and $Q(v)$ are prescribed,
they directly determine $\widetilde{T}_{vv}^{(+)}$, and $\widetilde{T}_{uu}^{(+)}=0$
from regularity. Then $\widetilde{T}_{vv}^{(-)}$ and $\widetilde{T}_{uu}^{(-)}$
are in principle determined by the evolution from the EH to the near-IH
region (with $\widetilde{T}_{vv}^{(+)}$ and $\widetilde{T}_{uu}^{(+)}=0$
serving as initial values) -- in a manner that depends on the sources
$Z_{R}$ and $Z_{S}$. Thus, these sources do affect $\widetilde{T}_{vv}^{(-)}$
and $\widetilde{T}_{uu}^{(-)}$ -- but otherwise the analytical approximation
is independent of these sources. Our goal in this section is to numerically
check this independence from the details of sources. 

To check this independence, we run the numerical code with three different
sets of sources which all lead to (approximately) the same values
of $\widetilde{T}_{vv}^{(-)}$ and $\widetilde{T}_{uu}^{(-)}$, and
we show that the results for $R_{,v}$,$R_{,u}$,$S_{,v}$,$S_{,u}$
are the same in all three cases -- meaning they are independent of
the specific choice of the sources. We use the parametrization given
in Eqs. (\ref{eq:RuvF1-1},\ref{eq:suv_F2-1}) for $Z_{R}$ and $Z_{S}$
in terms of the four parameters $Z_{R0}$, $Z_{Rr}$, $Z_{S0}$, $Z_{Sr}$.
In all three runs, we choose $\alpha=1$ for $\delta Q$, and $\frac{Q}{M}=0.8$.
The three sets of source parameters used in these three runs are given
in Table \ref{tab:Runs-with-equal-Tvv-Tuu}.

\begin{table}
\caption{Runs with equal $\widetilde{T}_{vv}$ and $\widetilde{T}_{uu}$}\label{tab:Runs-with-equal-Tvv-Tuu}

\begin{tabular}{|c|c|c|c|c|c|}
\hline 
run & $Z_{R0}$ & $Z_{Rr}$ & $Z_{S0}$ & $Z_{Sr}$ & $Z_{S0}+Z_{Sr}-\frac{1}{2}Z_{Rr}$\tabularnewline
\hline 
\hline 
run1 & 1 & 1 & 1 & 1 & 1.5\tabularnewline
\hline 
run2 & 1 & -1 & 0 & 1 & 1.5\tabularnewline
\hline 
run3 & 1 & 1 & 0 & 2 & 1.5\tabularnewline
\hline 
\end{tabular}
\end{table}

We found empirically that what matters for the final values of $\widetilde{T}_{vv}$
and $\widetilde{T}_{uu}$ in region 3 is the combination $Z_{S0}+Z_{Sr}-\frac{1}{2}Z_{Rr}$,
as displayed at the last column. (This empirical expression may be
explored analytically, but this is beyond the scope of this research.)
In the first stage we shall check that in region 3 we have the same
$\widetilde{T}_{uu}$ and $\widetilde{T}_{vv}$ for all three sets
of parameters.
\begin{figure*}
\caption{$\widetilde{T}_{uu}$ for the three different runs on the last u ray}\label{fig:Tuu-for-equivalent-runs}

\includegraphics{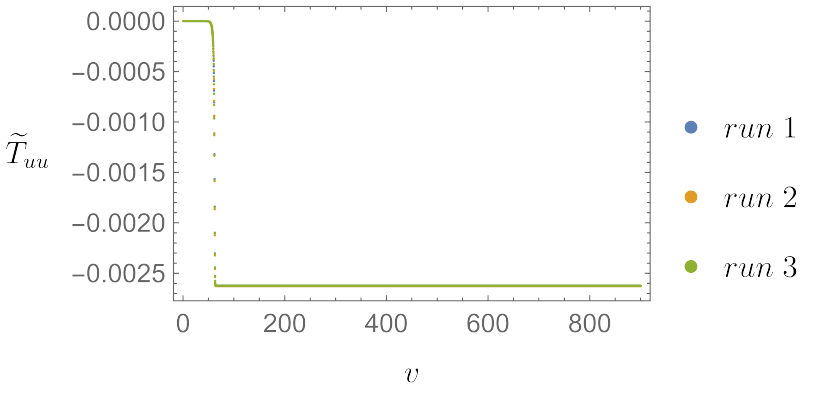}
\end{figure*}
Figure \ref{fig:Tuu-for-equivalent-runs} shows that $\widetilde{T}_{uu}$
is approximately equal for the different runs (the curves overlap).

\begin{figure}
\caption{$\widetilde{T}_{uu}$ for the three different runs, in region 3, on
the last u ray}\label{fig:Tuu-for-equivalent-runs-1}

\includegraphics[width=1\linewidth]{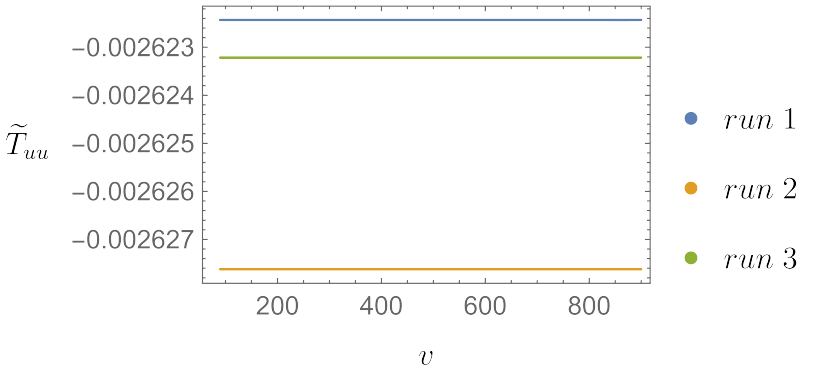}
\end{figure}
In the next Figure \ref{fig:Tuu-for-equivalent-runs-1} we zoom in
to region 3 where we compare $\widetilde{T}_{uu}$ for the three different
runs, and we can see they are pretty close to each other, with difference
of order $\sim10^{-5}$, to be compared with the $\sim10^{-3}$ value
of $\widetilde{T}_{uu}$.
\begin{figure*}
\caption{$\widetilde{T}_{vv}$ for the three different runs on the last u ray}\label{fig:Tvv-for-equivalent-runs}

\includegraphics{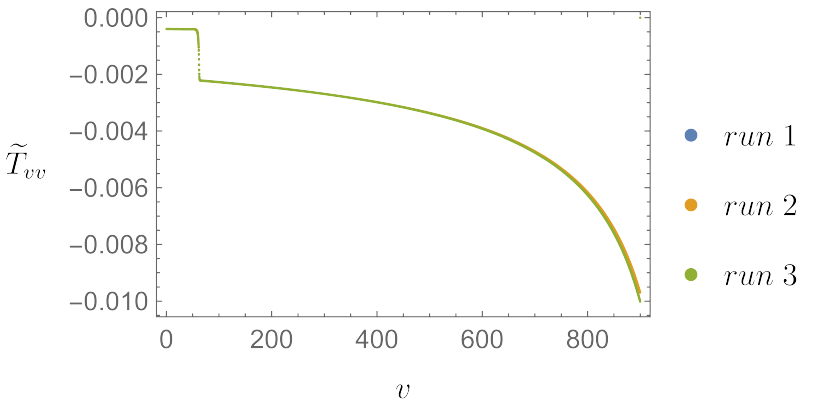}
\end{figure*}
In the following Figure \ref{fig:Tvv-for-equivalent-runs}, we show
that $\widetilde{T}_{vv}$ is also approximately equal.
\begin{figure*}
\caption{$S_{,v}$ for the three different runs, compared to $-\kappa_{-}(v)$
on the last u ray}\label{fig:s,v-for-equivalent-runs}

\includegraphics{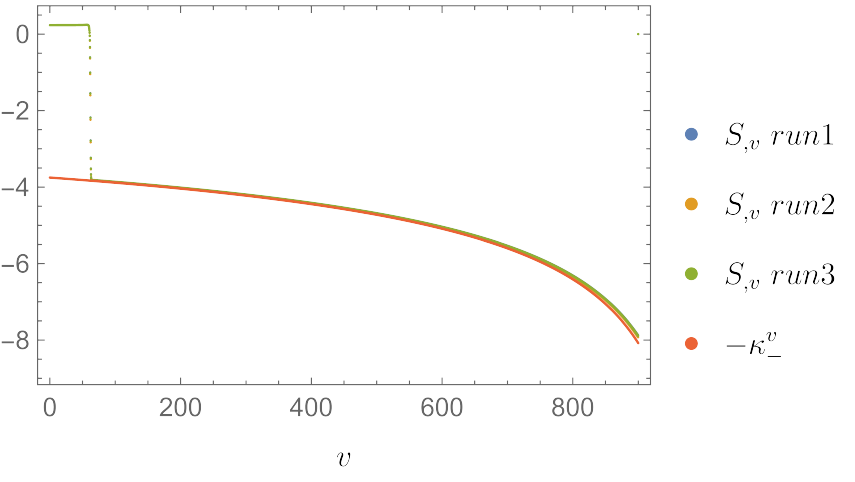}
\end{figure*}
We move on to show $S_{,v}$ along with its expected value $-\kappa_{-}(v)$
in Figure \ref{fig:s,v-for-equivalent-runs}. The analytical approximation
works well for all runs.
\begin{figure*}
\caption{$S_{,v}$ for the three different runs, compared to $-\kappa_{-}(v)$,
zoomed in on the last u ray}\label{fig:s,v zoomed-for-equivalent-runs}

\includegraphics{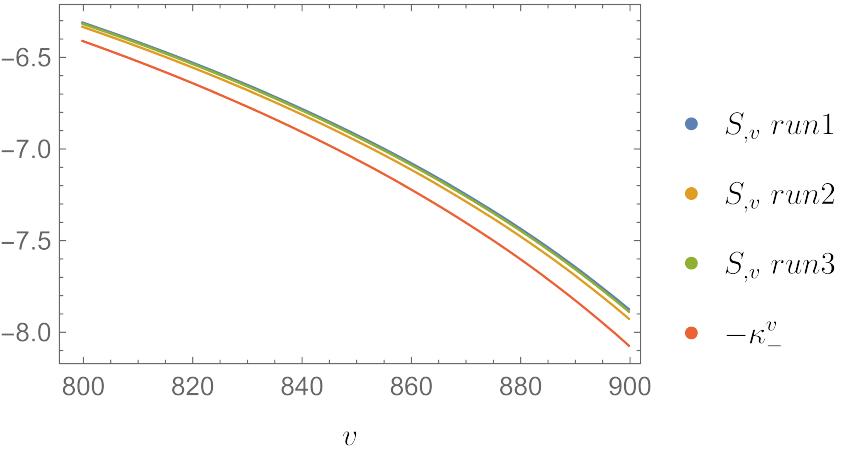}
\end{figure*}
We zoom in to the region $800\leq v\leq900$ in Figure \ref{fig:s,v zoomed-for-equivalent-runs}
to show how close are the three different runs.
\begin{figure*}
\caption{$S_{,u}$ for the three different runs, compared to $-\kappa_{-}^{u}$
on the last v ray}\label{fig:s,u-for-equivalent-runs}

\includegraphics{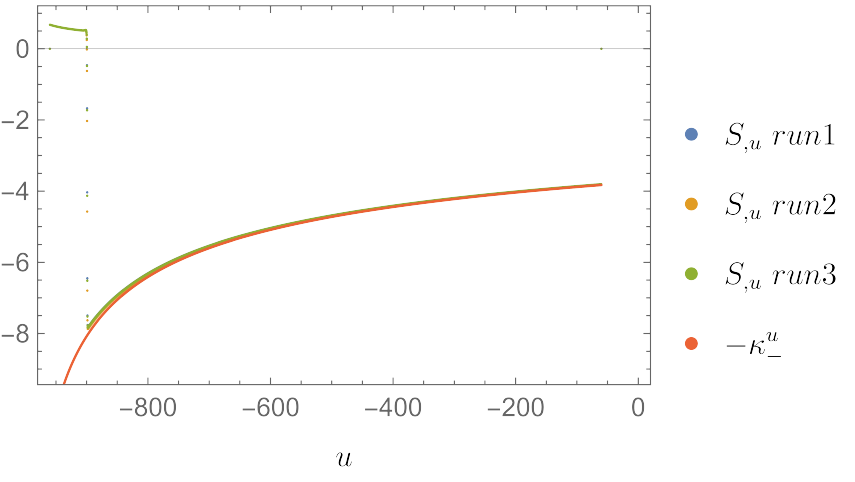}
\end{figure*}
We continue to show $S_{,u}$ and its expected value $-\kappa_{-}^{u}(u)$
in Figure \ref{fig:s,u-for-equivalent-runs}. All the three equivalent
runs are close together in region 3, near their expected value of
$-\kappa_{-}^{u}(u)$.

\begin{figure*}
\caption{$R_{,v}$ for the three different runs, compared to $-\frac{\widetilde{T}_{vv}}{\kappa_{-}^{v}}$
on the last u ray}\label{fig:R,v-for-equivalent-runs}

\includegraphics{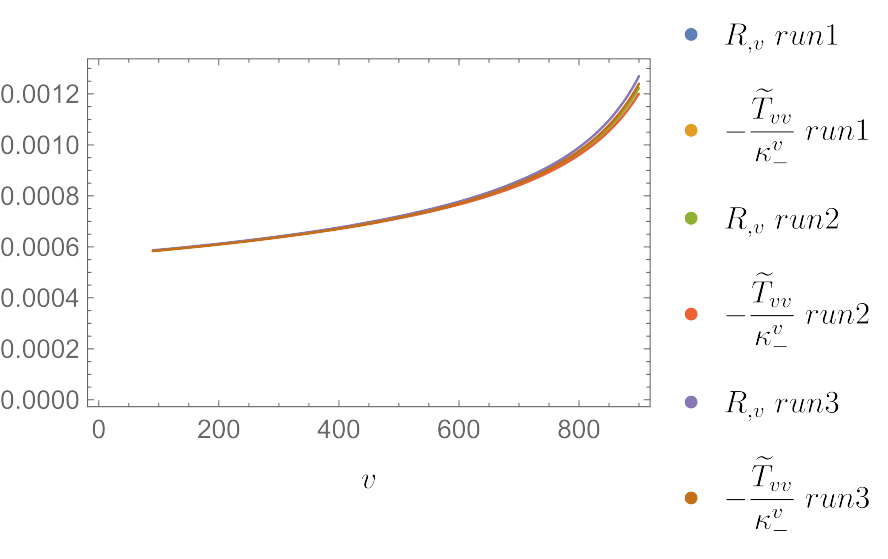}
\end{figure*}
We turn now to explore the derivatives of $R$. Figure \ref{fig:R,v-for-equivalent-runs}
shows both $R_{,v}$ and its analytical approximation $-\frac{\widetilde{T}_{vv}}{\kappa_{-}^{v}}$
for each run. All the three runs and their approximations are close
together.

\begin{figure*}
\caption{$R_{,u}$ for the three different runs, compared to $-\frac{\widetilde{T}_{uu}}{\kappa_{-}^{u}}$
on the last v ray}\label{fig:R,u-for-equivalent-runs}

\includegraphics{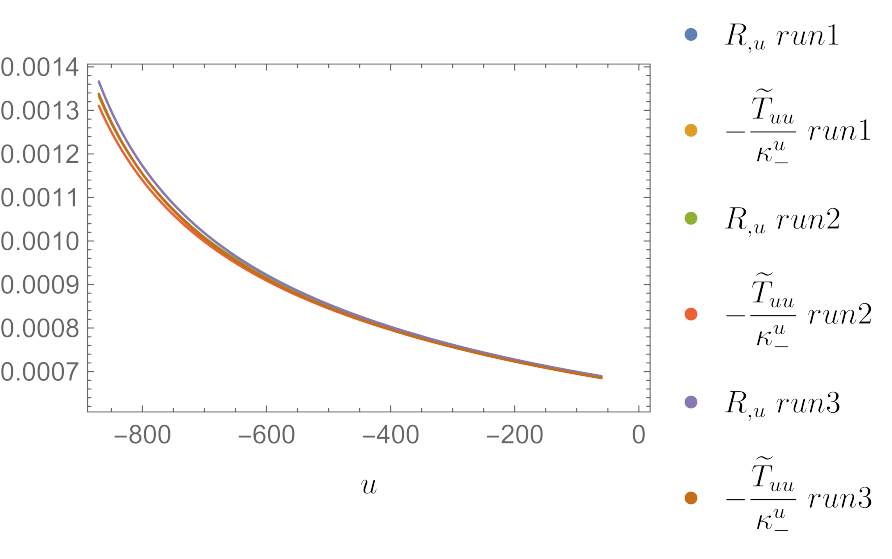}
\end{figure*}
Subsequently, we show in Figure \ref{fig:R,u-for-equivalent-runs}
both $R_{,u}$ and its analytical approximation $-\frac{\widetilde{T}_{uu}}{\kappa_{-}^{u}}$
for each run. All curves are close together, showing that the approximation
is good.

In conclusion, we have demonstrated that for all quantities we explored
here ($R_{,v}$,$R_{,u}$,$S_{,v}$,$S_{,u}$), the three different
runs indeed yielded approximately the same results regardless of which
specific sources were used -- and that the corresponding analytical
approximation is good in all cases. 

\section{Three additional runs with different sources}\label{sec:runs2-to-run4}

We shall now extend the validity test of the analytical approximation
to three additional sets of source parameters -- this time we chose
parameter sets which lead to \emph{different} values of $\widetilde{T}_{vv}$
and $\widetilde{T}_{uu}$ in region 3 (unlike what we did in section
\ref{sec:Independence-from-the-detail-of-sources}). Table \ref{tab:Runs}
details the various source parameters ($Z_{R0}$, $Z_{Rr}$, $Z_{S0}$,
$Z_{Sr}$,$\alpha$) and $\frac{Q}{M}$ used in each run. Run0 was
detailed in section \ref{sec:Numerical-results-run0}. Run1 was detailed
in section \ref{sec:Numerical-results-run1}. Our focus here is runs
2 through 4.

In all three runs we shall display $R(v)$,$R(u)$, and then the fluxes
$\widetilde{T}_{vv}(v)$ and $\widetilde{T}_{uu}(u)$, followed by
the 4 derivatives $S_{,v}(v)$,$S_{,u}(u)$ and $R_{,v}(v)$,$R_{,u}(u)$.
All these quantities will be displayed along either the last $u=const$
or $v=const$ rays depending on their arguments ($u$ or $v$). As
we shall see in all these 3 runs, the analytical approximation works
well.

\begin{table}
\caption{Runs with equal $\widetilde{T}_{vv}$ and $\widetilde{T}_{uu}$}\label{tab:Runs}

The signs of $\widetilde{T}_{vv}$ and $\widetilde{T}_{uu}$ are shown

\begin{tabular}{|c|c|c|c|c|c|c|c|c|c|}
\hline 
run & $Z_{R0}$ & $Z_{Rr}$ & $Z_{S0}$ & $Z_{Sr}$ & $\alpha$ & $\frac{Q}{M}$ & $Z_{S0}+Z_{Sr}-\frac{1}{2}Z_{Rr}$ & $\widetilde{T}_{vv}$ & $\widetilde{T}_{uu}$\tabularnewline
\hline 
\hline 
run0 & 0 & 0 & 0 & 0 & 0 & 0.8 & 0 & + & 0\tabularnewline
\hline 
run1 & 1 & 1 & 1 & 1 & 1 & 0.8 & 1.5 & -- & --\tabularnewline
\hline 
run2 & 1 & 1 & -1 & -1 & $\frac{1}{2}$ & 0.8 & -2.5 & + & --\tabularnewline
\hline 
run3 & -1 & -1 & 1 & 1 & 1 & 0.6 & 2.5 & -- & --\tabularnewline
\hline 
run4 & 1 & $-\frac{1}{2}$ & 1 & $-\frac{1}{2}$ & $-\frac{1}{2}$ & 0.8 & 0.75 & + & +\tabularnewline
\hline 
\end{tabular}
\end{table}

\subsection{Run2}

This run is done with sources equal to $Z_{R0}=$$Z_{Rr}=1$, $Z_{S0}$$=Z_{Sr}$$=-1$,
and $\alpha=\frac{1}{2}$ for $\delta Q$ and with $\frac{Q}{M}=0.8$.

\begin{figure*}
\caption{R for run2}

\subfloat[$R(v)$ for run2 on last u ray\label{fig:R(v)-for-run2}]{

\includegraphics{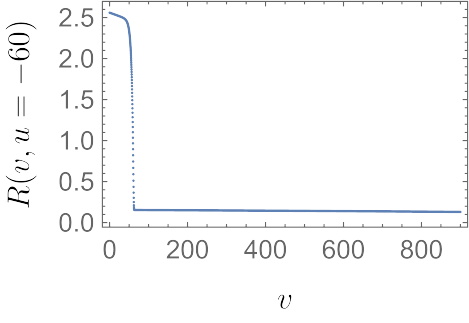}}\subfloat[$R(u)$ for run2 on last v ray\label{fig:R(u)-for-run2}]{

\includegraphics{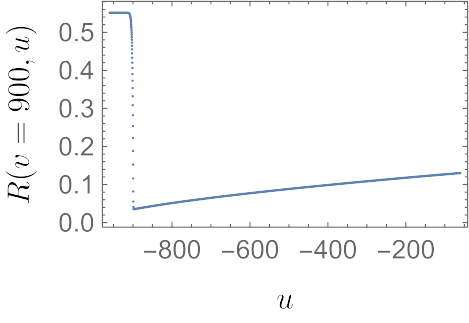}}
\end{figure*}

\begin{figure*}
\caption{$\widetilde{T}_{vv}$ and $\widetilde{T}_{uu}$ for run2}

\subfloat[$\widetilde{T}_{vv}$ for run2 on last u ray\label{fig:Tvv-for-run2}]{

\includegraphics{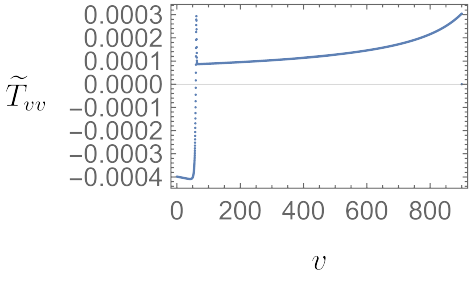}}\subfloat[$\widetilde{T}_{uu}$ for run2 on last v ray\label{fig:Tuu-for-run2}]{

\includegraphics{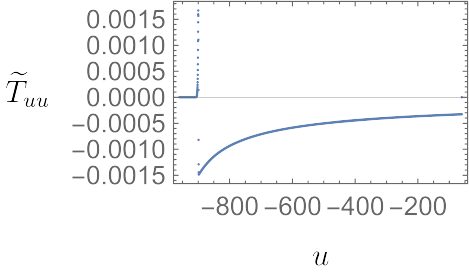}}
\end{figure*}

We can see in Figure \ref{fig:Tvv-for-run2} that $\widetilde{T}_{vv}>0$
in region 3, and from Figure \ref{fig:Tuu-for-run2} $\widetilde{T}_{uu}<0$
in region 3. This explains why in Figure \ref{fig:R(v)-for-run2}
$R$ is decreasing with $v$ ($R_{,v}<0$) in region 3, while in Figure
\ref{fig:R(u)-for-run2} $R$ is increasing with $u$ ($R_{,u}>0$)
in region 3. Figure \ref{fig:S,v-for-run2} shows the value of $S_{,v}$
is close to its analytical approximation. Similarly, in Figure \ref{fig:S,u-for-run2}
we can see that $S_{,u}$ is also close to $-\kappa_{-}^{u}$, as
expected from the analytical approximation. Figure \ref{fig:R,v-for-run2}
compares the value of $R_{,v}$ to the analytical approximation, and
demonstrates that they are close. Figure \ref{fig:R,u-for-run2} compares
the value of $R_{,u}$ to the analytical approximation, showing they
are close too. 


\begin{figure*}
\caption{$S_{,v}$ and $S_{,u}$ and their analytical approximation for run2}
\subfloat[$S_{,v}$ and $-\kappa_{-}$ for run2 on last u ray\label{fig:S,v-for-run2}]{

\includegraphics{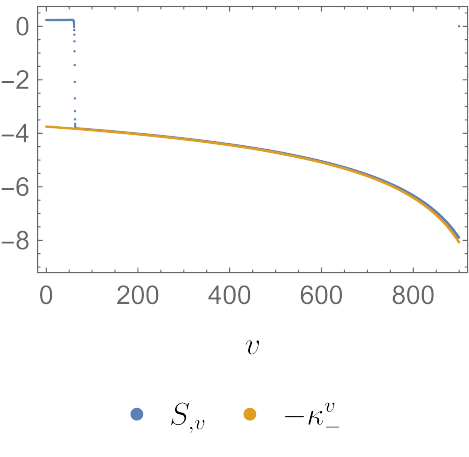}}\subfloat[$S_{,u}$ and $-\kappa_{-}^{u}$ for run2 on last v ray\label{fig:S,u-for-run2}]{

\includegraphics{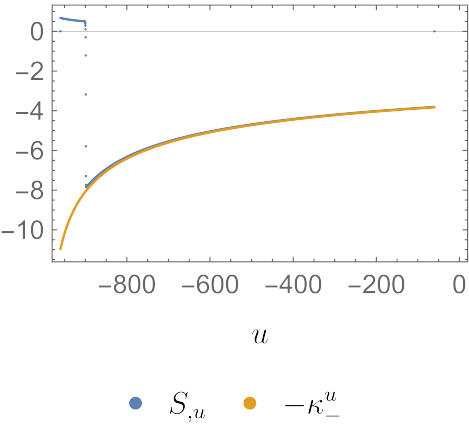}}
\end{figure*}

\begin{figure*}
\caption{$R_{,v}$ and $R_{,u}$ and their analytical approximation for run2}
\subfloat[$R_{,v}$ and $-\frac{\widetilde{T}_{vv}}{\kappa_{-}}$ for run2 in
region 3 on last u ray\label{fig:R,v-for-run2}]{

\includegraphics{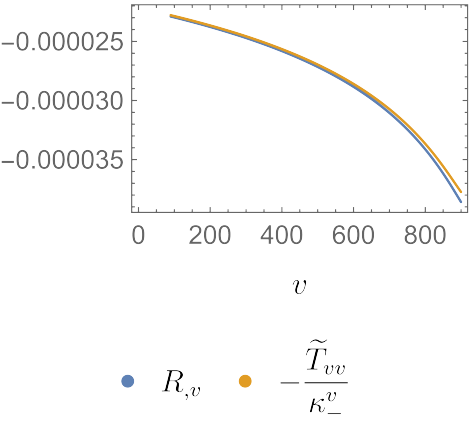}}\subfloat[$R_{,u}$ and $-\frac{\widetilde{T}_{uu}}{\kappa_{-}^{u}}$ for run2
in region 3 on last v ray\label{fig:R,u-for-run2}]{

\includegraphics{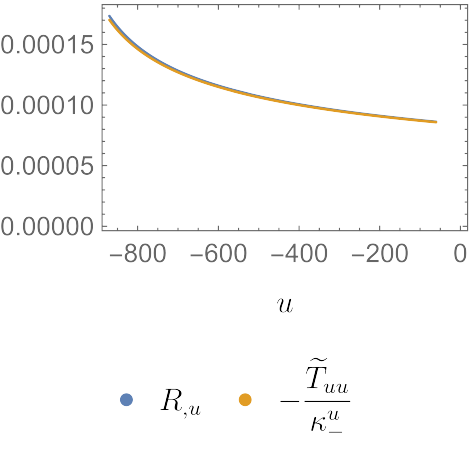}}
\end{figure*}

\subsection{Run3}

This run is done with sources equal to $Z_{R0}=$$Z_{Rr}=-1$, $Z_{S0}$$=Z_{Sr}$$=1$,
and $\alpha=1$ for $\delta Q$. This time (and unlike all other cases)
we choose $\frac{Q}{M}=0.6$.

\begin{figure*}
\caption{R for run3}

\subfloat[$R(v)$ for run3 on last u ray\label{fig:R(v)-for-run3}]{

\includegraphics{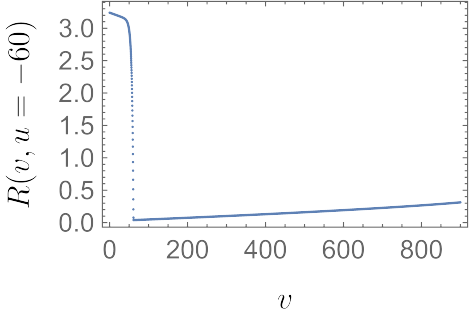}}\subfloat[$R(u)$ for run3 on last v ray\label{fig:R(u)-for-run3}]{

\includegraphics{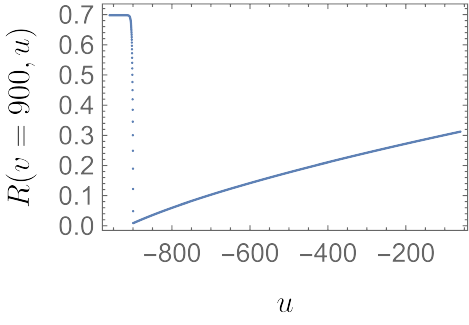}}
\end{figure*}

\begin{figure*}
\caption{$\widetilde{T}_{vv}$ and $\widetilde{T}_{uu}$ for run3}
\subfloat[$\widetilde{T}_{vv}$ for run3 on last u ray\label{fig:Tvv-for-run3}]{

\includegraphics{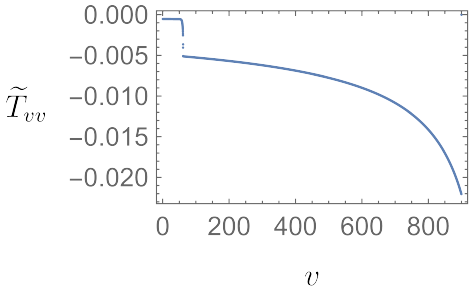}}\subfloat[$\widetilde{T}_{uu}$ for run3 on last v ray\label{fig:Tuu-for-run3}]{

\includegraphics{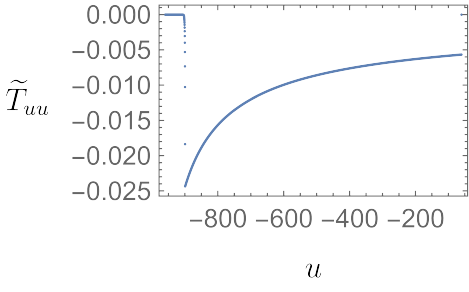}}
\end{figure*}

As can be seen in Figure \ref{fig:R(v)-for-run3}, $R$ reaches its
minimum value at the end of region 2 and then drifts upwards in region
3. This minimal value of $R=r_{-}^{2}=0.04$ is quite small (compared
to e.g. in Figure \ref{fig:R(v)-for-run2}) for a simple reason: Because
$\frac{Q}{M}$ is now smaller, $r_{-}$ is smaller than in previous
runs when $\frac{Q}{M}$ was chosen to be 0.8. Figures \ref{fig:Tvv-for-run3}
and \ref{fig:Tuu-for-run3} respectively show that both $\widetilde{T}_{vv}$
and $\widetilde{T}_{uu}$ are $<0$. Because of that, $R_{,u}>0$
and also $R_{,v}>0$ in region 3, as can be seen by the increasing
$R$ in Figures \ref{fig:R(v)-for-run3}, \ref{fig:R(u)-for-run3}.Figures
\ref{fig:S,v,S,u-for-run3},\ref{fig:R,v,R,u-for-run3} show the quantities
$S_{,v}$,$S_{,u}$ and $R_{,v}$,$R_{,u}$ respectively, compared
with their analytical approximation values. It can be seen that the
analytical approximation is close to the numerically computed values
for all these quantities.

\begin{figure*}
\caption{$S_{,v}$ and $S_{,u}$ and their analytical approximation for run
3}\label{fig:S,v,S,u-for-run3}
\subfloat[$S_{,v}$ and $-\kappa_{-}$ for run3 on last u ray\label{fig:S,v-for-run3}]{

\includegraphics{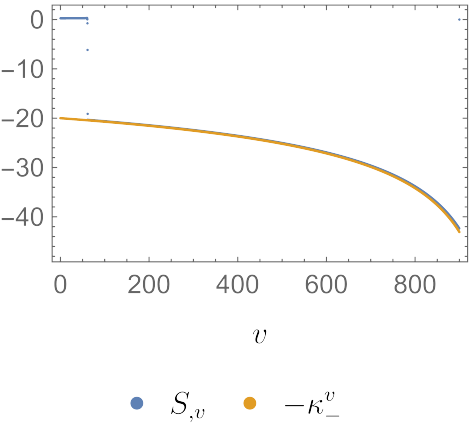}}\subfloat[$S_{,u}$ and $-\kappa_{-}^{u}$ for run3 on last v ray\label{fig:S,u-for-run3}]{

\includegraphics{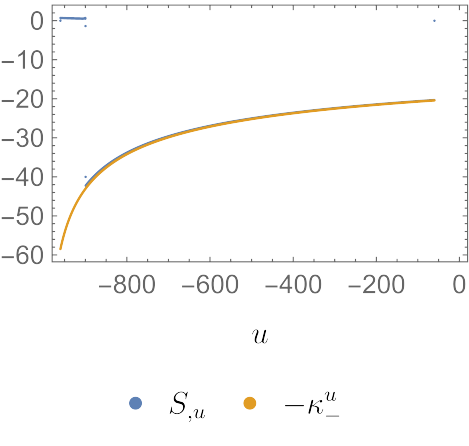}}
\end{figure*}
\begin{figure*}
\caption{$R_{,v}$ and $R_{,u}$ and their analytical approximation for run3}\label{fig:R,v,R,u-for-run3}
\subfloat[$R_{,v}$ and $-\frac{\widetilde{T}_{vv}}{\kappa_{-}}$ for run3 on
last u ray\label{fig:R,v-for-run3}]{

\includegraphics{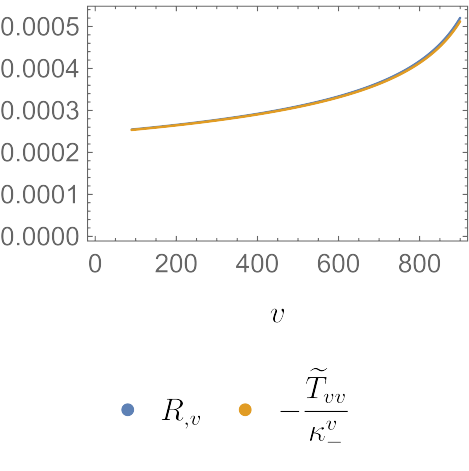}}\subfloat[$R_{,u}$ and $-\frac{\widetilde{T}_{uu}}{\kappa_{-}^{u}}$ for run3
on last v ray\label{fig:R,u-for-run3}]{

\includegraphics{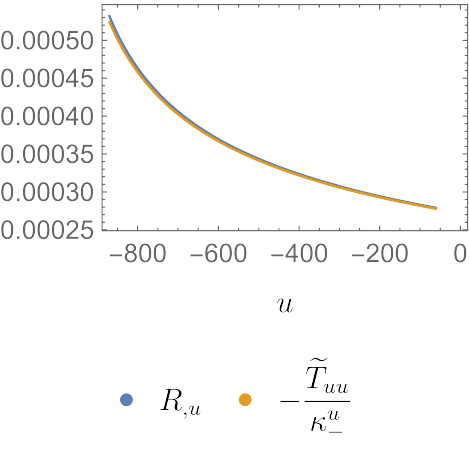}}
\end{figure*}

\subsection{Run4}

This run is done with source parameters taken to be $Z_{R0}=Z_{S0}$$=1$,
$Z_{Rr}$$=Z_{Sr}$$=-\frac{1}{2}$, and $\alpha=-\frac{1}{2}$ for
$\delta Q$ and with $\frac{Q}{M}=0.8$. In this run, unlike all other
runs, the maximum $v$ is taken to be 482 (compared with 900 in all
previous runs) to avoid (or at least reduce) complications that may
arise from the singularity that is going to form in this case (see
below) -- except for Figure \ref{fig:R(v=00003D600,u)-for-run4}
where $v$ is taken to be 600 in order to show where the singularity
forms on this null ray. 

\begin{figure}
\caption{$R(v)$ for run4 on last u ray}\label{fig:R(v)-for-run4}

\includegraphics{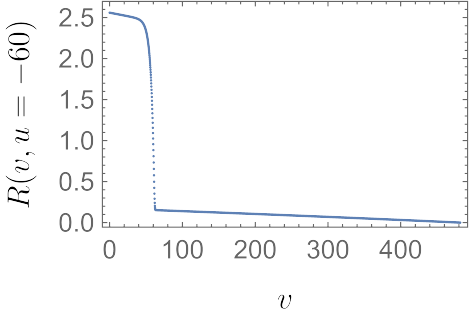}
\end{figure}

\begin{figure*}
\caption{$\widetilde{T}_{vv}$ and $\widetilde{T}_{uu}$ for run4}
\subfloat[$\widetilde{T}_{vv}$ for run4 on last u ray\label{fig:Tvv-for-run4}]{

\includegraphics{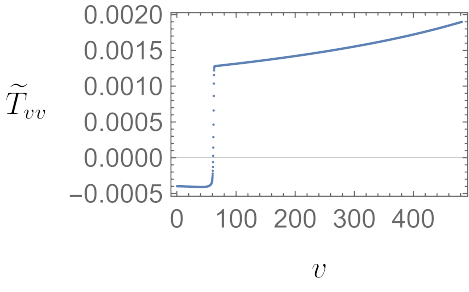}}\subfloat[$\widetilde{T}_{uu}$ for run4 on last v ray\label{fig:Tuu-for-run4}]{

\includegraphics{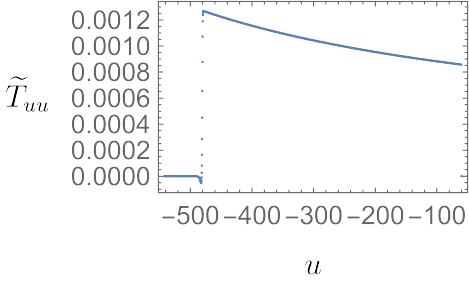}}
\end{figure*}

Figures \ref{fig:Tvv-for-run4} and \ref{fig:Tuu-for-run4} respectively
show that this time, both $\widetilde{T}_{vv}$ and $\widetilde{T}_{uu}$
are $>0$ in region 3. Because of that, both $R_{,v}$ and $R_{,u}$
are negative there. Correspondingly, Figure \ref{fig:R(v)-for-run4}
shows that, at $u=-60$, $R$ is decreasing and reaching zero at $v=482$.
Likewise, Figure \ref{fig:R(v=00003D600,u)-for-run4} shows, at $v=600$,
that $R$ is also decreasing as a function of $u$, reaching zero
at $u=-274$, which is marked by a short red line. This vanishing
of $R$ -- seen in all these figures -- expresses the formation
of a spacelike $R=0$ singularity, which we shall further discuss
at the end of this subsection. We also show Figure \ref{fig:R(v=00003D482,u)-for-run4},
which shows that at $v=482$, $R$ is decreasing all the way to zero
at $u=-60$. We chose to show quantities on this ray of $v=482$ for
the rest of this case to avoid crossing the singularity, as this is
the value of $v$ at which $R$ vanishes along the last-$u$ ray $u=-60$.
Figure \ref{fig:S,v-for-run4} and \ref{fig:S,u-for-run4} respectively
show $S_{,v}$ and $S_{,u}$ versus their analytical approximations
$-\kappa_{-}(v)$ and $-\kappa_{-}^{u}(u)$, again showing the good
quality of the analytical approximation. Figure \ref{fig:R,v-for-run4}
shows $R_{,v}$ compared with its analytical approximation, and they
are close but have a small gap which is this time bigger relative
to the previous runs. A similar gap is also seen in Figure \ref{fig:R,u-for-run4}
which shows $R_{,u}$ compared with its analytical approximation.
We shall now explore this gap, and show that this deviation from the
analytical approximation is a finite-$\varepsilon_{0}$ deviation
(which in this specific case happens to be bigger than in the previous
cases).

\begin{figure*}
\caption{$R(u)$ for run4 along two $v=const$ rays}

\subfloat[$R(v=482,u)$ for run4\label{fig:R(v=00003D482,u)-for-run4}]{

\includegraphics{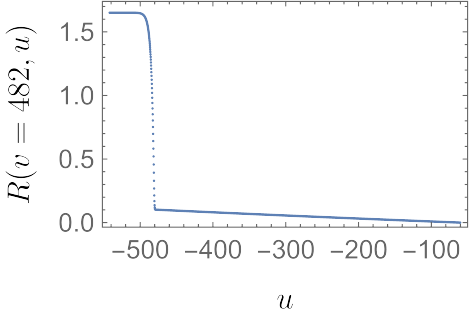}}\subfloat[$R(v=600,u)$ for run4\label{fig:R(v=00003D600,u)-for-run4}]{

\includegraphics{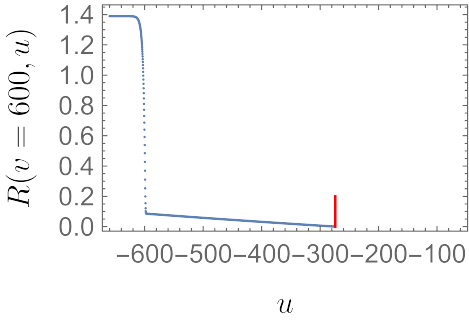}}

The vertical red line in Panel b denotes the $u$-value where $R=0$
\end{figure*}

\begin{figure*}
\caption{$S_{,v}$ and $S_{,u}$ and their analytical approximation for run4}
\subfloat[$S_{,v}$ and $-\kappa_{-}$ for run4 on last u ray\label{fig:S,v-for-run4}]{

\includegraphics{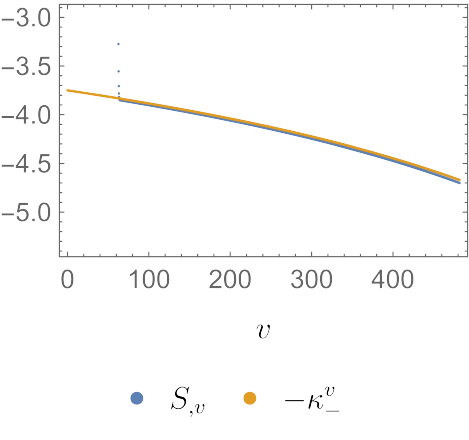}}\subfloat[$S_{,u}$ and $-\kappa_{-}^{u}$ for run4 on last v ray\label{fig:S,u-for-run4}]{

\includegraphics{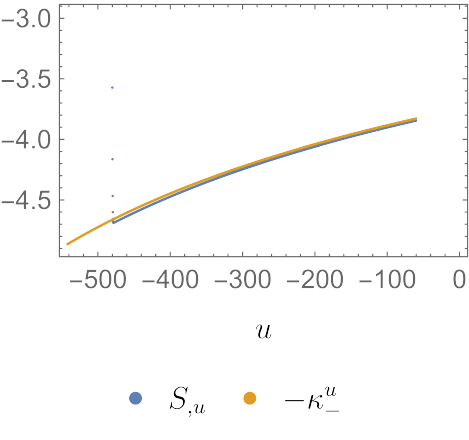}}
\end{figure*}

\begin{figure*}
\caption{$R_{,v}$ and $R_{,u}$ and their analytical approximation for run4}
\subfloat[$R_{,v}$ and $-\frac{\widetilde{T}_{vv}}{\kappa_{-}}$ for run4 on
last u ray\label{fig:R,v-for-run4}]{

\includegraphics{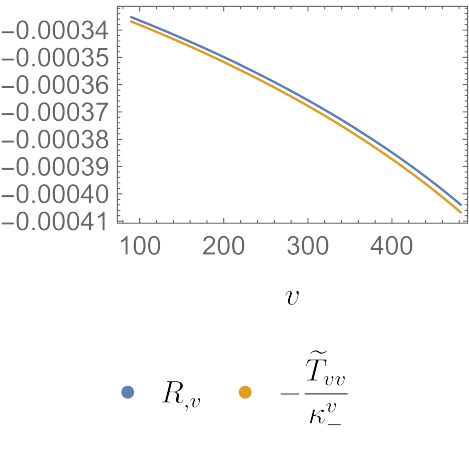}}\subfloat[$R_{,u}$ and $-\frac{\widetilde{T}_{uu}}{\kappa_{-}^{u}}$ for run4
on last v ray\label{fig:R,u-for-run4}]{

\includegraphics{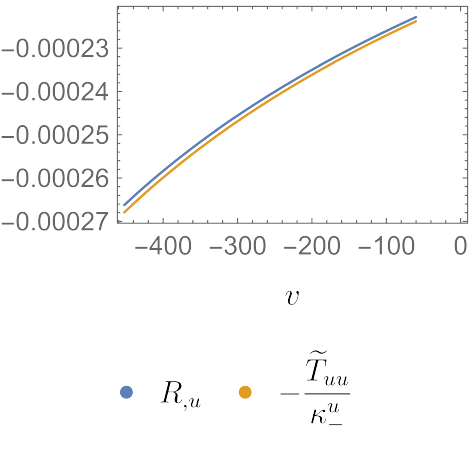}}
\end{figure*}

\begin{figure*}
\caption{$R_{,v}+\frac{\widetilde{T}_{vv}}{\kappa_{-}}$ with two $\varepsilon_{0}$
and scaled by $\frac{1}{4}$ in run4 on last u ray}\label{fig:Tvv-approx epsilon comparison x4-run4}

\includegraphics[scale=0.8]{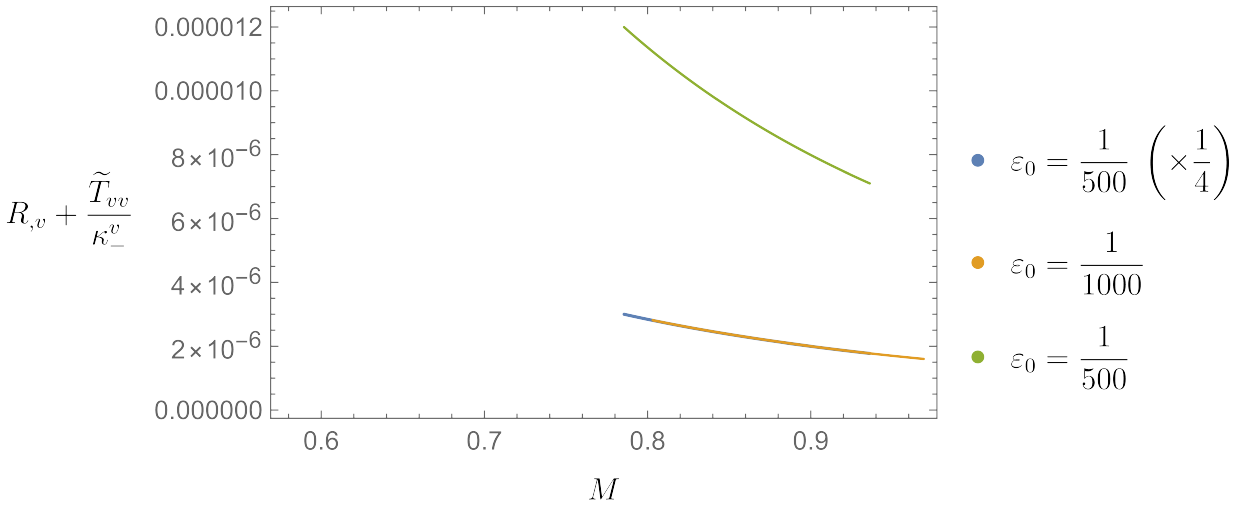}
\end{figure*}

To this end, we shall run the same case with $\varepsilon_{0}=\frac{1}{500}$
as well, and compare the deviations from the analytical approximation
obtained in both values of $\varepsilon_{0}$, namely $\frac{1}{500}$
and $\frac{1}{1000}$, similar to subsection \ref{subsec:Deviation-in-R,v}.
Figure \ref{fig:Tvv-approx epsilon comparison x4-run4} presents this
deviation in $R_{,v}$ for these two values of $\varepsilon_{0}$,
with the $\varepsilon_{0}=\frac{1}{500}$ case scaled by $\frac{1}{4}$.
The blue and orange curves overlap, demonstrating that this deviation
from the analytical approximation scales as $\varepsilon_{0}^{2}$
-- indicating that this is indeed a finite-$\varepsilon_{0}$ deviation.

\begin{figure*}
\caption{Spacetime diagram for $R=0$ singularity}\label{fig:run4_singulairty_diagram}

\includegraphics[scale=0.75]{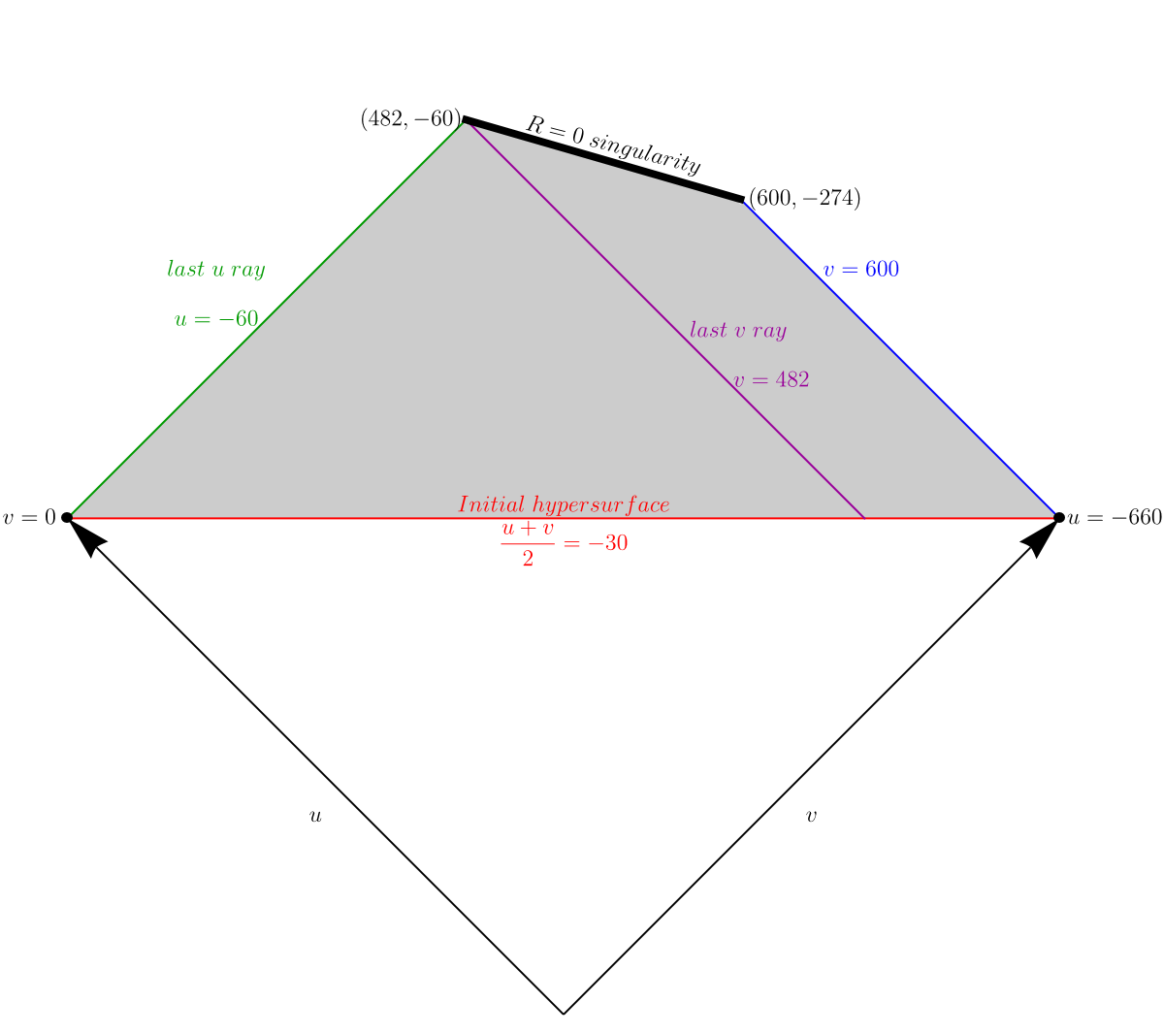}
\end{figure*}
In this case (and unlike the previous cases we explored above), because
both $\widetilde{T}_{vv}$ and $\widetilde{T}_{uu}$ are positive,
a spacelike $R=0$ singularity forms at the middle of the numerical
domain, blocking access to the would-be CH. The spacetime diagram
corresponding to this situation is shown in Figure \ref{fig:run4_singulairty_diagram}.This
$r=0$ spacelike singularity that forms inside the semiclassical charged
black hole resembles the spacelike singularity found in Ref \citep{brady1995black},
although it must be recalled that they studied classical perturbations
whereas here we consider the backreaction of a quantum field. \footnote{There is a possibility that the two fluxes will have opposite signs
in region 3. In this case, as it turns out, what matters is the sign
of $\protect\widetilde{T}_{uu}$: if it is negative (as it is e.g.
in run 2 above), then a CH will form, and if it is positive, a spacelike
$R=0$ singularity will form instead.}

\section{Conclusion}

The main research question that motivated us here is to understand
the backreaction of the metric due to the semiclassical energy fluxes
inside an evaporating charged black hole. Recently an analytical approximation
was developed for this semiclassical metric, with emphasis on the
near-IH region -- which we refer to as \emph{region 3}. In this work,
we sought to numerically validate this analytical approximation. To
this end we developed a numerical code which evolves the semiclassical
Einstein equation, from near the EH up to the very neighborhood of
the IH.

We began with analyzing the Einstein equations and the Maxwell equations
under spherical symmetry -- from which we derived a convenient form
of the evolution and constraint Einstein's equations for the two metric
variables $R(u,v)$ and $S(u,v)$ in double-null coordinates $u$,$v$.
A finite-difference numerical code was developed to evolve the above
mentioned evolution equations for $R$ and $S$ on a discrete grid.
This code was constructed to be of second-order accuracy in the discretization
parameter $du=dv$. 

The dimensionless evaporation rate parameter is $\varepsilon_{0}$
($\propto\frac{\hbar}{M^{2}}$). It is assumed to be $\ll1$, corresponding
to a macroscopic black hole. For astrophysical black holes, $\varepsilon_{0}$
is typically of order $10^{-82}$ or smaller, but in our numerical
simulation due to obvious technical constraints we took it to be $\varepsilon_{0}=10^{-3}$
(if we wish to significantly evaporate the black hole within the numerical
simulation, the grid size has to be of order $\frac{1}{\varepsilon_{0}}$
-- which is totally infeasible for $\varepsilon_{0}\sim10^{-82}$).

We checked the truncation error arising from the finite-$du$ grid
size and found it to scale as $du^{2}$, as expected from our second-order
numerical code. Overall the numerical error is small enough to allow
us to accurately compare the numerical results with the analytical
approximation. 

In our numerical test of the analytical approximation, we focused
primarily on the four derivatives $R_{,v}$,$R_{,u}$,$S_{,v}$,$S_{,u}$.
The numerical results confirmed the accuracy of the analytical approximation:
we checked that the deviations between the numerical result and the
analytical approximation for $S_{,u}$, $S_{,v}$ scale as $\varepsilon_{0}$,
as expected. The deviations for $R_{,u}$,$R_{,v}$ scale as $\varepsilon_{0}^{2}$,
because $R_{,v}$ and $R_{,u}$ themselves are of order $\varepsilon_{0}$
(unlike $S_{,v}$ and $S_{,u}$ which scale as $(\varepsilon_{0})^{0}$). 

As mentioned above, the analytical approximation for the metric inside
the evaporating black hole is based on four main input functions (to
be obtained from the semiclassical theory): $M(v)$ and $Q^{v}(v)$
at the event horizon, and $\widetilde{T}_{uu}^{(-)}(v),$$\text{\ensuremath{\widetilde{T}_{vv}^{(-)}}}(v)$
at the inner horizon. In order to run the simulation, we have to assume
a specific form for the functions $T_{uv}^{(SC)}(u,v)$ and $T_{\theta\theta}^{(SC)}(u,v)$.
(These functions, too, are to be determined from semiclassical gravity.)
We found it more convenient to express these two stress-energy components
via the source terms $Z_{R}$, $Z_{S}$. These source terms do affect
the evolving flux components $T_{vv}$ and $T_{uu}$ -- however they
do not directly affect the evolving metric near the IH (provided that
$\widetilde{T}_{vv}^{(-)}$ and $\widetilde{T}_{uu}^{(-)}$ have been
specified). One of our goals was to directly test this independence.
We verified this by running simulations with three different configurations
of sources $Z_{R}$ and $Z_{S}$ which all lead to the same $\widetilde{T}_{vv}^{(-)}$,
$\widetilde{T}_{uu}^{(-)}$. We found that all three source configurations
produced the same results for the evolving metric in region 3 (up
to $O(\varepsilon_{0})$ corrections) -- as predicted by the analytical
approximation.

 In the next stage we run the code with three additional different
configurations of the sources which lead to different $\widetilde{T}_{vv}^{(-)}$
and $\widetilde{T}_{uu}^{(-)}$ values, and have shown that the analytical
approximation again gives good approximation for the metric variables
in all these cases. 

In one of the cases tested, $\widetilde{T}_{vv}^{(-)}$ and $\widetilde{T}_{uu}^{(-)}$
turned out to be positive. The analytical approximation then predicts
that $R_{,v}$ and $R_{,u}$ should be negative throughout region
3. Therefore $R$ decreases to zero and a spacelike $R=0$ singularity
should form. The numerical simulation confirmed this prediction. We
presented the spacetime diagram in this case (see Figure \ref{fig:run4_singulairty_diagram})
illustrating the formation of an $R=0$ singularity. 

In conclusion, we numerically tested the validity of the analytical
approximation for the backreaction metric inside an evaporating charged
black hole in various situations. All these numerical simulations
confirmed the validity of this analytical approximation.

Throughout this research we took the RSET to be of the form obtained
by assuming RN background metric (in Unruh state). This was motivated
by the expectation (or hope) that this RN-based RSET should well-approximate
the RSET of the \emph{self-consistent} solution of the semiclassical
Einstein equation. It is not yet known if this expectation is realized
or not. In any case, the backreaction metric constructed here nevertheless
represents the result of the first stage in an iteration scheme aimed
to construct the mentioned self-consistent solution. A natural future
research direction would be to calculate the RSET in this backreaction
metric, and in particular to see how similar or different is it from
the RSET considered in this work. 

\subsection{Acknoledgments}

I would like to thank my advisor Amos Ori. I would also like to thank
Noa Zilberman. The financial help of the Technion is greatly appreciated.

\section{Appendix A - irregularity of ``naive $g_{1}$ metric''}

We shall show here that the geometry given by the ``naive metric'',

\begin{equation}
ds^{2}=\left(1-\frac{2M(v)}{r}+\frac{Q(v)^{2}}{r^{2}}\right)du\;dv+r^{2}\left(d\theta^{2}+\sin^{2}\theta d\phi^{2}\right)\label{eq:naivemetric}
\end{equation}
is inherently irregular at the event horizon. 

First of all, the determinant of the metric Eq. (\ref{eq:naivemetric})
is zero at $r=r_{+}(v)$ because

\begin{equation}
1-\frac{2M(v)}{r_{+}(v)}+\frac{Q(v)^{2}}{r_{+}(v)^{2}}=0
\end{equation}
there. Unlike the standard Reissner--Nordström geometry in double
null Eddington coordinates, in our case (where $M$ and $Q$ depend
on v) this irregularity of the metric cannot be fixed by Kruskalization,
as we shall now show.

Start by expanding the metric around $r_{+}$ (at fixed $v$):

\begin{equation}
\left.\frac{\partial}{\partial r}\left(1-\frac{2M(v)}{r}+\frac{Q(v)^{2}}{r^{2}}\right)\right|_{r=r+}=\frac{2M}{r_{+}^{2}}-\frac{2Q^{2}}{r_{+}^{3}}=2\kappa_{+}^{v}(v)
\end{equation}
so around $r_{+}$the metric is approximated (at leading order in
$r-r_{+}(v)$) by:

\begin{equation}
ds^{2}=2\kappa_{+}^{v}(v)\left(r-r_{+}(v)\right)du\;dv+r^{2}\left(d\theta^{2}+\sin^{2}\theta d\phi^{2}\right)
\end{equation}
Around $r_{+},$along a $v=const$ line, $r_{*}$is approximated by
(using Eq. (\ref{eq:rstar naive metric})):

\begin{equation}
\frac{u+v}{2}\equiv r_{*}=\frac{1}{2\kappa_{+}^{v}(v)}\log\left|r-r_{+}\right|+const(v)
\end{equation}
Solving for $r-r_{+}$:

\begin{equation}
r-r_{+}=e^{2\kappa_{+}^{v}(v)r_{*}}const(v)=e^{\kappa_{+}^{v}(v)v+\kappa_{+}^{v}(v)u}const(v)\label{eq:r minus r plus at EH-1}
\end{equation}
Mark $const(v)\equiv h_{0}(v)$:
\begin{equation}
r-r_{+}=e^{\kappa_{+}^{v}(v)v+\kappa_{+}^{v}(v)u}h_{0}(v)\label{eq:r minus r plus at EH}
\end{equation}
So $g_{uv}$ is equal to:

\begin{equation}
g_{uv}=\kappa_{+}^{v}(v)\left(r-r_{+}\right)=\kappa_{+}^{v}(v)e^{\kappa_{+}^{v}(v)v+\kappa_{+}^{v}(v)u}h_{0}(v)
\end{equation}
and the overall metric is:

\begin{equation}
ds^{2}=2\kappa_{+}^{v}(v)h_{0}(v)\;e^{\kappa_{+}^{v}(v)v}dv\;e^{\kappa_{+}^{v}(v)u}du+r^{2}\left(d\theta^{2}+\sin^{2}\theta d\phi^{2}\right)
\end{equation}
The issue will arise from the factor $e^{\kappa_{+}^{v}(v)u}$, which
vanishes at the EH (where $u\rightarrow-\infty$). We shall therefore
have to transform $u$ into a corresponding Kruskal-like coordinate
$U$. But first we'll transform $v$ to $V$ Kruskal coordinate for
convenience. To this end, we need to absorb the exponent $e^{\kappa_{+}^{v}(v)v}$
into the definition of Kruskal coordinate, that is, we want to have: 

\begin{equation}
dV=e^{\kappa_{+}^{v}(v)v}dv
\end{equation}
Namely, $\frac{dV}{dv}=e^{\kappa_{+}^{v}(v)v}$, and by integrating
we get:

$V(v)=\int e^{\kappa_{+}^{v}(v)v}dv$. We get the line element:

\begin{equation}
ds^{2}=2h_{0}(v)\;\kappa_{+}^{v}(v)e^{\kappa_{+}^{v}(v)u}du\;dV+r^{2}\left(d\theta^{2}+\sin^{2}\theta d\phi^{2}\right)
\end{equation}

Note that $\kappa_{+}^{v}(v)$ is a function of $v$ and therefore
it is a function of $V$. 

We'll now build the Kruskal coordinate $U$. $\frac{\partial U}{\partial u}$
needs to absorb the problematic term $e^{\kappa_{+}^{v}(v)u}$ in
the metric. That means that $\frac{\partial U}{\partial u}=C(v)e^{\kappa_{+}^{v}(v)u}$
for some coefficient $C(v)$. From this we get by integration that
$U=\frac{C(v)}{\kappa_{+}^{v}(v)}e^{\kappa_{+}^{v}(v)u}+Const(v)$.
It turns out that both coefficients $C(v)$ and $Const(v)$ do not
affect the regularity, so we choose 
\begin{equation}
U\equiv e^{\kappa_{+}(v)u}
\end{equation}
We get that:
\begin{equation}
\frac{\partial U}{\partial u}=\kappa_{+}^{v}(v)e^{\kappa_{+}^{v}(v)u}
\end{equation}
$u$ satisfies:

\begin{equation}
u=\frac{\log U}{\kappa_{+}^{v}(v)}
\end{equation}

We'll now do the coordinate transformation from $g:(u,V)$$\rightarrow$$\widehat{g}:(U,V)$.
The derivative of $u$ with respect to $V$ isn't zero, which is why
this coordinate $U$ isn't null:

\begin{equation}
\frac{\partial u}{\partial V}=\frac{-1}{\kappa_{+}^{v}(v)^{2}}\;\frac{d}{dV}\left(\kappa_{+}^{v}(v)\right)\log U=h_{1}(V)\log U
\end{equation}
where we have defined 
\begin{equation}
h_{1}(V)\equiv\frac{-1}{\kappa_{+}^{v}(v)^{2}}\;\frac{d}{dV}\left(\kappa_{+}^{v}(v)\right)
\end{equation}
We calculate $g_{VV}$:

\begin{align}
\widehat{g}_{VV} & =g_{uV}\frac{\partial u}{\partial V}\frac{\partial V}{\partial V}=\kappa_{+}^{v}(v)h_{0}(v)Uh_{1}(V)\log U\\
 & =\kappa_{+}^{v}(v)h_{0}(v)h_{1}(V)U\log U=h_{2}(v)U\log U
\end{align}
where we have defined $h_{2}(v)$ as:

\begin{equation}
h_{2}(v)\equiv\kappa_{+}^{v}(v)h_{0}(v)h_{1}(V)
\end{equation}
Take the derivative of $\widehat{g}_{VV}$ with respect to $U$:

\begin{equation}
\widehat{g}_{VV,U}=h_{2}(v)\left(\log U+1\right)
\end{equation}
Note that as $u\rightarrow-\infty$ at the event horizon, $U\rightarrow0$,
$\log U\rightarrow-\infty$, so $\widehat{g}_{VV,U}\rightarrow-\infty$.
The $UV$ component of the new metric is:

\begin{equation}
\widehat{g}_{UV}=g_{uV}\frac{\partial u}{\partial U}\frac{\partial V}{\partial V}=\kappa_{+}^{v}(v)h_{0}(v)U\frac{1}{U\kappa_{+}^{v}(v)}=h_{0}(v)
\end{equation}
Now the metric is:
\begin{equation}
ds^{2}=h_{2}(V)U\log UdV^{2}+2h_{0}(v)dU\;dV+r^{2}\left(d\theta^{2}+\sin^{2}\theta d\phi^{2}\right)
\end{equation}
We see that while the metric is continuous and non-singular (namely
$\det\widehat{g}\neq0$) at the EH, the derivative of the metric diverges
there. 

To see that this divergence reflects a true geometrical singularity
and not just coordinate singularity, we look at the scalar curvature
of the metric. Using the \emph{Ricci package} we get:\begin{widetext}

\begin{align}
R_{curvature} & =\frac{h_{2}(v)}{Uh_{0}(v)^{2}}+\frac{4h_{2}(v)r\;r_{,U}\log U}{r^{2}h_{0}(v)^{2}}\\
 & +\frac{1}{r^{2}h_{0}(v)^{2}}\left(2h_{0}(v)^{2}+4h_{2}(v)r\;r_{,U}-4h_{0}(v)r_{,V}r_{,U}+\right)\\
 & +\frac{1}{r^{2}h_{0}(v)^{2}}\left(2Uh_{2}(v)\kappa_{+}(v)r_{,U}^{2}\log U-8h_{0}(v)r\;r_{,UV}+4Uh_{2}(v)r\;r_{,UU}\log U\right)
\end{align}
\end{widetext}We wish to show that this expression diverges as $U\rightarrow0$
at the EH. To do that, we calculate from Eq. (\ref{eq:r minus r plus at EH})
(the approximation of $r-r_{+}$ near the EH) the derivatives of $r$:

\begin{equation}
r-r_{+}=e^{\kappa_{+}^{v}(v)v+\kappa_{+}^{v}(v)u}h_{0}(v)\equiv Uh_{3}(v)
\end{equation}
where we defined $h_{3}$ as:

\begin{equation}
h_{3}(v)\equiv h_{0}(v)e^{\kappa_{+}^{v}(v)v}
\end{equation}
Therefore, the derivative $r_{,U}$ is equal to:

\begin{equation}
r_{,U}=h_{3}(v)
\end{equation}
And the derivative $r_{,V}$ is equal to:

\begin{equation}
r_{,V}=Uh_{3}'(v)
\end{equation}
The second derivatives are ($r_{,VV}$ is not needed for this analysis):

\begin{equation}
r_{,UV}=h_{3}'(v),\hfill r_{,UU}=0
\end{equation}
Substituting these derivatives into the expression for the scalar
curvature $R$ we obtain:\begin{widetext}

\begin{align}
R_{curvature} & =\frac{h_{2}(v)}{Uh_{0}(v)^{2}}+\frac{4h_{2}(v)rh_{3}(v)\log U}{r^{2}h_{0}(v)^{2}}\\
 & +\frac{1}{r^{2}h_{0}(v)^{2}}\left(2h_{0}(v)^{2}+4h_{2}(v)rh_{3}(v)-4h_{0}(v)Uh_{3}'(v)h_{3}(v)\right)\\
 & +\frac{1}{r^{2}h_{0}(v)^{2}}\left(2Uh_{2}(v)\kappa_{+}(v)h_{3}(v)^{2}\log U-8h_{0}(v)rh_{3}'(v)\right)
\end{align}

\end{widetext}As $U\rightarrow0$, the first term diverges like $\frac{1}{U}$,
the second term diverges weaker as $\log U$, and the rest is finite.
We conclude that the scalar curvature $R_{curvature}$ diverges like
$\frac{1}{U}$ at the EH, implying there is an inherent curvature
singularity there.

\section{Appendix B - determination of the initial values from $R_{,u}$,
$S_{,u}$}

We know the geometry in the near EH region 1, it is given by the $g_{1}$
metric Eq. (\ref{eq:g2_metric},\ref{eq:f g2}). We use this metric
to construct the initial conditions for the numerically evolving metric
in region 2 (and subsequently region 3). In principle, this requires
the initial values of the unknowns $R$,$S$ and their time derivatives
on a hypersurface of constant $T\equiv\frac{u+v}{2}$, namely on $T=T_{0}$.
We denote $Y\equiv\frac{v-u}{2}$. Thus the standard initial value
problem would in principle require the four functions $R$, $R_{,T}$,
$S$, $S_{,T}$ to produce a unique solution (where the partial derivatives
are taken while keeping $Y$ fixed). Because $R_{,u}=\frac{1}{2}R_{,T}-\frac{1}{2}R_{,Y}$,
and $R_{,Y}$ is in principle known along the initial hypersurface
(from the initial function $R(Y)$), we can also use $R_{,u}$ for
the initial values instead of $R_{,T}$. We prefer using $R_{,u}$
instead of $R_{,T}$ because it's easier to calculate, as $M(v)$
is constant along lines of constant $v.$ 

\begin{figure*}
\caption{Initial values rows}\label{fig:Initial-values-rows}
\includegraphics{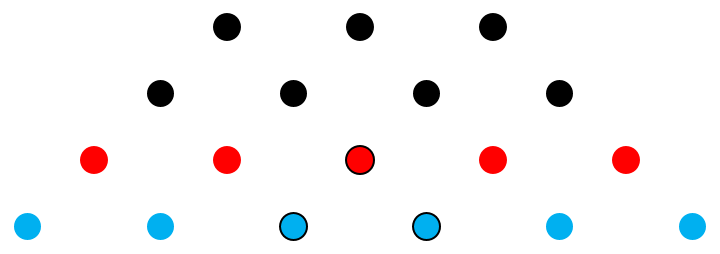}
\end{figure*}

In practice, our numerical algorithm requires two adjacent rows of
initial values of $R$ and $S$ (but not of $R_{,T}$ and $S_{,T}$),
namely the row $T=T_{0}$ and the subsequent row. We show in Figure
\ref{fig:Initial-values-rows} the two rows of initial values, $T=T_{0}$
in blue and the next one in red, which are required for evolution
into the black points. We are given $R_{,u}$ and $S_{,u}$ at $T=T_{0}$
directly from the initial values of $g_{1}$ metric, and need to translate
it into values of $R$ and $S$ on the red row. 

\begin{figure}
\caption{Initial values points}\label{fig:Initial-values-points}
\includegraphics[scale=0.5]{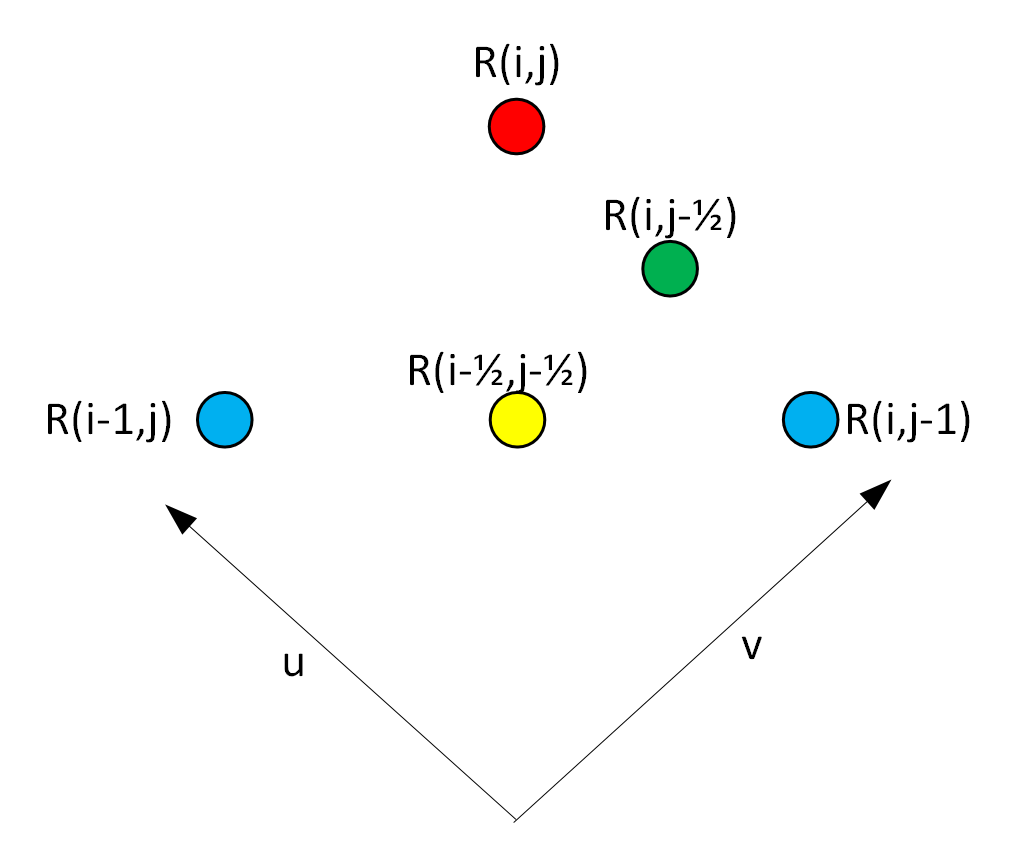}
\end{figure}

We proceed to calculate $R$ at the red point $(i,j)$ as shown in
Figure \ref{fig:Initial-values-points} by calculating $R_{,u}$ at
the green point $(i,j-\sfrac{1}{2})$, and use that, and the value
of $R$ at the blue point $(i,j-1)$, to obtain $R$ at the red grid
point $(i,j)$ (see Eq. (\ref{eq:R(i,j)=00003DR(i,j-1)+du R_,u})
below). To obtain $R_{,u}(i,j-\sfrac{1}{2})$, we first calculate
$R_{,u}$ at the yellow point $R(i-\sfrac{1}{2},j-\sfrac{1}{2})$
by averaging: \\
\begin{equation}
R_{,u}(i-\sfrac{1}{2},j-\sfrac{1}{2})=\frac{1}{2}\left(R_{,u}(i-1,j)+R_{,u}(i,j-1)\right)
\end{equation}
We know $R_{,u}(i-1,j)$ and $R_{,u}(i,j-1)$ because they are given
by $R_{,u}$ in the metric $g_{1}$. Then we calculate $R_{,u}$ at
the green point $(i,j-\sfrac{1}{2})$ by using $R_{,uv}:$\begin{widetext}

\begin{equation}
R_{,u}(i,j-\sfrac{1}{2})=R_{,u}(i-\sfrac{1}{2},j-\sfrac{1}{2})+\frac{1}{2}dvR_{,uv}(i-\sfrac{1}{2},j-\sfrac{1}{2})
\end{equation}
\end{widetext}in which we get $R_{,uv}$ by using the evolution equations.
Finally, we can calculate $R(i,j)$ from $R_{,u}$$(i,j-\sfrac{1}{2})$
and $R(i,j-1)$:

\begin{equation}
R(i,j)=R(i,j-1)+du\;R_{,u}(i,j-\sfrac{1}{2})\label{eq:R(i,j)=00003DR(i,j-1)+du R_,u}
\end{equation}
To summarize, we got $R(i,j)$ at the new red row from $R$ and $R_{,u}$
(and $R_{,uv}$) at the previous blue row:\begin{widetext}

\begin{equation}
R(i,j)=R(i,j-1)+du\left(\frac{1}{2}\left(R_{,u}(i-1,j)+R_{,u}(i,j-1)\right)+\frac{1}{2}dvR_{,uv}(i-\sfrac{1}{2},j-\sfrac{1}{2})\right)
\end{equation}
\end{widetext}The same procedure is applied to obtain $S$ at the
new red row, using $S_{,u}$ (and $S_{,uv}$). This procedure eventually
provides $R$ and $S$ (in the whole numerical grid) in the required
precision, which is second order in $du$.

\bibliographystyle{apsrev4-2}
\bibliography{article}

\end{document}